\documentclass{MITcsail}
\usepackage{caption}
\usepackage{subcaption}
\usepackage{wrapfig}
\usepackage{hyperref}
\usepackage{comment}
\usepackage{tcolorbox}
\usepackage{longtable}
\usepackage[stable]{footmisc}
\tcbuselibrary{breakable}

\newcommand{\pglobal}{p_\text{global}}
\newcommand{\plocal}{p_\text{local}}
\newcommand{\Nc}{\langle N(c) \rangle}
\newcommand{\Ncz}{\langle N(c_0) \rangle}

\definecolor{figblue}{HTML}{4169E1} 

\title{The Look-Elsewhere Effect in Searches for New Physics:\\ from Classical Methods to Machine Learning}
\author
{
Gaia Grosso~$^{1,2,3}$\footnote{Leading author, Correspondence E-mail: \href{mailto:gaiag795@mit.edu}{gaiag795@mit.edu}}, 
Sergio Rodríguez-Benítez~$^{4}$,
Louis Lyons~$^{5,6}\footnote{Correspondence E-mail: \href{mailto:l.lyons@physics.ox.ac.uk}{l.lyons@physics.ox.ac.uk}}$\\
\vspace{1em} 
\normalfont{\small $^{1}$NSF AI Institute for Artificial Intelligence and Fundamental Interactions}\\
\normalfont{\small $^{2}$MIT Laboratory for Nuclear Science, Cambridge, MA}\\
\normalfont{\small $^{3}$School of Engineering and Applied Sciences, Harvard University, Cambridge, MA}
\\
\normalfont{\small $^{4}$Instituto de Física Teórica UAM/CSIC, Campus de Cantoblanco, Madrid, Spain}
\\
\normalfont{\small $^{5}$Department of Physics, Blackett Laboratory, Imperial College London, UK}
\\
\normalfont{\small $^{6}$Department of Physics, Oxford University, UK}
}
\begin{document}

\maketitle
\thispagestyle{firstpagestyle} 

\begin{abstract}

The look-elsewhere effect (LEE) arises when the location of an effect in parameter space is not specified in advance: scanning many locations inflates the significance of the largest excess, and the local $p$-value must be corrected into a global one. A well-established framework, rooted in the work of Davies and operationalised for particle physics by Gross and Vitells, provides this correction for scans over continuous parameters. Some aspects, however, remain unsettled. The definition of ``elsewhere'' is ambiguous and ranges from a single distribution to an entire experimental programme. Moreover, the analytical shortcuts of the classical framework do not carry over to searches based on machine learning (ML), which are affected by an implicit post-selection bias that depends on the model's resolution and expressiveness.
This review covers the LEE in searches for new physics at colliders, from its statistical foundations to these modern challenges. We discuss the corrections for continuous scans and for multiple discrete tests, illustrated with simple numerical examples, and the interaction between ML and the LEE, which we connect to the broader literature on selective inference. We highlight open problems, including the computational cost of empirical calibration for complex models and the development of analytical shortcuts for ML test statistics, and offer recommendations for reporting discovery significance in both classical and ML-based searches in particle physics.
\end{abstract}
\tableofcontents
\section{Introduction}

In experimental science, the strength of evidence against a hypothesis is commonly quantified by a \emph{p-value}: the probability, under the assumption that the hypothesis is true, of observing data at least as extreme as what was actually observed. A small p-value indicates that the data are unlikely under the hypothesis, and is taken as evidence against it. By convention, the hypothesis being tested is the \emph{null hypothesis} $H_0$ --- typically the prediction of a known theory or the absence of a new effect --- and a sufficiently small p-value is interpreted as a \emph{discovery}: evidence that the null is wrong and something new is present.

When the location of the new effect in some parameter space is known in advance --- for example, if theoretical physics predicts a particle at a specific mass --- the p-value is computed at that mass location and the interpretation is straightforward. But when the location is not known, the analyst must scan over many possible values. The \emph{local p-value} ($\plocal$), computed at the most significant location within the scan, overstates the evidence, because even under the null hypothesis, scanning many locations makes it likely that at least one will produce a small p-value by chance. The \emph{global p-value} ($\pglobal$) corrects for this by answering the question: what is the probability, under the null, of finding an excess \emph{at least as significant as the one observed anywhere in the scan range}?

The problem is known by different names in different communities. Physicists call it the \emph{look-elsewhere effect} and quantify it through a trial factor. Statisticians frame the same phenomenon as \emph{hypothesis testing when a nuisance parameter is present only under the alternative}~\cite{davies1977, davies1987} or, more broadly, as \emph{selective inference}~\cite{lee2016exact, berk2013valid}: any situation in which the data are used both to \emph{select} what to test and to \emph{perform} the test. The LEE is one instance of this broader class of post-selection problems --- the selection being ``pick the parameter value that maximises the test statistic.'' Other instances arise across the sciences: variable selection in regression (which features are significant?~\cite{freedman1983note}), subgroup analysis in clinical trials (which patient groups benefit from treatment?~\cite{assmann2000subgroup}), model selection (which model fits best?~\cite{burnham2002model}), and region selection in neuroimaging (which brain areas are active?~\cite{kriegeskorte2009circular}). In all cases the statistical pathology is the same --- the more places you look, the more likely you are to find a significant-looking statistical fluctuation --- and the solutions (conditioning on the selection, splitting the data, or constraining the selection's expressiveness) are shared.

In the physical sciences, the problem was first addressed systematically in the context of signal detection. Davies derived sharp bounds on the global p-value in terms of the level crossings of a stochastic process~\cite{davies1977, davies1987}, and Gross and Vitells (GV) translated this into a practical recipe for high-energy physics, enabling the trial factor to be estimated from a modest number of pseudo-experiments~\cite{gross2010}. These tools are now standard in collider searches for narrow resonances, and variants have been adopted in astrophysics (including searches for dark matter annihilation lines in gamma-ray spectra~\cite{Algeri:2016gtj} and, with a two-dimensional extension~\cite{Vitells:2011da}, neutrino point sources in the sky), and in cosmology (including searches for gravitational wave signals from compact object mergers~\cite{abbott2016gw150914}, for signatures of inflation in Planck data~\cite{Bayer:2020pva}, and for exoplanets~\cite{Bayer:2021lhk}). The LEE has been an important part of the statistical discussion surrounding major discoveries: the Higgs boson was reported by both ATLAS~\cite{ATLAS:2012yve} and CMS~\cite{CMS:2012qbp} using local significances of $5.9\sigma$ and $5.0\sigma$ respectively, with global p-values quoted separately as supporting information.

Yet the problem is not fully settled, for at least three reasons.

First, the definition of ``elsewhere'' is ambiguous. A search for a resonance at an unknown mass involves a LEE within a single distribution, but what about a programme of searches across many final states, or across all analyses at a laboratory? Different definitions of the search space lead to different trial factors, and the choice involves scientific judgment as much as statistics. The general searches conducted at the Tevatron~\cite{CDF:2008voc} and the LHC~\cite{ATLAS:2018zdn, CMS:2020zjg} have confronted this question operationally, scanning hundreds of final states and thousands of kinematic distributions for deviations from the Standard Model. Their solution --- running the full analysis pipeline on background-only pseudo-experiments across all channels simultaneously --- calibrates the global p-value exactly but at substantial computational cost.
The scope of ``elsewhere'' also has a temporal dimension: as an experiment accumulates data over time, an excess can appear at different times as well as at different locations, a ``look-elsewhen'' effect that a single scan of the final dataset does not capture. The e-value framework~\cite{grunwald2020safe, wang2022false} addresses this axis directly, providing inference that remains valid at any stopping time by construction (Section~\ref{sec:evalues}).

Second, the classical framework assumes specific structure that may not hold in general. The Davies bound requires the test statistic to be a Gaussian or $\chi^2$ process; the Gross-Vitells extrapolation relies on the analytical scaling of this process at high significance. When these assumptions hold, the resulting corrections are tight and computationally cheap. When they do not --- as is the case for many machine learning-based test statistics --- one must fall back on cruder tools such as the Bonferroni correction, which controls the family-wise error rate but is needlessly conservative for correlated tests, or the Benjamini-Hochberg procedure~\cite{bh_1995} and its Benjamini-Yekutieli extension to arbitrary dependence~\cite{benjamini2001control}, which control the false discovery rate under weaker assumptions but have seen little adoption in particle physics, where selections are conventionally characterised by signal efficiency and background mis-identification rate rather than by false-discovery control.

Third, machine learning is transforming the way searches are conducted, and it interacts with the post-selection problem in a particularly acute way. Classical searches use a fixed, pre-specified signal template that does not adapt to the data, so the selection (where to look) is independent of the test (how significant is the excess). Signal-agnostic machine learning-based searches break this independence by fitting a flexible model globally to the full phase space in a single pass. The model is richer and potentially more sensitive, but its flexibility means it can overfit to statistical fluctuations anywhere in the data, producing an implicit LEE whose intensity depends on the model's resolution, regularisation, and architecture, none of which are accounted for by the Davies/GV framework. The standard remedy --- calibrating the p-value via pseudo-experiments that run the full pipeline on background-only data --- remains valid but becomes computationally expensive when each pseudo-experiment requires retraining the model. Approximations such as train/test splitting and $k$-folding have been proposed~\cite{hein2025}, but their sensitivity cost relative to full calibration has not been established.

This paper reviews the look-elsewhere effect as it arises in searches for new physics at colliders, from its classical statistical foundations through its modern incarnations in multiple testing theory, and signal-agnostic machine learning-based searches. The statistical problem is not specific to particle physics, and throughout the review we point to its counterparts in astrophysics, cosmology, genomics and clinical trials where they clarify the discussion. The conventions, the illustrative examples and the recommendations target the high-energy physics community.

Section~\ref{sec:elsewhere} discusses the ambiguity of ``elsewhere'' and the different levels at which the LEE arises. The classical continuous and discrete correction methods are reviewed in Sections~\ref{sec:classical} and~\ref{sec:multipletesting}. Section~\ref{sec:evalues} then introduces e-values as an alternative framework. We turn in Section~\ref{sec:ml} to the specific challenges posed by machine learning methods: the implicit LEE that arises when a flexible model is fitted to the data, the role of model resolution, calibration strategies, and the additional LEE incurred when several models are applied to the same data. Section~\ref{sec:experiments} proposes a few illustrative examples of the LEE problem and compares different methods to account for it. Finally, Section~\ref{sec:summary} collects recommendations and open problems.

\section{Where is ``Elsewhere''?}
\label{sec:elsewhere}

The look-elsewhere effect corrects for the fact that a signal could have appeared somewhere other than where it was observed. But this immediately raises a question: what counts as ``somewhere''? Different choices of the search space lead to different trial factors, and the choice is not always obvious. 
It is useful to distinguish several nested levels of ``elsewhere'', each with different practical implications. Below, we describe just three.

\subsection{Case (i): within a single distribution}
\label{sec:case_i}

This is the standard Davies/Gross-Vitells setting that will be described in Section~\ref{sec:classical}: one scans a single kinematic distribution (typically an invariant mass spectrum) over a range of values, and the LEE arises from the freedom to find the most significant excess anywhere in that range. Even in this simplest case, the definition of the scan range involves choices that affect the trial factor.

The upper end of the range should be set by where the analysis loses sensitivity. In the Higgs search, a useful criterion is the highest mass at which a $5\sigma$ peak would still be considered a plausible signal. For a Standard Model Higgs search, this means the mass at which the predicted cross-section becomes too small for the analysis to have any sensitivity.

The lower end raises a different issue: should previously excluded regions contribute to the LEE? For the Higgs boson discovery, the LEP experiments had excluded $m_H < 114.4$~GeV at 95\% confidence. Should the LHC search range start at 114.4~GeV, thereby reducing the trial factor? There is an argument either way. If a significant signal were observed in the LEP-excluded region, most physicists would still want to claim it as a discovery --- the LEP exclusion was only at 95\%, not $5\sigma$. Including the LEP-excluded region in the scan range is therefore the conservative and arguably a better choice. The same logic applies more strongly to previous exclusions from the LHC experiments themselves: using last year's exclusion to shrink this year's scan range amounts to ignoring the new data in those mass regions, which is hard to justify.
The same question arises in any scan-based search: in a neutrino point-source analysis, should the scan cover the full sky or only the Galactic plane? In a genome-wide association study (GWAS), which tests millions of single-nucleotide polymorphisms (SNPs) --- individual positions in the genome where individuals differ by a single DNA base --- for association with a trait or disease, should previously tested SNPs be included in the current study's multiple testing correction?

\subsection{Case (ii): across multiple analyses}
\label{sec:case_ii}

When many analyses are conducted in parallel --- for example, all SUSY searches at a given experiment, or searches for the same resonance in different decay channels (diphoton, four-lepton, $b\bar{b}$, etc.) --- a significant result in any one of them must in principle be penalised for the total number of places where a fluctuation could have appeared. This is a different kind of LEE than case~(i): the ``elsewhere'' is not within a single distribution but across an ensemble of distributions or final states. For searches in different decay channels, this across-channel penalty applies only when the relative signal rates (branching ratios) are unknown and each channel is tested independently; if they are fixed by the signal model, the channels are instead combined in a single fit with their relative normalisations constrained, and there is no multiplicity over channels to correct for.

This is also the setting addressed by the model-independent general searches that have been carried out at the Tevatron and the LHC. The CDF Global Search~\cite{CDF:2008voc} partitioned high-$p_T$ events into 399 exclusive final states and scanned 5036 invariant-mass distributions using three complementary algorithms: Vista (a global comparison of event populations and kinematic shapes), the BumpHunter~\cite{Choudalakis:2011qn} (a scan for narrow resonances in invariant mass distributions), and SLEUTH~\cite{knuteson2001sleuth} (a scan for excesses in the high-$\sum p_T$ tails). The ATLAS experiment developed a similar strategy~\cite{ATLAS:2018zdn}, where events were classified into 704 exclusive event classes, 686 of which were retained for the statistical analysis, and two kinematic distributions, the effective mass and the total invariant mass, were scanned in each. The CMS MUSiC analysis~\cite{CMS:2020zjg} follows the same philosophy, searching hundreds of final states and multiple kinematic distributions for deviations from the Standard Model prediction.

In all of these analyses, the case-(ii) LEE is handled by the same mechanism as case~(i): background-only pseudo-experiments that run the \emph{full} pipeline across \emph{all} channels simultaneously. The BumpHunter computes a local p-value for the most significant bump in each mass distribution, then generates pseudo-experiments in which the same scanning algorithm is applied to every distribution, producing a null distribution of the most significant p-value across all channels. The global p-value is read off from the tail of this distribution. MUSiC follows a two-level procedure: for each distribution, a local p-value $\tilde{p}$ is computed via $\mathcal{O}(10^5)$ pseudo-experiments that account for the within-distribution LEE (case~(i)); then a second layer of pseudo-experiments asks how often \emph{any} distribution produces a $\tilde{p}$ as small as the observed minimum, accounting for the across-distribution LEE (case~(ii)).

Choudalakis~\cite{Choudalakis:2011qn} formalised this nested structure through the concept of ``hypertests'': a hierarchy of hypothesis tests in which the test statistic at each level is the most significant result from the level below. The trial factor at each level is estimated by pseudo-experiments at that level. This framework makes explicit that the case-(ii) LEE is not fundamentally different from case~(i) --- it is simply an additional layer of the same maximisation-over-nuisance-parameters problem, with the ``nuisance parameter'' being the identity of the channel rather than the mass within a channel.

The computational cost of this approach scales with the number of channels times the number of mass bins times the number of pseudo-experiments. For the CDF search (399 final states, $\sim$5000 mass variables, $\mathcal{O}(10^5)$ pseudo-experiments), this was already demanding; for future searches with ML-based anomaly detection across high-dimensional feature spaces, it may become prohibitive without approximation schemes analogous to the Gross-Vitells extrapolation.

Outside particle physics, the canonical case-(ii) setting is the genome-wide association study, where millions of genetic variants are tested for association with a disease and the LEE spans all variants simultaneously~\cite{storey2003statistical}.

\subsection{Case (iii): across all analyses at a laboratory}
\label{sec:case_iii}

The more expansive definition of ``elsewhere'' would include all analyses conducted at a given experiment or laboratory. If an experiment runs hundreds of searches for different new physics scenarios, a $5\sigma$ fluctuation in any one of them is less surprising than a $5\sigma$ fluctuation in a single pre-specified search. 
From a purely statistical standpoint, any anomaly found anywhere in a large experimental programme should be penalised for the total number of places where it could have been found.

In practice, nobody applies a formal LEE at this level. The implicit assumption is that each analysis has a specific physics motivation that distinguishes it from a random scan: a search for stop quarks is not the same as a search for dark photons, even if both scan a mass spectrum. 
The $5\sigma$ discovery threshold serves as a large safety margin that implicitly absorbs this unquantified case-(iii) LEE (see Section~\ref{sec:thresholds}).

However, model-agnostic ML searches blur the boundary between targeted and untargeted analyses. A ML anomaly detector that scans the full phase space across all observables simultaneously is operationally closer to the case-(iii) extreme than to a single targeted search. It is searching for ``anything unusual anywhere'', and the LEE should in principle reflect this. The general searches described in Section~\ref{sec:case_ii} partially address this by scanning many final states, but they still rely on pre-defined event classifications and kinematic variables that are typically analysed in univariate setting. ML methods that learn features directly from multidimensional data go further, scanning an implicitly defined and potentially much larger space of possible anomalies. In this sense, ML-based anomaly detection makes the case-(iii) LEE quantitatively relevant for the first time: the effective search space of a single ML analysis may rival that of an entire programme of targeted searches, and the appropriate trial factor correction is an open problem that we address in Section~\ref{sec:ml}.\\

The difference among the three cases can be understood in terms of \emph{whose} LEE is being computed. A graduate student performing a single analysis defines ``elsewhere'' as the scan range of that analysis. The convener of a search programme (e.g., all SUSY searches at CMS) might define it as the union of all analyses in the programme. The spokesperson of the experiment might consider all analyses conducted by the collaboration. The director-general of the laboratory might, in principle, consider all analyses at all experiments. Each of these defines a different and progressively larger search space, leading to a progressively larger trial factor for the same local excess. The same $3\sigma$ fluctuation that is noteworthy from the graduate student's perspective may be entirely unremarkable from the spokesperson's.

In practice, the convention in particle physics is to report the global p-value using the analyst's own definition of the search range. This is the narrowest and most precisely defined scope, corresponding to case~(i) or a well-delineated instance of case~(ii). It is reproducible, unambiguous, and directly tied to the analysis as published. Broader corrections --- for the full programme of searches, or for the experiment as a whole --- are not applied formally, but are implicitly absorbed by the $5\sigma$ discovery threshold (Section~\ref{sec:5sigma}).

What is essential, regardless of the convention adopted, is that the analyst \emph{states explicitly which definition of ``elsewhere'' was used} when reporting a global p-value. A global p-value without this specification is incomplete: the reader cannot assess whether the correction is appropriate for their own level of scepticism. This is especially important for ML-based anomaly detection, where the effective search space is typically much larger than in a classical single-channel scan, and the implicit case-(iii) LEE may no longer be negligible.

\subsection{The time axis: ``look-elsewhen''}
\label{sec:elsewhen}

The three cases above concern the \emph{spatial} scope of the search, namely how much of the data space counts as ``elsewhere.'' A distinct dimension of the LEE arises from the \emph{temporal} evolution of the analysis. As a collider accumulates data over time, its sensitivity crosses the $5\sigma$ threshold at different masses at different times. In principle, at each moment when sensitivity reaches $5\sigma$ for a given mass, there is a chance of a $5\sigma$ fluctuation right there. The time axis introduces an additional dimension of ``elsewhere'' that is not accounted for by a single scan of the final dataset.

The issue has been discussed informally within the HEP community --- sometimes under the name ``keep-looking bias''~\cite{dorigo2010keeplooking} --- motivated by cases where a marginal signal grew to $5\sigma$ with surprisingly little additional data. It is the classical problem of \emph{repeated significance testing}, or \emph{sequential monitoring}, and the theory of sequential analysis quantifies it: by the law of the iterated logarithm (see Ref.~\cite{demortier2014testing} for a discussion in the particle-physics context), the maximum significance observed by monitoring a cumulative dataset up to sample size $n$ grows only as $\sqrt{2\log\log n}$, which is extremely slow. With a $5\sigma$ discovery threshold, the inflation is therefore modest, consistent with the conclusion reached in internal studies within the CDF and CMS collaborations, and in current practice it is further limited because analyses are performed on a few discrete datasets (``Run~1,'' ``Run~2'') rather than continuously. The effect is nonetheless nonzero, and it can matter for lower thresholds or for searches where the significance stays near the boundary for extended periods.

Where a correction is needed, the standard solutions come from the clinical trials literature: group sequential methods~\cite{pocock1977group, o1979multiple}, which pre-specify a small number of interim analyses with adjusted thresholds, and alpha-spending functions~\cite{lan1983discrete}, which allocate the type~I error budget over time without fixing the analysis schedule in advance.

Although the look-elsewhen effect is small in current practice, two developments may make it more relevant. First, ML-based anomaly detection enables continuous monitoring of data streams, opening the path to more systematic sequential searches, in which the number of looks is no longer limited to a few discrete datasets. Second, the e-value framework~\cite{vovk2021values} provides inference that is valid at any stopping time by construction, so the look-elsewhen problem does not arise, without the pre-specified schedule or error budget that the $p$-value corrections require. We discuss e-values and their applicability to physics in Section~\ref{sec:evalues}.

\subsection{Discovery and exclusion thresholds}
\label{sec:thresholds}
The definition of ``elsewhere'' directly affects how significant an observed excess must be before it is considered a discovery, and whether a correction is needed at all when setting exclusion limits. This subsection discusses the conventions used in particle physics for both cases, and the reasoning behind the asymmetry between them.
Throughout, we quote the significance of an excess in two interchangeable ways: as a p-value $p$ --- the probability under $H_0$ of a fluctuation at least as extreme --- and as the equivalent Gaussian significance $Z = \Phi^{-1}(1-p)$, where $\Phi$ is the cumulative distribution function of the standard normal, expressed in units of $\sigma$. Unless stated otherwise, significances are one-sided.

\subsubsection{Discovery: the 5$\sigma$ convention}
\label{sec:5sigma}
The $5\sigma$ threshold for claiming a discovery in particle physics is not derived from first principles; it is a convention that reflects accumulated experience with trial factors and systematic uncertainties. The trial-factor argument goes back to Rosenfeld's 1968 survey of claimed meson resonances~\cite{rosenfeld1968there}, written when new hadrons were sought as bumps in effective-mass histograms from bubble-chamber experiments. Rosenfeld estimated that the field was producing ``at least 100{,}000 potential bumps per year''. Since a one-sided $4\sigma$ fluctuation has a probability of about 1 in 32{,}000, one ``should expect several $4\sigma$ and hundreds of $3\sigma$ fluctuations'' every year --- enough to account for the many claimed resonances that had failed to be confirmed. The ``elsewhere'' was the entire output of the community, every histogram of every experiment inspected by eye, and it was further inflated by choices made after looking at the data: a $2\sigma$ or $3\sigma$ fluctuation ``can easily be amplified to `$4\sigma$' or `$5\sigma$'; all it takes is a little enthusiasm''. Alvarez made the same point in his Nobel lecture~\cite{alvarez1969recent}, describing G.~Lynch's program GAME, which hid the real histogram among a hundred simulated from a smooth curve fitted to the data: in its first use, the experimenter picked out two of the simulated histograms as significant and dismissed all the others, his own included. Rosenfeld's conclusion was that the theorist should ``wait for nearly $5\sigma$ effects'', and that for the experimenter ``any bump less than about $5\sigma$ constitutes only a call for a repeat of the experiment''.

In today's vocabulary, this is a look-elsewhere correction applied to the field as a whole, at the broadest level of our case (iii) (Section~\ref{sec:case_iii}). A local significance of $5\sigma$ corresponds to a p-value of $2.87 \times 10^{-7}$: with $10^5$ potential bumps per year, a $5\sigma$ fluctuation was expected only about once every 35 years. 
The high threshold was motivated by the asymmetric cost of errors. A false discovery is scientifically catastrophic: a new particle is announced, the community reorganises its theoretical and experimental programme around it, and it turns out to be a fluctuation. 
In the 60's the trial factor was understood qualitatively but could not be computed for an individual result, and a large margin on the local significance was the only safeguard available. 

Today, the trial factor at a search-level, our case (i) (Section~\ref{sec:case_i}), can be computed explicitly (see Section~\ref{sec:classical}), but the threshold is still applied to the significance prior to this correction, the local significance. At the analysis-level, trial factors of several hundred are common, and $5\sigma$ local can correspond to $3$--$4\sigma$ analysis-level global. For the Higgs discovery, for instance, the ATLAS experiment reported a local significance of $5.9\sigma$ ($\plocal = 1.7 \times 10^{-9}$) and a global significance of $5.1\sigma$ ($\pglobal = 1.7 \times 10^{-7}$) over the search range 110--600~GeV~\cite{ATLAS:2012yve}, corresponding to a trial factor of approximately 100.  
While the case-(i) LEE can now be controlled, the case-(ii) and case-(iii) have not become smaller or easier to quantify since Rosenfeld's time, still justifying a $5\sigma$ threshold after the analysis-level correction is performed. For this reason, we argue in Section~\ref{sec:summary} that the discovery threshold is better applied to the \emph{global} significance of the analysis, so that the trial factor is corrected explicitly rather than absorbed into the margin, and the margin is the same for every BSM analysis.

Moreover, it is worth noting that the threshold can depend on context. When the existence of a process is already predicted by the Standard Model and the question is whether it has been observed --- as in the case of Higgs boson pair production (HH) --- a lower threshold may be considered appropriate since there is generally no LEE. Observing HH is confirming a predicted process, not discovering a new particle: the prior expectation that HH exists (given the established Higgs mechanism) is very high, whereas the prior expectation of an arbitrary new resonance at an arbitrary mass is low. For such confirmatory measurements, $3\sigma$ is often accepted as ``evidence'' even if a $5\sigma$ ``discovery'' has not yet been reached. Even though it is expressed in frequentist language, the distinction is ultimately a Bayesian one, as the threshold reflects a judgment about prior plausibility.

Another case that requires different treatment is anomaly detection, which serves a different scientific function than a targeted search. Anomaly detection is a \emph{screening} tool, rather than a \emph{discovery} tool. Its role is to identify anomalies without a signal hypothesis, and to warrant targeted follow-up searches with a pre-specified signal hypothesis.
The follow-up is then a standard case-(i) analysis with a well-defined and smaller LEE, performed on independent data\footnote{otherwise the screening step is itself a selection that the follow-up must account for (Section~\ref{sec:postselection})} and held to the usual $5\sigma$ standard on its global p-value. We return to this two-tier approach in Section~\ref{sec:summary}.

The question of which significance level to adopt is not specific to particle physics. Fisher presented the $5\%$ level as a matter of convenience rather than principle~\cite{Fisher1930}, and the Neyman--Pearson framework makes the choice of level explicitly a trade-off between the two types of error, with weights that depend on the problem at hand~\cite{NeymanPearson1933}; the American Statistical Association's 2016 statement on p-values reached the same conclusion and declined to recommend any universal threshold~\cite{Wasserstein2016}. 

\subsubsection{Exclusion: why 95\%, and why no LEE?}
\label{sec:exclusion}
Exclusion limits in particle physics are conventionally set at 95\% confidence level, corresponding to $1.645\sigma$ for a one-sided test. This is far less stringent than the $5\sigma$ required for discovery, and the asymmetry reflects the different consequences of being wrong. A false exclusion --- declaring ``there is no signal at mass $m_0$'' when in fact there is one --- is an error of omission: someone else will find the signal later, or the same experiment will find it with more data. The scientific programme is not damaged; one simply needs to look again. 
The 95\% confidence level is not specific to HEP; it is the standard convention across most of experimental science, originating with Fisher's conventional choice of $\alpha = 0.05$~\cite{fisher1970statistical} and long predating the modern limit-setting machinery. The technique used to construct exclusion limits in particle physics is the CLs method of Junk~\cite{junk1999confidence} and Read~\cite{read2002presentation}, developed at LEP and since used in essentially all collider searches; CLs limits are conventionally reported at 95\%, but the method itself is independent of the chosen confidence level.

Importantly, the LEE applies to \emph{discovery} but not to \emph{exclusion}. This asymmetry was discussed by Lyons~\cite{lyons2010} and given a formal treatment by Lockhart~\cite{lockhart2010}, who proved that when testing multiple hypotheses, the relevant trial factor is at most $k$, where $k$ is the number of null hypotheses that can be simultaneously true. For a particle that can exist at only one mass, $k = 1$ and no multiplicity correction arises. The intuition is straightforward: when setting an upper limit at a specific mass $m_0$, one is testing the hypothesis ``there is a signal at $m_0$ with cross-section $\sigma$'' against the null, and the test is conducted at that specific mass. The fact that the signal could have been at a different mass is irrelevant to the question of whether it is excluded at $m_0$. Each mass point is tested independently, and the resulting exclusion contour does not require a trial factor correction. As Cowan noted in his CERN lectures~\cite{cowan2012}, there is however an analogous issue of ``spurious exclusion'' (excluding signal models even in the presence of a signal) but this is a distinct problem from the LEE.

This asymmetry sometimes causes confusion: the same analysis may report both a local excess (subject to the LEE) and an exclusion limit (not subject to the LEE) at different mass points, and the two statements carry different statistical meanings. The LEE penalises the discovery claim precisely because the analyst did not specify the mass in advance; the exclusion is valid at each mass individually regardless of what happens elsewhere.
\section{Single search with a scan parameter}
\label{sec:classical}

\subsection{The problem: nuisance parameter present only under the alternative}

Consider a hypothesis test in which the null hypothesis $H_0$ (e.g., background only) is tested against a family of alternatives indexed by a continuous parameter $\theta$ (e.g., the mass of a new resonance). Under $H_0$, the data do not depend on $\theta$, so the parameter is meaningless when no signal is present. Under the alternative, $\theta$ specifies where the signal appears. This asymmetry --- a parameter that enters the model only under the alternative --- is the defining feature of the look-elsewhere problem, first formalised by Davies~\cite{davies1977, davies1987}.

If $\theta$ were known in advance, one would compute the appropriate test statistic $S(\theta)$ and compare it to its known null distribution. 
For a profile likelihood ratio test, Wilks' theorem~\cite{wilks1938large} guarantees that $S(\theta)$ follows a $\chi ^{2}$ distribution asymptotically provided standard regularity conditions are met: the nested hypotheses are well-defined, the log-likelihood function is sufficiently smooth, the true parameter values lie in the interior of the parameter space, and the parameters are identifiable under the null.
The problem is that $\theta$ is unknown: one must scan over all possible values of $\theta$ and report the most significant result. The test statistic actually used is therefore the supremum
\begin{equation}
M = \sup \{ S(\theta) :\, L \leq \theta \leq U \}\,,
\label{eq:Msup}
\end{equation}
where $[L,\,U]$ is the range of values over which $\theta$ is scanned. The distribution of $M$ under $H_0$ is not $\chi^2$, because maximising $S(\theta)$ over $\theta$ inflates its value relative to any fixed-$\theta$ test. Wilks' theorem does not apply when the parameter being maximised over is unidentifiable under the null: the maximum likelihood estimator of $\theta$ has no consistent target under $H_0$ and simply follows the noise.

The naive p-value --- treating $M$ as if $\theta$ had been fixed a priori --- is the local p-value $\plocal = P(S(\theta_0) > M \mid H_0)$ for some reference point $\theta_0$. This underestimates the true false positive rate because $M$ is the maximum of many correlated values. The correct quantity is the global p-value
\begin{equation}
\pglobal = P\!\left(\sup_\theta S(\theta) > M \,\big|\, H_0\right)\,.
\label{eq:pglobal}
\end{equation}
The look-elsewhere effect is the ratio between the two: $\pglobal > \plocal$ whenever the scan range admits more than one effectively independent value of $\theta$.

\subsection{The Davies bound and the upcrossing formula}
\label{subsec:davies}
Davies~\cite{davies1977} derived a sharp upper bound on the global p-value by connecting the problem to the theory of level crossings of stochastic processes. As $\theta$ varies continuously over $[L,\,U]$, the test statistic $S(\theta)$ traces a random curve. Under $H_0$ this curve is a Gaussian process (in the normal case) or a chi-squared process (in the likelihood ratio case). An \emph{upcrossing} of the level $c$ occurs each time $S(\theta)$ crosses from below $c$ to above $c$ as $\theta$ increases. Each upcrossing represents an effectively independent chance for a background fluctuation to mimic a signal. The bound is
\begin{equation}
\pglobal \;\leq\; \plocal \;+\; \Nc\,,
\label{eq:davies_bound}
\end{equation}
where $\Nc$ is the expected number of upcrossings of the level $c$ by $S(\theta)$ over the scan range. The bound is tight for large $c$, which is the regime of interest for significance claims.

In the normal case, where $S(\theta)$ has a standard normal distribution for each fixed $\theta$, the 1977 paper gives the explicit formula
\begin{equation}
P\!\left(\sup_\theta S(\theta) > c\right) \;\leq\; \Phi(-c) \;+\; \frac{e^{-c^2/2}}{2\pi} \int_{L}^{U} \!\sqrt{-\rho_{11}(\theta)}\; d\theta\,,
\label{eq:davies_normal}
\end{equation}
where $\Phi$ is the standard normal cumulative distribution function, $\rho(\theta',\theta) = \mathrm{corr}\left[S(\theta'),S(\theta)\right]$ is the autocorrelation of the process, and $\rho_{11}(\theta) = [\partial^2\rho(\theta',\theta)/\partial\theta'^2]_{\theta'=\theta}$ measures how rapidly this correlation decays (i.e., how fast the test statistic changes as $\theta$ varies). The integral $\int\!\sqrt{-\rho_{11}}\,d\theta$ encodes the total ``oscillation budget'' of the process over the scan range.\\

The 1987 paper~\cite{davies1987} extends the result to the chi-squared case, where $S(\theta) = Z_1^2(\theta) + \cdots + Z_s^2(\theta)$ is a sum of squared Gaussian components with $s$ degrees of freedom. The bound takes the same form as~\eqref{eq:davies_bound} but with $\Nc$ expressed in terms of the eigenvalues $\lambda_1(\theta),\ldots,\lambda_s(\theta)$ of a matrix that captures the conditional variability of the derivative of the process given its current value. These eigenvalues generalise $-\rho_{11}$ to the vector setting. The 1987 paper also introduces a practical approximation: one can estimate the integral in~\eqref{eq:davies_normal} directly from the observed curve $S(\theta)$ by computing its \emph{total variation}
\begin{equation}
V = \int_{L}^{U} \!\left|\frac{\partial S(\theta)}{\partial\theta}\right| d\theta = |S(\theta_1) - S(L)| + |S(\theta_2) - S(\theta_1)| + \cdots + |S(U) - S(\theta_n)|\,,
\label{eq:total_variation}
\end{equation}
where $\theta_1,\ldots,\theta_n$ are the successive turning points of $S(\theta)$. The global p-value estimate for the normal case is then
\begin{equation}
\pglobal \;\approx\; \Phi(-M) \;+\; \frac{V\, e^{-M^2/2}}{\sqrt{8\pi}}\,,
\label{eq:davies_approx}
\end{equation}
where $M$ is the observed maximum of $S(\theta)$. This requires no analytical knowledge of the autocorrelation function: one simply evaluates $S(\theta)$ on a grid, reads off $M$ and $V$, and plugs them in. Davies verified through simulations that~\eqref{eq:davies_approx} provides an excellent approximation, and that the ``naive test'' --- which ignores the scan and uses standard normal critical values --- is catastrophically miscalibrated (e.g., a nominal 5\% test produces a 20.8\% false positive rate in the two-phase regression example of~\cite{davies1987}).

\subsection{The Gross \& Vitells procedure}
\label{sec:gv}

Gross and Vitells~\cite{gross2010} translated Davies' framework into a practical recipe for HEP bump hunts, where the test statistic is the profile likelihood ratio $q(\theta)$ scanned over a mass range. Rather than computing $\Nc$ analytically --- which requires knowing the signal template shape and the autocorrelation function of $q(\theta)$ --- they estimate it empirically at a low reference level $c_0$ from a modest number of background-only pseudo-experiments. For each pseudo-experiment the full analysis is run: the $q(\theta)$ curve is computed across the scan range, and the number of upcrossings above $c_0$ is counted. Averaging over $\mathcal{O}(100)$ pseudo-experiments gives a stable estimate of $\Ncz$, because at a low level (Gross and Vitells use $c_0 = 0.5$) almost every pseudo-experiment has several upcrossings.

The key insight from Davies is that the \emph{shape} of how $\Nc$ depends on $c$ is known analytically, even when the normalisation is not. For a $\chi^2$ process with $s$ degrees of freedom, the scaling is
\begin{equation}
\Nc \;=\; \Ncz \cdot \left(\frac{c}{c_0}\right)^{(s-1)/2} \exp\!\left(-\frac{c - c_0}{2}\right)\,.
\label{eq:gv_extrapolation}
\end{equation}
This allows extrapolation from the easily estimated $\Ncz$ to any higher level $c$, including levels corresponding to $5\sigma$ where direct estimation would require $\mathcal{O}(10^7)$ pseudo-experiments. 
In the large-$c$ limit, $P(\chi^2_{s+1} > c) \propto c^{(s-1)/2}e^{-c/2}$, so one can rewrite Davies' bound (Eq.~\eqref{eq:davies_bound}) using Eq.~\eqref{eq:gv_extrapolation}\footnote{\label{fn:convention}We follow the convention of Ref.~\cite{gross2010}, where the signal strength can take any value and $q(\theta)$ is sensitive to both excesses and deficits. When only excesses are considered ($\mu \ge 0$, as in Section~\ref{sec:experiments}), the first term of Eq.~\eqref{eq:gv_asymptotic} becomes $\tfrac{1}{2}P(\chi^2_1 > c)$ and $\mathcal{N}$, which is determined by the expected number of upcrossings of the statistic actually scanned, is halved; the trial factor is unchanged, and the coefficient in Eq.~\eqref{eq:trial_factor} becomes $\sqrt{2\pi}$.} as
\begin{equation}
P(q(\hat \theta) > c) \;\le\; P(\chi^2_s > c) \;+\; {\cal N} \cdot P(\chi^2_{s+1} > c)\,,
\label{eq:gv_asymptotic}
\end{equation}
where $q(\hat\theta)=\max_\theta q(\theta)$ over the mass range. Eq.~\eqref{eq:gv_asymptotic} has a simple interpretation. The first term is the local $p$-value at a fixed mass, where $q$ follows a $\chi^2_s$ distribution, with one degree of freedom for each fitted signal parameter. The second term accounts for letting the mass float. Over a mass interval of the order of the signal width, $q(\theta)$ varies smoothly, and moving the mass to the position that maximises $q$ acts, for large $q$, like fitting one additional parameter: it adds one more squared standard normal variable to the test statistic, turning $\chi^2_s$ into $\chi^2_{s+1}$.\footnote{For $s=1$, write $q = Z^2$, with $Z(\theta)$ a smooth Gaussian process of unit variance. Near a large value of $Z$ the curvature is $Z'' \approx -\lambda Z$, with $\lambda = \mathrm{Var}[Z']$, and maximising $Z + Z'\delta + \tfrac12 Z''\delta^2$ over the shift $\delta$ gives $\max q \approx Z^2 + Z'^2/\lambda$, the sum of two independent squared standard normal variables.}
Beyond that interval, $q(\theta)$ becomes effectively independent of its earlier values, so the scan range contains about $\cal N$ such intervals, each offering a separate chance of a large fluctuation.
These intervals refer to the set of effectively independent mass hypotheses.
In the example presented in Ref.~\cite{gross2010}, using $c_0=0.5$ and 100 pseudo-experiments gives $\Ncz = 4.34 \pm 0.11$, corresponding to an effective number of independent search regions ${\cal N} = 5.58$. This single number suffices to compute the global p-value at any significance level.
Moreover, because for large $c$ the exponential factor of  a $\chi^2_k$ tail probability is the same for every $k$, each additional degree of freedom contributes one more factor of $\sqrt{c}$. 
For $s=1$, $\sqrt{c} = Z_{\mathrm{local}}$, so the ratio of the second term of Eq.~\eqref{eq:gv_asymptotic} to the first --- the trial factor minus one --- grows linearly with the local significance:
\begin{equation}
\text{trial factor} \;\approx\; 1 + {\cal N} \sqrt{\pi/2}\; Z_\mathrm{local} \qquad (s = 1)\,.
\label{eq:trial_factor}
\end{equation}
This linear growth has practical consequences: a $3\sigma$ local excess requires roughly three times more global correction than a $1\sigma$ excess. The Higgs boson discovery at the LHC illustrates the procedure: the ATLAS experiment reported a local significance of $5.9\sigma$ at $m_H \approx 126$~GeV, which was corrected to a global significance of $5.1\sigma$ over the search range 110--600~GeV, corresponding to a trial factor of roughly 100~\cite{ATLAS:2012yve}.

\subsection{Multi-dimensional searches}
\label{sec:multidim}

Many searches involve more than one unknown parameter. A resonance may have both an unknown mass and an unknown width; an astrophysical source may be located at an unknown position in the two-dimensional sky. Vitells and Gross~\cite{Vitells:2011da} extended the one-dimensional framework to multi-dimensional search spaces using the theory of random fields and the expected Euler characteristic of excursion sets.

In one dimension, the relevant topological quantity is the number of upcrossings --- the number of times the process $S(\theta)$ enters the excursion set $\{S > c\}$. In two or more dimensions, the excursion set $\{S(\theta) > c\}$ is a region of the parameter space, and its Euler characteristic $\phi(A_c)$ generalises the upcrossing count: for a $\chi^2$ random field with one degree of freedom in two search dimensions, the expected Euler characteristic takes the form
\begin{equation}
\mathbb{E}[\phi(A_c)] \;=\; P(\chi^2_1 > c) \;+\; e^{-c/2}\!\left(N_1 + \sqrt{c}\; N_2\right)\,,
\label{eq:euler_2d}
\end{equation}
where $N_1$ and $N_2$ are coefficients that encode the geometry of the search region and the correlation structure of the field. A typical example is a search for a resonance $X \to YH$, in which the two masses $m_X$ and $m_Y$ are the scan parameters, undefined under the background-only hypothesis, and the signal strength is the parameter of interest. The test statistic in this case is a $\chi^2_1$ over a two-dimensional parameter space. When both masses are scanned on a grid finer than the corresponding resolutions, the problem is formally identical to the sky search of Ref.~\cite{Vitells:2011da} and the Euler-characteristic formula, Eq.~\ref{eq:euler_2d}, applies. When $m_Y$ is instead tested only at a few values separated by more than its resolution, the search falls in the partly continuous, partly discrete regime discussed below.  Many such searches have been performed at the LHC; in the ATLAS diphoton search with the 2015 data~\cite{ATLAS:2016gzy}, for example, the global significance was evaluated over the full mass–width search region using pseudo-experiments, and cross-checked with the method of Ref.~\cite{Vitells:2011da}.
As in the one-dimensional case, these coefficients can be estimated from a small number of pseudo-experiments --- Vitells and Gross use as few as 20 --- by computing the Euler characteristic of excursion sets at two low reference levels and solving for $N_1$ and $N_2$. The resulting formula provides the global p-value at arbitrary significance, and was validated against $2\times 10^5$ full pseudo-experiments for the IceCube neutrino point-source search~\cite{Vitells:2011da}.\\

The multi-dimensional extension is relevant to a broad class of searches across experimental physics. In astrophysics, searches for neutrino point sources scan the two-dimensional celestial sphere for an excess above the atmospheric background~\cite{Vitells:2011da}. 
In collider physics, searches for long-lived particles (LLPs) provide a natural example of three or more search dimensions: the signal hypothesis is typically parameterised by the mass of the particle produced in the collision that decays into the LLP, the mass of the LLP itself, and its proper decay length $c\tau$, defining a three-dimensional search space. The sensitivity of the analysis varies strongly across this space --- different detector subsystems dominate at different lifetimes, and the signal topology changes qualitatively between prompt and highly displaced decays~\cite{Alimena:2019zri}. The LEE must account for the full multi-dimensional search volume, not just each axis independently. Although mass and lifetime are both continuous parameters in principle, LLP analyses in practice often scan the mass continuously while testing only a small discrete set of benchmark $c\tau$ values, because signal simulation and reconstruction efficiency vary strongly with lifetime and a continuous signal model across the full range is rarely available. When a continuous parameter is treated as a discrete grid, the trial factor along that axis is bounded between 1 and the number of values tested, whereas the continuously scanned dimensions contribute a Davies/GV-type correction; the joint correction across all dimensions then requires either the Euler characteristic approach of~\cite{Vitells:2011da} or brute-force pseudo-experiments over the full parameter space. This intermediate regime --- partly continuous, partly discrete --- is not fully covered by either the Davies/GV framework or standard multiple testing corrections, and represents an open problem. More generally, any search whose signal model involves continuous parameters beyond a single mass --- such as width, coupling strength, mixing angle, sky position, or lifetime --- faces a multi-dimensional LEE that grows with the dimensionality of the search space.

\subsection{A Bayesian view of the trial factor}
\label{sec:bayes}

The trial factor, $p_{\rm global}/p_{\rm local}$, is a well defined quantity for a given test statistic, scan range and observed significance. However, when the scanned parameter is continuous, there is not a unique way to estimate it: brute-force pseudo-experiments, the Davies bound of Eq.~\eqref{eq:davies_bound}, and the Gross--Vitells extrapolation in Eq.~\eqref{eq:gv_extrapolation}, are different routes to the same quantity, with different costs and assumptions. Bayer and Seljak~\cite{Bayer:2020pva} adopt a hybrid Bayesian--frequentist approach: they use the Bayesian evidence to define the trial factor of a frequentist test statistic whose global $p$-value requires no pseudo-experiments. Bayesian reasoning defines the trial factor as the ratio of prior to posterior volume, but the output is a frequentist $p$-value. We outline the main idea in the following.

\paragraph{The Occam factor.}
A Bayesian compares $H_1$ and $H_0$ through the Bayes factor $B_{10}=p(x|H_1)/p(x|H_0)$, the ratio of the probabilities that the two hypotheses assign to the observed data. Unlike the typical frequentist likelihood-ratio, in which each hypothesis is evaluated at its best-fit parameters, here each term is a marginal likelihood, i.e.\ the likelihood \emph{averaged} over the prior of all the parameters of that hypothesis. Parameters common to both hypotheses, such as those describing the background, contribute similar factors to numerator and denominator, which approximately cancel in the ratio; the net effect is governed by the parameters present only under $H_1$. Consider one such parameter $\theta$, with a flat prior $\pi(\theta)=1/\Delta$ over a range $\Delta=U-L$. Then
\begin{equation}
  p(x|H_1) = \int p(x|\theta,H_1)\,\pi(\theta)\,d\theta
         = \frac{1}{\Delta}\int_{L}^{U} p(x|\theta,H_1)\,d\theta .
\end{equation}
If the data determine $\theta$ with a precision $\sigma \ll \Delta$, the likelihood is non-negligible only within a narrow peak around its maximum $\hat\theta$, which we approximate by a Gaussian, $p(x|\theta,H_1)\approx p(x|\hat\theta,H_1)\,e^{-(\theta-\hat\theta)^2/2\sigma^2}$. 
The integral is then the area of this peak, $\sqrt{2\pi}\,\sigma\,p(x|\hat\theta,H_1)$, so that the average is the maximum likelihood multiplied by $\sqrt{2\pi}\,\sigma/\Delta$, the fraction of the prior range that the data still allow. In this step, known as the Laplace approximation, the Gaussian is integrated over the whole $\mathbb{R}$ rather than over the prior $[L,U]$; this is accurate when $\hat\theta$ lies more than a few $\sigma$ inside the range and the likelihood elsewhere in the range is negligible compared with the peak.
Writing the maximum likelihood ratio as $e^{Z^2/2}$, with $Z$ the significance of the excess in standard deviations, this gives
\begin{equation}
  \ln B_{10} \;\approx\; \frac{Z^2}{2} \;-\; \ln\frac{\Delta}{\sqrt{2\pi}\,\sigma},
  \label{eq:trotta}
\end{equation}
equivalent to Eq.~(26) of Ref.~\cite{trotta2008bayes}, where it is derived for a Gaussian prior.
The first term rewards the better fit of $H_1$; the second, often called the Occam factor,
penalises $H_1$ for the size of the parameter space it had to spread its prior over.

\paragraph{Application to the look-elsewhere effect.}
In a resonance search, the parameter present only under $H_1$ is the location $\theta$ of the signal. Its prior range is the scan range, $\Delta = U-L$, and the width $\sigma_\theta$ of the likelihood peak, measured from the peak curvature, is the precision with which the data localise the excess. Treating the signal strength as fixed for the moment, Eq.~(\ref{eq:trotta}) reads
\begin{equation}
  B_{10} \;\approx\; \frac{e^{Z^2/2}}{N_B},
  \qquad
  N_B \;=\; \frac{U-L}{\sqrt{2\pi}\,\sigma_\theta}.
\end{equation}
The evidence for a signal is the local likelihood ratio divided by $N_B$, the number of peak
widths that fit within the scan range. Bayer and Seljak~\cite{Bayer:2020pva} identify $N_B$ with the trial factor. $N_B\,p_{\rm local}$ is then the expected number of background fluctuations, anywhere in the scan range $\Delta$, at least as large as the observed one. If we divide the range into $K$ small regions, each with probability $N_B\,p_{\rm local}/K$ of containing such a fluctuation (assumed largely independent of each other), we can compute the \v{S}id\'ak correction as a way to quantify the global p-value from a scan over discrete regions (see Eq.~\eqref{eq:sidak} in Section~\ref{sec:bonferroni}): $1-(1-N_B\,p_{\rm local}/K)^K$. Taking the limit for $K\to\infty$ the latter becomes
\begin{equation}\label{eq:pglobal_bayer}
  p_{\rm global} \;\approx\; 1-\exp\left(-N_B\,p_{\rm local}\right).
\end{equation}
Since $N_B$ depends on the width $\sigma_\theta$ of the observed peak, and $Z$ on its height, Eq.~\eqref{eq:pglobal_bayer} depends on both, and is therefore the $p$-value of a statistic combining the two. Bayer and Seljak define this statistic as
\begin{equation}\label{eq:qS}
q_S\equiv-2\ln(N_B\,p_{\rm local})=Z^2-2\ln N_B+\ln(2\pi Z^2),
\end{equation}
where the second equality uses the asymptotic local $p$-value, $p_{\rm local}\approx e^{-Z^2/2}/(\sqrt{2\pi}\,Z)$. Since multiplying a $p$-value by $N_B$ is equivalent to lowering $Z^2$ by $2\ln N_B$, $q_S$ is the local test statistic $Z^2$ penalised by the trial factor; the penalty $2\ln N_B$ is the Occam factor of Eq.~(\ref{eq:trotta}) in the same units.
Substituting $N_B\,p_{\rm local}=e^{-q_S/2}$ into Eq.~\eqref{eq:pglobal_bayer} gives $p_{\rm global}\approx1-\exp(-e^{-q_S/2})$. Usually, the global $p$-value requires the null distribution of the highest peak, which depends on the scan range and resolution of each search and is obtained with pseudo-experiments. Here, instead, range and resolution enter through $N_B$, inside $q_S$, and the null distribution of $q_S$ is the same in every search, so the global $p$-value follows from the observed $q_S$ alone. When the scan shows several peaks, $q_S$ is computed for each from its own height and width, and the global $p$-value is that of the peak with the largest $q_S$; this need not be the highest peak, since a broader peak has a smaller $N_B$.
The width $\sigma_\theta$ measured on the data is set, on average, by how quickly the signal template changes with $\theta$ (Section~\ref{sec:model_dependence}), and shrinks as the excess becomes more significant, so that $N_B$ grows linearly with $Z$, as in Eq.~\eqref{eq:trial_factor}; to leading order it coincides with the Davies estimate. The authors verify with background-only pseudo-experiments that the resulting p-values are calibrated across scan ranges, template widths and up to three scanned dimensions.
The method relies on the Laplace approximation (a smooth, approximately Gaussian peak), which typically holds with enough data; , by the asymptotic normality of the likelihood. When it does not, the same construction applies with the Bayes factor computed numerically: in exoplanet transit searches, where the Laplace approximation is poor, Robnik and Seljak~\cite{Robnik} integrate the evidence around the peak by quadrature and recover $p$-values that agree with simulations. Unlike Gross--Vitells, $N_B$ takes the width of the observed peak as representative of the whole range, so the resolution should not vary strongly across it.
A follow-up study~\cite{Bayer:2021lhk} removes even the curvature computation: since the number of noise peaks above a level $c$ in the observed scan estimates the expected number of upcrossings, the trial factor can be calibrated from the heights of the lower peaks in the same dataset. This is the Gross--Vitells extrapolation of Eq.~\eqref{eq:gv_extrapolation}, with the pseudo-experiments that fix $\langle N(c_0)\rangle$ replaced by a count on the data itself, at the cost of a $\sim$10--20\% statistical uncertainty on $p_{\rm global}$, from the finite number of noise peaks counted in a single dataset, and the assumption that the lower peaks are background fluctuations.

The signal strength $\mu$ requires a separate remark. A fully Bayesian treatment also averages over a prior on $\mu$, which produces a second Occam factor of the form of Eq.~(\ref{eq:trotta}), with  $\Delta$ now the prior range of $\mu$ and $\sigma$ its measurement uncertainty. This factor has no counterpart in the frequentist LEE: it depends on an arbitrary prior choice and, for a sufficiently broad prior, favours $H_0$ whatever the data (the Jeffreys--Lindley paradox). It is the location factor, not this one, that corresponds to the trial factor. For this reason Bayer and Seljak choose the prior on $\mu$ so that it introduces no look-elsewhere penalty: its width is set to the fitted signal strength, $\sqrt{q}\,\sigma_\mu\approx\hat\mu$, so that its Occam factor only reproduces the normalisation of the local $p$-value. As the authors note, a strict Bayesian could object to a prior that depends on the observed signal; they justify it as the narrowest prior consistent with the measured amplitude. Robnik and Seljak~\cite{Robnik} extend this reasoning to searches in which the priors are known for some parameters but not for others, such as the amplitude: the Bayes factor computed with the known priors is used as the test statistic, and its significance is quoted as a frequentist $p$-value, which is less sensitive than the Bayes factor itself to the poorly known prior.

\paragraph{An intrinsic property of the search.}
It is not a coincidence that the Bayesian Occam factor and the frequentist trial factor agree.
Indeed, starting from the prior-to-posterior volume of Ref.~\cite{Bayer:2020pva} and counting peaks above a threshold, Ref.~\cite{Bayer:2021lhk} recovers the up-crossing extrapolation of Gross and Vitells.
The look-elsewhere effect is an intrinsic property of the question being asked---is there a signal \emph{somewhere} in this range?---and of the experiment's ability to distinguish one location from another, and it appears across inference frameworks. In both cases the penalty is set by the number of signal hypotheses that the data can tell apart: the scan range divided by the resolution. The frequentist pays it at the end through the distribution of the maximum under $H_0$; the Bayesian pays it at the start, by spreading prior probability over the range. Neither framework can avoid it, and in both the size of the penalty depends on how ``elsewhere'' is defined (Section~\ref{sec:elsewhere}).

\subsection{What determines the trial factor?}
\label{sec:model_dependence}

A point that is often underappreciated in physics analyses is that the trial factor is set by the width of the signal template used in the fit. By template we mean the expected signal shape after detector effects, i.e.\ the intrinsic line shape convolved with the experimental resolution. Its width determines how rapidly $S(\theta)$ decorrelates as $\theta$ varies (Eq.~\ref{eq:davies_normal}): a narrow template produces a rapidly oscillating $S(\theta)$, with many upcrossings per unit range of $\theta$, and a wide template a slowly varying one, with fewer.
For a narrow resonance, the width is dominated by the experimental mass resolution, so in the most common case the trial factor is effectively a property of the detector and the scan range. Since the resolution usually degrades with increasing mass, fluctuations decorrelate more slowly at high mass, as in the example of Ref.~\cite{gross2010}. When the intrinsic width is comparable to or larger than the resolution, or when the width is itself a hypothesis chosen or scanned by the analyst, the trial factor also depends on the signal model: two analyses scanning the same data with different assumed signal widths obtain different trial factors, even though the data and the scan range are identical.

This is sometimes a source of confusion, since the pseudo-experiments used for calibration are generated under the null hypothesis and do not depend on the signal hypothesis. The trial factor nevertheless does, because the signal template defines the test statistic $S(\theta)$: it is smaller for a wide resonance not because the background fluctuates less, but because the test is less sensitive to localised fluctuations. In the Gross \& Vitells framework this dependence is absorbed into $\Ncz$, since the pseudo-experiments are fitted with the same template. The common shorthand ``trial factor $\approx$ range / resolution'' is correct when ``resolution'' means the width of the signal template, and not the bin width.

This observation becomes central when moving to machine learning-based searches, where the ``signal template'' is replaced by a flexible, data-adaptive model whose effective resolution is not specified a priori and may not even be known to the analyst. We return to this point in Section~\ref{sec:ml}.
\section{Multiple Searches}
\label{sec:multipletesting}

In a search over several channels, final states or analyses, the null hypothesis can be written as the intersection of per-channel null hypotheses, ``no signal in channel $k$'', and rejecting it amounts to rejecting at least one of them. In other words, the signal is present in at least one of the channels.
The methods reviewed below provide the corresponding correction when the channels are discrete and approximately independent, a situation in which the continuous-parameter machinery of Davies and GV does not apply.

Whether a multi-channel search allows for multiple sources of signal at all depends on what the signal model fixes. If the relative signal rates across channels are predicted, for example by the branching ratios of the particle being sought, the channels can be combined into a single likelihood with one signal-strength parameter. The result is a single test, and no correction for the number of channels arises, although a scan over unknown parameters such as the mass still requires one. 
In contrast, if the relative rates are free, either each channel is tested separately and the result is corrected for the number of channels, or all signal strengths are fitted jointly, which gives a single test with more degrees of freedom. In either case, the additional freedom costs significance.

The per-channel formulation becomes essential when a separate decision is required for each channel, i.e.\ when the question is not only whether the null fails but where. Two features then distinguish the problem from a mass scan. First, several channels can carry a signal at the same time, whereas a single resonance appears at only one mass, so that at most one of the point-nulls of a scan can be false (Section~\ref{sec:exclusion}).
Second, one must choose which error rate to control across the set of decisions.

That choice depends on where in an analysis the multiple test sits. At an early stage, multiple tests often act as a classifier that selects events or objects for further study, such as jets containing $b$ hadrons. Some false positives are then expected and acceptable, because the downstream analysis measures and corrects for them; what matters is the fraction of the selected sample that is falsely tagged, which is the quantity controlled by the false discovery rate (Section~\ref{sec:bh}).\footnote{In practice, HEP characterises a tagger by its signal efficiency $\epsilon_s = P(\text{tag} \mid \text{signal})$ and background mis-identification rate $\epsilon_b = P(\text{tag} \mid \text{background})$ rather than by the FDR. Being conditional on the true class, $\epsilon_s$ and $\epsilon_b$ are properties of the classifier alone, whereas the FDR $\approx \epsilon_b/[\epsilon_b + \epsilon_s (N_s/N_b)]$ also depends on the signal fraction $N_s/N_b$ in the sample. The same tagger can thus show a worse FDR simply because the signal fraction is lower --- as for $b$-tagging at LEP2 relative to the $Z$ pole --- so efficiency and mis-identification rate give a characterisation that transfers across analyses.}
At the end of an analysis, the multiple test instead underlies a physics conclusion such as a discovery claim, where the aim is not to be wrong at all. The appropriate criterion is then the probability of any false rejection, the family-wise error rate (Section~\ref{sec:bonferroni}).

The methods below are thus the natural tools for case~(ii) of Section~\ref{sec:case_ii} and, in the context of machine learning, for combining results from several models applied to the same data (Section~\ref{sec:multiple_ml}). Applying them to a single continuous scan is an approximation that ignores the correlation between nearby mass points, and is generally inferior to the Davies/GV approach when the latter is applicable.

\subsection{Bonferroni, \v{S}id\'{a}k, and family-wise error rate}
\label{sec:bonferroni}

The simplest multiple testing corrections control the \emph{family-wise error rate} (FWER): the probability of making at least one false discovery across all $N$ tests.

The Bonferroni bound uses the union bound: if $N$ tests are performed and the smallest observed p-value is $p_{\min}$, then
\begin{equation}
\pglobal \;\leq\; N \cdot p_{\min}\,.
\label{eq:bonferroni}
\end{equation}
This is valid under \emph{arbitrary} dependence between the tests --- no assumption about their correlation structure is needed. Its conservatism is the price of this generality: when tests are positively correlated (as they typically are in HEP, where overlapping signal templates induce correlated excesses), the true FWER is much smaller than $N \cdot p_{\min}$.

The \v{S}id\'{a}k correction is the exact result when the tests are \emph{independent}:
\begin{equation}
\pglobal \;=\; 1 - (1 - p_{\min})^N\,.
\label{eq:sidak}
\end{equation}
For small $p_{\min}$ (the regime of interest near discovery), the two are numerically indistinguishable: $1 - (1 - p)^N \approx Np$ when $Np \ll 1$. At low significance they differ meaningfully only with many tests, where the \v{S}id\'{a}k correction is tighter. Since HEP tests are rarely independent --- overlapping mass bins, shared systematic uncertainties, correlated channels --- the Bonferroni bound is the more broadly applicable of the two, and the \v{S}id\'{a}k correction should be used only when independence can be justified.

Both corrections are almost always too conservative for HEP searches within a single distribution, for the reasons already noted: the effective number of independent tests is much smaller than $N$ due to positive correlations, and the Davies/GV framework accounts for this. However, Bonferroni retains value in two situations: (a)~when combining genuinely distinct analyses (case~(ii), where different final states may be approximately independent), and (b)~when the test statistic has a non-standard null distribution, so that the $\chi^2$ assumptions required by Davies/GV do not hold --- as is often the case for ML-based test statistics.

\subsection{Benjamini-Hochberg and false discovery rate}
\label{sec:bh}
The Benjamini-Hochberg (BH) procedure~\cite{bh_1995} was developed to address a problem that arises naturally in fields where thousands of hypotheses are tested simultaneously. The canonical example is the genome-wide association study (GWAS), in which millions of genetic variants are tested for association with a disease~\cite{storey2003statistical}. In this setting, some variants are genuinely associated with the disease and many are not, and the goal is to identify as many true associations as possible while controlling the fraction of false positives among those identified. Requiring that \emph{no} false positive be made (FWER control, as in Bonferroni) is far too conservative when millions of tests are performed: it demands such stringent per-test thresholds that most genuine associations are missed. What is needed instead is a guarantee on the \emph{proportion} of false positives among the discoveries --- a quantity called the \emph{false discovery rate} (FDR).

Formally, suppose $N$ tests are performed and $R$ of them are rejected (declared significant); among those $R$ rejections, let $V$ be the number of false rejections (nulls that were actually true). The FDR is defined as $\text{FDR} = \mathbb{E}[V/R]$, with the convention $V/R = 0$ when $R = 0$. FWER controls $P(V \geq 1)$ --- the probability of \emph{any} mistake. FDR controls $\mathbb{E}[V/R]$ --- the expected \emph{fraction} of mistakes among claims. The distinction matters when $R$ is large: if 20 hypotheses are rejected and one is a false positive, the FDR is 5\% even though the FWER is 100\%.

The BH procedure works as follows. Sort the $N$ p-values in increasing order: $p_{(1)} \leq p_{(2)} \leq \cdots \leq p_{(N)}$. Find the largest index $k$ for which
\begin{equation}
p_{(k)} \;\leq\; \frac{k}{N}\,\alpha\,.
\label{eq:bh}
\end{equation}
Then reject the hypotheses corresponding to $p_{(1)},\ldots,p_{(k)}$. This controls the FDR at level $\alpha$ when the tests are independent. Benjamini and Yekutieli~\cite{benjamini2001control} proved that the same procedure also controls the FDR under \emph{positive dependence} (specifically, the positive regression dependence on subsets condition). 
The same paper also introduced a modified procedure---with the threshold in Eq.~\eqref{eq:bh} rescaled by the harmonic factor $c(N) = \sum_{i=1}^{N} 1/i \approx \ln N$---that controls the FDR under \emph{arbitrary} dependence, including negatively correlated tests, at the cost of a $\ln N$ loss in power.
In GWAS this positive-dependence condition is satisfied because nearby genetic variants on the same chromosome are correlated through linkage disequilibrium; in HEP it is satisfied because overlapping signal templates or shared systematic uncertainties induce positive correlations between test statistics in nearby bins.

The BH procedure has seen widespread adoption in genomics, neuroimaging, and other fields where large-scale multiple testing is routine, but remains rare in particle physics--- even though FDR-\emph{style} reasoning is in fact routine there under other names, as in the measurement and subtraction of fake rates in a selected event sample, where a controlled level of contamination is expected and corrected for rather than eliminated. The reason is partly cultural --- the $5\sigma$ discovery convention targets the FWER, not the FDR --- and partly structural: HEP discovery typically involves \emph{at most one} signal, and the goal is to determine whether it exists at all, which is a question better addressed by FWER control or the Davies/GV framework. In the genomics setting, by contrast, many of the $N$ nulls are expected to be false (many genes are differentially expressed), and the goal is to identify as many as possible while keeping the false discovery fraction under control--- much as, within HEP, the early selection stage tolerates and corrects a controlled false-positive fraction rather than eliminating it.

Where BH becomes natural in HEP is at the \emph{screening} stage of a large-scale search: when scanning thousands of channels or final states, the question shifts from ``is there exactly one signal?'' to ``which channels are worth following up?'' This is structurally the same problem as GWAS --- many hypotheses, some fraction expected to be interesting, tolerance for a controlled rate of false leads --- and BH provides a principled way to address it.\\

As discussed at the beginning of this section, the choice between FWER and FDR control reflects \emph{where} in an analysis the multiple test sits. The same logic recurs for machine-learning searches (Section~\ref{sec:ml}): a signal-agnostic scan is a screening stage suited to FDR control, while the targeted follow-up that produces a discovery claim is held to the stricter FWER-style threshold.

\section{E-values and anytime-valid inference}
\label{sec:evalues}
\label{sec:evalues_motivation}\label{sec:what_evalues}\label{sec:evalue_decisions}
\label{sec:combining_evalues}\label{sec:monitoring}\label{sec:lee_evalues}
\label{sec:evalues_physics}\label{sec:evalue_power}

The methods of Sections~\ref{sec:classical} and~\ref{sec:multipletesting} assume that the significance level and the analysis plan are fixed in advance. Examining the data at interim stages, or deciding to collect more after an inconclusive result, invalidates the type~I error guarantee unless it is explicitly corrected for, as in the group sequential and $\alpha$-spending methods mentioned in Section~\ref{sec:elsewhen}. \emph{E-values}~\cite{grunwald2020safe, vovk2021values, ramdas2023game, ramdas2025ebook} were developed for fields in which this rigidity is costly, such as clinical trials, online A/B testing (randomized controlled experiments) and financial risk monitoring, and they provide inference that remains valid however often the data are examined. They have so far seen little use in fundamental physics. We give only a brief account here, restricted to the aspects that bear on the look-elsewhere effect, and refer to the monograph of Ramdas and Wang~\cite{ramdas2025ebook} for proofs and a systematic treatment.

\paragraph{Definition and basic properties.}
An e-value for a null hypothesis $H_0$ is a non-negative random variable $E$ satisfying
\begin{equation}
\mathbb{E}[E \mid H_0] \;\leq\; 1\,,
\label{eq:evalue_def}
\end{equation}
so that large observed values are evidence against the null. The simplest example is the likelihood ratio $E = p(x \mid H_1)/p(x \mid H_0)$ against a specific alternative $H_1$, whose expectation under $H_0$ is $\int p(x \mid H_1)\,dx = 1$. Markov's inequality gives $P(E \geq 1/\alpha \mid H_0) \leq \alpha$, so rejecting when $E \geq 1/\alpha$ is a valid level-$\alpha$ test that requires no knowledge of the null distribution of $E$. This threshold is conservative; when the null distribution of $E$ is known, or can be estimated from background-only pseudo-experiments as is routine in HEP, the threshold can instead be placed at the upper $\alpha$-quantile of that distribution, as for any other test statistic~\cite[\S2.2]{ramdas2025ebook}. Two combination rules follow directly from Eq.~\eqref{eq:evalue_def}: the product of independent e-values is an e-value for the intersection of their null hypotheses, and a weighted average $\sum_i w_i E_i$, with $w_i \geq 0$ and $\sum_i w_i = 1$, is an e-value under arbitrary dependence between the $E_i$, without any need to model their correlation.

\paragraph{Anytime validity.}
The property that motivated the framework concerns data that accumulate over time. An \emph{e-process} is a non-negative sequence $(M_t)_{t \geq 1}$, updated as data arrive, such that $M_\tau$ is an e-value for $H_0$ at every stopping time~$\tau$; the running product of likelihood-ratio e-values computed on successive independent batches of data is the standard example. In the betting interpretation of Ref.~\cite{ramdas2023game}, $M_t$ is the capital of a gambler wagering against $H_0$ in a game that is fair or unfavourable whenever $H_0$ is true. Ville's inequality bounds the probability that such a process \emph{ever} reaches $1/\alpha$:
\begin{equation}
P\!\left(\sup_{t \geq 1} M_t \geq 1/\alpha \;\Big|\; H_0\right) \;\leq\; \alpha\,.
\label{eq:ville}
\end{equation}
The process can therefore be inspected after every new batch of data, and the null rejected as soon as $M_t \geq 1/\alpha$, with no pre-specified stopping rule and no $\alpha$-spending. This is the framework's answer to the look-elsewhen effect of Section~\ref{sec:elsewhen}: repeated looks need no correction, because they do not invalidate the test. Conventional fixed-sample p-values do not have this property, although p-values that do can be derived from an e-process as $p_t = \min\big(1,\, 1/\max_{s \leq t} M_s\big)$~\cite{ramdas2023game}.

\paragraph{The spatial look-elsewhere effect.}
Anytime validity removes the temporal axis of the LEE, but not the spatial one. Suppose that for each candidate mass $\theta \in [L,\,U]$ one has an e-value $E_\theta$ for the background-only null, for instance a likelihood ratio against a signal template at mass~$\theta$. The maximum $\sup_\theta E_\theta$ is not an e-value, but for any probability density $\pi$ on $[L,\,U]$ chosen in advance the mixture
\begin{equation}
E_\pi \;=\; \int_L^U E_\theta\, \pi(\theta)\, d\theta
\label{eq:mixture_evalue}
\end{equation}
is one, since $\mathbb{E}[E_\pi \mid H_0] = \int_L^U \mathbb{E}[E_\theta \mid H_0]\,\pi(\theta)\,d\theta \leq 1$; a weighted sum plays the same role for a discrete set of channels. No correction is applied afterwards. When the $E_\theta$ are likelihood ratios, $E_\pi$ is the Bayes factor for a signal whose mass has prior $\pi$, which connects this construction directly to the hybrid Bayesian-frequentist treatment of the LEE in Ref.~\cite{Bayer:2020pva}. The cost of searching is now carried by $\pi$: for a signal near $\hat\theta$, $E_\pi$ is approximately $E_{\hat\theta}$ multiplied by the prior mass that $\pi$ assigns to the peak of $E_\theta$ around $\hat\theta$. For a flat prior the resulting penalty is the ratio of the scan range to the width of that peak, which is set by the signal template and narrows as the signal becomes stronger; the penalty therefore grows with both the number of resolution elements in the scan range and the local significance, as the Gross--Vitells trial factor does (Section~\ref{sec:classical}). Bayer and Seljak~\cite{Bayer:2020pva} relate the trial factor to the ratio of prior to posterior volume. The two accountings differ in when the cost is paid: the classical procedure searches first and corrects afterwards, whereas the e-value procedure commits to a weighting first and never corrects. Because $\pi$ must be fixed in advance and appears explicitly in $E_\pi$, the construction forces the analyst to state what ``elsewhere'' means, turning the ambiguity discussed in Section~\ref{sec:elsewhere} into an explicit choice. Finally, a mixture of e-processes over $\theta$ is again an e-process, so Eq.~\eqref{eq:ville} covers the spatial and temporal axes of the LEE at once.

\paragraph{Applicability in physics and open problems.}
The guarantees above matter where a physics analysis is genuinely sequential: online data-quality monitoring, real-time anomaly detection, and the accumulation of evidence across successive data-taking periods. Two properties are also useful offline: the weighted average combines correlated tests validly without an estimate of their correlation, and the e-BH procedure~\cite{wang2022false}, the e-value analogue of Benjamini--Hochberg (Section~\ref{sec:bh}), controls the false discovery rate under arbitrary dependence with no adjustment to the procedure. For a standard analysis of a fixed dataset, calibrated with pseudo-experiments, none of these guarantees is exercised and there is little reason to leave the familiar p-value framework; we do not suggest that e-values replace p-values in such analyses. Two open problems are worth flagging for the searches discussed in this review. First, the e-values best suited to accumulating evidence are those that maximise the expected growth rate $\mathbb{E}[\log E \mid H_1]$ against a specified alternative~\cite{ramdas2025ebook}, so it is unclear how to construct powerful e-values for signal-agnostic searches, in which $H_1$ is deliberately left unspecified; the mixture of Eq.~\eqref{eq:mixture_evalue} is only a partial answer, since it requires a parametrised family to mix over. Second, systematic uncertainties make the null hypothesis composite, and a likelihood ratio computed against a single member of the null family need not be an e-value under the others; universal inference~\cite{wasserman2020universal} and the growth-optimal constructions of Gr\"unwald et al.~\cite{grunwald2020safe} are the natural starting points.
\section{Machine Learning and the Look-Elsewhere Effect}
\label{sec:ml}

The LEE in machine learning is not a new statistical problem. It is a specific instance of \emph{post-selection inference}~\cite{lee2016exact, berk2013valid}: using data to select what to test, then testing on the same data. This pattern, and the resulting bias, has been identified independently in several fields (Section~\ref{sec:postselection}).  
The classical LEE is itself a post-selection problem: the data select the mass $\hat\theta$ that maximises the test statistic, and evaluating the test at $\hat\theta$ as if that mass had been fixed in advance gives the local p-value. 
What makes it tractable is that the selection is explicit and low-dimensional. In an ML-based search, the selection is implicit (embedded in the model's weights), high-dimensional, and data-adaptive, but the statistical structure is identical.

The implicit LEE grows with how much of the alternative hypothesis a search leaves to be fitted to the data, which places searches on a continuum from the simple hypothesis test at one end to the goodness-of-fit test at the other. At the hypothesis-testing end, the alternative $H_1$ is fully specified: the test has nothing to adapt to the data, and no post-selection bias arises. A classical bump hunt sits near this end, its only data-driven freedom being the mass scan over a fixed family of alternatives. At the goodness-of-fit end, the alternative is not specified by any signal model but is a broad, flexible family (for an ML-based test, the set of functions the model can represent), and the member that best describes the data is selected entirely by the fit. The test can therefore flag departures of any shape anywhere in the phase space. In between lie parametrised models that carry a signal hypothesis but still fit part of their freedom to the data. 
This is why classical tools transfer cleanly only near the hypothesis-testing end: the Neyman-Pearson optimality and $\chi^2$ asymptotics behind Davies/GV assume a parametrised alternative, an assumption that weakens as a search moves toward the goodness-of-fit end, where the effective search space grows (the implicit case-(iii) LEE of Section~\ref{sec:elsewhere}). 
The supervised/weakly-supervised distinction that follows is a practical proxy for a search's position along this continuum, but not a sharp boundary.

Concretely, the supervised mode sits near the hypothesis-testing end: a classifier trained on a specific signal hypothesis is fixed before the test data are examined, so it introduces no post-selection bias and the LEE reduces to the parameter scan of the downstream analysis, where classical tools apply. Weakly supervised and unsupervised modes sit toward the goodness-of-fit end, because the model is trained on the observed data and its selection adapts to the specific fluctuations in the dataset being tested. This is the case both for \emph{resonant anomaly detection}~\cite{hein2025, grosso2025robust}, where a data-driven classifier selects events at each candidate mass (adding an ML-induced LEE on top of the classical mass scan), and for methods that \emph{learn the test statistic itself}~\cite{DAgnolo:2018cun, Letizia:2022xbe, grosso2024goodness} by fitting the full phase space globally --- where the model simultaneously selects where the anomaly is, what shape it has, and how significant it is, all from the same data.

The three classical solutions to post-selection inference map directly onto strategies for ML searches: account for the selection in the null distribution~\cite{lee2016exact,berk2013valid} corresponds to empirical calibration (Section~\ref{subsec:pseudoexp}); split the data~\cite{cox1975note} corresponds to train/test splitting (Section~\ref{subsec:splitting}); constrain the selection corresponds to model regularisation (Section~\ref{subsec:regularization}). The open question is not what to do but how to do it efficiently for complex, expensive-to-train models. The remainder of this section develops each strategy after examining the structural difference between classical and ML-based searches (Section~\ref{sec:global_template}) and the role of model resolution (Section~\ref{sec:resolution}).

\subsection{The LEE as a post-selection problem}
\label{sec:postselection}

The multiple testing methods of Section~\ref{sec:multipletesting} assume that the set of hypotheses to be tested is fixed in advance. In many practical settings, however, the data are used not only to \emph{test} hypotheses but also to \emph{select} which hypotheses to test, a problem known in the statistics literature as \emph{post-selection inference} or \emph{selective inference}. This is a more general framework that encompasses both the classical LEE (where the data select the most significant mass point) and the ML-LEE (where the data select the most anomalous region via a trained model).

The dangers of post-selection inference have been recognised for decades. Freedman~\cite{freedman1983note} showed that when a quantity is fitted as a linear combination of many input variables, none of which is actually related to it, keeping only the variables that look promising in a first fit and refitting reliably produces fitted weights that appear statistically significant, a pure artefact of the selection. In particle-physics terms, this is like fitting a distribution with many template shapes, each with a free normalisation, keeping only those whose normalisations look significant, and refitting: the surviving normalisations will appear significant even if the data contain no contribution from any of the templates. 
In clinical trials, \emph{subgroup analysis} poses the same problem: a treatment that shows no overall effect may appear effective in a data-selected subgroup, and testing the treatment effect in that subgroup using the same data inflates the significance~\cite{assmann2000subgroup}. In neuroimaging the same circularity is known as ``double dipping''~\cite{kriegeskorte2009circular}. 
Tukey~\cite{tukey1977exploratory} drew the foundational distinction between \emph{exploratory} analysis (generating hypotheses from data) and \emph{confirmatory} analysis (testing hypotheses on independent data), and argued that conflating the two is a fundamental source of irreproducibility.

The oldest and simplest solution is \emph{data splitting}: use one portion of the data to select, and an independent portion to test~\cite{cox1975note}. This is exactly the train/test split of Section~\ref{subsec:splitting}, and it predates machine learning by decades. Its limitation --- the loss of statistical power from using only a fraction of the data for each purpose --- is the same in both settings.

More recent work has developed formal frameworks for valid inference without data splitting. Berk et al.~\cite{berk2013valid} proposed \emph{post-selection inference} (PoSI): confidence intervals that are valid after \emph{any} model selection procedure, at the cost of wider intervals (analogous to the conservatism of Bonferroni). Lee et al.~\cite{lee2016exact} developed \emph{conditional} selective inference, computing p-values and confidence intervals conditional on the selection procedure --- for example, the distribution of the test statistic at mass $\hat m$, given that $\hat m$ was the mass that maximised the test statistic. The global $p$-value of Section~\ref{sec:classical} takes the first of these routes: Davies' bound controls the distribution of the maximum over all locations, a simultaneous guarantee closer in spirit to PoSI than to conditional inference.


\subsection{ML-based goodness-of-fit as a global template fit}
\label{sec:global_template}

Machine learning-based anomaly detection is not categorically different from classical template-based searches --- it is a continuous extension of the same idea. Both approaches fit a model to data and test for deviations from a reference; the difference lies in the flexibility of the fitted alternative.

In a classical resonance search, a fixed parametric model, typically a smooth background function plus a signal template of known shape, is fitted to the data, with the background constrained over a wide mass range, often the full spectrum. Only the signal component is localised: its position is scanned across the search range, one hypothesis at a time, while its shape is fixed by the template. Each position is therefore tested against a one-parameter family of alternatives, and the trial factor counts the effectively independent positions within the scan, the quantity $\cal N$ estimated by Gross and Vitells.

The structural difference is thus not the extent of the fit but the flexibility of the alternative. A classical template can absorb a fluctuation only if it has the template’s shape, and only at one position at a time, so the resulting miscalibration is fully captured by the scan over that position. An ML-based search can adapt to statistical fluctuations of any shape anywhere in the full phase space in a single training pass.
It is parametrically richer, but structurally it is still a template: it approximates the likelihood ratio $p(\text{data})/p(\text{reference})$ at every point in the feature space, just as the classical template approximates it within each window.

The single structural difference --- local sequential fit versus global simultaneous fit --- has a direct statistical consequence. A classical search can overfit at most locally, within the current window; the miscalibration is bounded and correctable window by window. An ML-based search can overfit to statistical fluctuations anywhere in the full phase space in a single training pass. The result is a variant of the LEE that is harder to quantify than in the classical case: the effective number of independent looks depends on how the model's sensitivity is distributed across the feature space, on the correlation structure of the overfitting, and on the model's resolution relative to the data's intrinsic scales.
Regularisation techniques promoting sparsity can help control how the model's sensitivity is distributed in the space. We will return to the topic of regularisation in Section~\ref{subsec:regularization}.

\subsection{The role of model resolution}
\label{sec:resolution}

In the Davies framework, the trial factor depends on how rapidly the test statistic decorrelates as the nuisance parameter varies, which is controlled by the signal template shape (Section~\ref{sec:model_dependence}). For an ML model, the analog is the model's \emph{intrinsic resolution}: how finely it can distinguish features across the phase space. A more expressive model --- a narrow kernel, a deep network with little regularisation --- probes finer scales and accesses more independent resolution elements across the full space, inflating the effective trial factor. A less expressive model --- a wide kernel, a heavily regularised network --- probes coarser scales, accesses fewer elements, and is less susceptible to fluctuations but also less sensitive to fine-scale signals.

In principle, there is an optimal model resolution that matches the signal's characteristic scale: a model that is too coarse misses the signal, while one that is too fine overfits to fluctuations. This is the same logic that governs window-width choice in classical searches. But in the ML setting this matching is obstructed on two fronts. First, for generic ML models such as neural networks and BDTs, resolution is not an explicit or directly tunable parameter --- it is an emergent property of the architecture, regularisation scheme, and training dynamics. One cannot simply ``set'' the resolution to a desired value the way one sets a kernel width. Second, in a signal-agnostic search the signal's characteristic scale is unknown by construction, so even if the model resolution were controllable there would be no target to match it to.

For kernel-based methods~\cite{Letizia:2022xbe}, the kernel width $\sigma$ is an exception: it is an explicit resolution parameter that can be varied and studied directly. This is precisely what Ref.~\cite{grosso2025multiple} exploits when studying the correlation structure of tests at different $\sigma$ values --- the kernel width provides a direct handle on the resolution axis.

\subsection{Strategies for handling the implicit LEE}
\label{sec:calibration}

The implicit LEE of an ML-based search can be addressed at three distinct strategies: by \emph{measuring and correcting} for its contamination of the p-value through empirical calibration, by \emph{reducing} its size through model regularisation, or by \emph{avoiding} it through data splitting. These strategies are complementary and can be combined.
\subsubsection{Measure and correct: empirical calibration}
\label{subsec:pseudoexp}

The most direct approach to handling the implicit LEE is to estimate the null distribution of the test statistic empirically and correct for it. The full analysis pipeline --- including model training, feature selection, and scoring --- is run on many datasets generated under the null hypothesis, and the observed test statistic is compared to the resulting distribution. This absorbs whatever LEE the model produces, automatically, because each null dataset undergoes the same implicit search as the real data. Three methods for generating the null datasets are available, each with different assumptions and domains of applicability.

\paragraph{Pseudo-experiments (toys).} The most common approach in HEP is to generate synthetic datasets from a known or estimated model of the background process, typically using Monte Carlo simulation. The full pipeline is run on each synthetic dataset, producing a null distribution of the test statistic. The p-value is defined as the fraction of pseudo-experiments that produce a test statistic at least as large as the observed one:
\begin{equation}
\pglobal = \frac{1}{|\mathcal{T}_R|} \sum_{t \in \mathcal{T}_R} \mathbf{1}[t > t_\mathrm{obs}]\,,
\label{eq:nplm_pvalue}
\end{equation}
where $\mathcal{T}_R$ is the set of test statistic values from pseudo-experiments generated under the reference hypothesis. This is exactly what Gross and Vitells do when they count upcrossings in background-only toys, and for ML anomaly detection it has been used in the NPLM methodology from its inception~\cite{DAgnolo:2018cun,d2021learning,Letizia:2022xbe}. The method requires a reliable generative model of the background. In HEP this comes from Monte Carlo simulation or, more often for searches, from a background model fitted to the data; such a model is not always available in other fields.

\paragraph{Permutation tests.} When the hypothesis test takes the form of a two-sample comparison the null hypothesis is that the two samples are drawn from the same distribution. Under this null, the sample labels are exchangeable: any permutation of the labels produces an equally valid dataset. The null distribution of the test statistic can therefore be estimated by repeatedly permuting the labels, recomputing the test statistic for each permutation, and comparing the observed value to this permutation distribution. This is the approach taken by MMDAgg~\cite{schrab2023mmd} and MMD-FUSE~\cite{biggs2023mmd} for kernel-based two-sample tests. The method requires no generative model and no simulation --- the observed data themselves provide the null reference --- and it is exact (not approximate) under the exchangeability assumption. Its limitation is that exchangeability may not hold when nuisance parameters (such as systematic uncertainties) break the symmetry between the two samples, or when the reference sample is generated from simulation rather than drawn from the same source as the data.

\paragraph{Bootstrap.} When neither a generative model nor an exchangeability assumption is available, the null distribution can be approximated by resampling the observed data with replacement. This is the parametric or nonparametric bootstrap, depending on whether a model is fitted to the data probability distribution before resampling. The bootstrap does not require simulation and is broadly applicable, but it has a conceptual limitation in the hypothesis testing context: it assumes the observed data are representative of the null hypothesis, which is problematic if a signal is present. In practice, a common approach is to fit the null model (background only) to the data and then resample from the fitted model --- a parametric bootstrap~\cite{davison1997bootstrap} that is closely related to the pseudo-experiment approach but uses the data-driven fit rather than an independent simulation as the generative model. 

\paragraph{When to use which.} The choice among these methods depends on what is available:
\begin{itemize}
\item If a reliable model of the background is available, \emph{pseudo-experiments} drawn from it give the most direct calibration. Simulation alone is often not accurate enough for this, so in practice the model is usually constrained by the data, for example a smooth functional form fitted to the spectrum or an estimate from control regions, with systematic uncertainties included as nuisance parameters. The calibration is only as good as this model.
\item If the test is a two-sample comparison and the two samples are exchangeable under the null, \emph{permutation tests} are preferred: they are exact, require no additional simulation.
\item If neither simulation nor exchangeability is available, the \emph{bootstrap} provides a fallback, with the caveat that the null model must be estimated from the data and the resulting calibration is approximate.
\end{itemize}
In all three cases the null distribution absorbs the full implicit LEE, regardless of the model complexity or the dimensionality of the search, and the method uses all the data for both training and scoring, avoiding the power loss from data splitting.

\paragraph{Computational cost.} Each pseudo-experiment, permutation, or bootstrap replicate requires retraining the ML model, which can be expensive for large networks. This is the genuinely new challenge that ML introduces relative to the classical setting --- not the conceptual problem of miscalibration (which is the same as it has always been) but the computational cost of the remedy. The Gross-Vitells trick of estimating $\Ncz$ at a low level and extrapolating analytically (Eq.~\ref{eq:gv_extrapolation}) cannot be applied here, because the null distribution of the ML test statistic is generally not $\chi^2$ and the analytical scaling does not hold. Full tail sampling is therefore needed, and at $5\sigma$ this requires $\mathcal{O}(10^7)$ pseudo-experiments --- each with a full retraining --- which is prohibitive for most current models.

The cost of generating each toy depends further on the dimensionality of the search. For a low-dimensional test --- even an ML-based one applied to a single observable or a handful --- toys can be drawn cheaply by resampling a fitted background model, exactly as in a classical bump hunt. When the search is genuinely high-dimensional and no low-dimensional background surrogate is available, each toy instead requires a full synthetic event sample through the simulation chain, a cost that compounds with the per-toy retraining discussed below. The same high dimensionality thus acts twice: it enlarges the effective trial factor (Section~\ref{sec:resolution}) and it raises the cost of the calibration needed to correct for it.
\subsubsection{Reduce: model regularisation}
\label{subsec:regularization}
Regularisation constrains the model's capacity, namely the complexity of the function it can learn from the data, and thereby controls the number of independent statistical fluctuations the model can capture. A more regularised model probes fewer structures in the feature space, which reduces the implicit trial factor at the cost, sometimes, of sensitivity to very localised signals. Several forms of regularisation are relevant:

\begin{itemize}
\item \emph{Smoothness constraints.} Weight decay (L2 penalty), weight clipping, and spectral normalisation all limit the magnitude or variation of the learned function, preventing the model from fitting sharp, localised fluctuations. In the NPLM framework, Ref.~\cite{d2021learning} introduced weight clipping as a means of enforcing Lipschitz continuity of the learned density ratio, which bounds the sensitivity of the test to pointwise fluctuations and provides a connection to robust hypothesis testing: a smoother density ratio corresponds to a test that is less powerful against highly localised alternatives but more robust against statistical noise.

\item \emph{Sparsity constraints.} Loss penalties, pruning, and architectural bottlenecks limit the number of active parameters or features, restricting the directions in the feature space that the model can probe simultaneously. Ref.~\cite{grosso2025sparse} introduced an anomaly detection model based on self-organising ensembles of a small number of Gaussian kernels, enforcing an aggressive sparsity constraint. This design effectively bounds the implicit LEE by construction: with only a handful of kernels, each covering a localised region of the feature space, the model cannot resolve more independent regions than it has kernels. The model was shown to perform powerful tests in problems with thousands of input dimensions, suggesting that sparsity can mitigate the resolution inflation that would otherwise make high-dimensional searches intractable.

\item \emph{Early stopping.} Halting training before convergence prevents the model from fitting fine-scale noise in the later stages of optimisation. This is an implicit form of regularisation whose effect on the resolution is harder to quantify than explicit constraints, but it is widely used in practice.

\end{itemize}

Generally, all forms of regularisation reduce the model's effective resolution, which reduces the implicit trial factor. The cost is always a loss of sensitivity to signals whose characteristic scale is finer than the regularised resolution. The optimal regularisation strength is therefore signal-dependent --- and in a signal-agnostic search, unknown. Moreover, it should not be assumed that all regularisation methods are equally effective at mitigating the implicit LEE. Different forms of regularisation constrain the model in different ways --- smoothness, sparsity, capacity --- and their effect on the number of independent directions the model can probe may differ substantially from their effect on standard measures of generalisation performance. A systematic study of which regularisation strategies most effectively reduce the implicit trial factor, and at what cost to sensitivity across different signal morphologies, is an important open problem.

\subsubsection{Avoid: train/test splitting and $k$-folding}
\label{subsec:splitting}

A different strategy is to ensure that the LEE does not contaminate the global p-value. Train/test splitting and $k$-folding achieve this by separating the data used for the search (training) from the data used for evaluation (testing). The model is free to be as expressive as it wants on its training set --- it may overfit aggressively, producing a large implicit LEE --- but the test statistic is computed on held-out data that the model has never seen, so the overfitting does not inflate the reported significance.

Ref.~\cite{hein2025} studies this approach for weakly supervised resonant anomaly detectors in the CWoLa framework. The paper compares two strategies: (a)~train/test splitting, which restores calibration by ensuring independence between the search and the scoring but uses only half the data for each; and (b) $k$-folding, which splits the data into $k$ parts and, for each part, trains a model on the other $k-1$ and uses it to score the held-out part. Every event is thus scored by a model that did not see it during training, and the scored events from all folds are pooled, so that the test statistic is computed on the full dataset. Since the $k$ models share most of their training data, their training and final configuration are not independent and as a result the $k$-fold p-values remain slightly biased.  The authors apply an empirical calibration fit to correct the residual bias. This correction is feasible because the $k$-fold bias is mild and smooth --- amenable to fitting with a moderate number of pseudo-experiments. In the original non calibrated test, the bias is too severe for the same approach: the test statistic is so data-specific that a smooth calibration function cannot capture it without direct tail sampling.

It is important to recognise that splitting and $k$-folding do not reduce the LEE --- they prevent it from affecting the evaluation by ensuring that the scored data is independent of the search. The implicit trial factor of the model may still be large; it simply does not contaminate the test statistic on the held-out fold. The cost is a reduction in the data available for scoring: a 50/50 split uses only half the data for evaluation, and even with $k$-folding each fold scores on only $1/k$ of the data. This leads to a loss of statistical power relative to using the full dataset.

\subsubsection{Calibration versus power}
\label{subsec:calibration_vs_power}

The strategies described above should be evaluated on two independent axes: \emph{calibration} (are the p-values correct under the null?) and \emph{power} (can the method detect a real signal after the LEE correction?). A method that produces perfectly calibrated p-values but has very low power is statistically valid but scientifically useless --- it never finds anything.

The three strategies occupy different positions in this two-dimensional space. Pseudo-experiment calibration is exact (the null distribution absorbs the full LEE by construction) and retains maximum power, because all the data are used for both training and scoring --- the test statistic is evaluated on the same data that defined the search, and the calibration corrects for whatever bias this introduces. Train/test splitting is also exact (the test data are independent of the search) but sacrifices power, because only a fraction of the data are used for scoring. $k$-folding interpolates: it uses more data for scoring than a single split, but the calibration is approximate and requires an empirical bias correction whose reliability at high significance has not been established. 
Regularisation operates on a different axis entirely: it does not calibrate the p-value but reduces the implicit LEE by constraining the model. This helps power when the signal’s characteristic scale is at least as coarse as the finest scale the regularised model can resolve: the signal is still captured, while the null distribution of the test statistic is lowered, and the threshold for a global significance with it. It hurts power when the signal is finer than that scale, since the model can no longer resolve it.

This framing clarifies why pseudo-experiment calibration should be regarded as the gold standard when computationally feasible: it is the only strategy that achieves exact calibration without sacrificing power. The sensitivity cost of the alternatives --- how much power $k$-folding or splitting loses relative to full calibration, and how this cost depends on the signal morphology and the dataset size --- is one of the most important open questions in this area.

\subsubsection{Combining strategies}
\label{subsec:combining}

The three strategies are complementary and can be combined to mutual advantage. Regularisation reduces the size of the implicit LEE, making the null distribution of the test statistic less heavy-tailed and therefore cheaper to calibrate via pseudo-experiments. This is essentially the approach taken by weight clipping a MLP model in Ref.~\cite{d2021learning}: the hard cut on the values that the trainable weights can take keeps the implicit LEE manageable, and the pseudo-experiment calibration makes the p-values exact. The combination is both cheaper (fewer pseudo-experiments needed because the tail falls faster) and more powerful (all the data is used for scoring) than splitting or $k$-folding alone.

In practice, the choice among strategies depends on the computational budget and the model complexity. For lightweight models where retraining is cheap (e.g., small networks, Boosted Decision Trees or Nystr\"om subsampled kernel methods), pseudo-experiment calibration is the gold standard and should be used directly. For expensive models where retraining is prohibitive (e.g., large transformers, deep generative models), $k$-folding with empirical bias correction may be the only feasible option --- but it should be understood as an approximation whose sensitivity cost relative to full calibration has not been systematically established. 
In all cases, regularisation should be as strong as the signals of interest allow. It reduces the difficulty of whichever calibration method is used downstream, but once the model can no longer resolve the finest signal scale the search aims to be sensitive to, further regularisation only costs power. In a signal-agnostic search, that scale is itself a choice the analyst must state (Section~\ref{sec:resolution}).

\subsection{Multiple ML tests on the same data}
\label{sec:multiple_ml}

A different facet of the LEE arises when multiple ML models --- differing in hyperparameters, architecture, or training procedure --- are applied to the same dataset. Each model defines a different test, but all tests act on the same data, so the test statistics are correlated. Reporting the most significant result across all models without correction inflates the false positive rate, just as scanning over mass does in a classical search.

\paragraph{Multiple tests via hyper-parameter scan.}

The problem of combining multiple kernel-based tests on the same data has been studied both in the statistics and in the HEP literature, with closely related but independently developed solutions.

In the statistics literature, the problem arises in two-sample tests based on the maximum mean discrepancy (MMD), a kernel-based measure of the difference between two distributions. The kernel bandwidth $\sigma$ sets the scale of the differences the test is sensitive to, so a single bandwidth can miss a signal at another scale. MMDAgg~\cite{schrab2023mmd} addresses this by running one test per bandwidth and rejecting if any of them rejects. The threshold of each test is set from the permutation distribution of its statistic, and the thresholds are calibrated jointly so that the combined test has the desired overall level. Because the permuted datasets go through the same multi-bandwidth computation as the real data, the correlation between bandwidths is accounted for automatically. MMD-FUSE~\cite{biggs2023mmd} replaces the ``any test rejects'' rule with a smooth combination of the normalised MMD values, a weighted soft maximum, again calibrated by permutations; the kernels can themselves be chosen from the data, without data splitting, in a way that keeps the permutation calibration valid. In both cases, the power adapts to the unknown scale of the signal without a Bonferroni-like penalty for the number of bandwidths.

In the HEP literature, Ref.~\cite{grosso2025multiple} studies the same problem in the context of NPLM, where different choices of the kernel width $\sigma$ define different tests on the same data. A key observation is that $\sigma$ directly controls the model's intrinsic resolution: tests with similar $\sigma$ values are highly correlated (they probe the same scales), while tests with well-separated $\sigma$ values are less correlated (they probe different scales).
In the resonant anomaly detection setting, Ref.~\cite{grosso2025robust} applies a related strategy: NPLM tests with several kernel widths and selection thresholds are run at each candidate resonance mass, and their results are combined to achieve sensitivity that is uniform across signal morphologies. The mass scan thus adds a continuous LEE on top of a discrete multiple-testing LEE, since only a small set of widths is used, each requiring its own training and calibration; the two must be corrected jointly. Both Ref.~\cite{grosso2025multiple} and Ref.~\cite{grosso2025robust} handle multiple testing by calibrating each kernel configuration individually via pseudo-experiments, and Ref.~\cite{grosso2025robust} finally corrects for the mass scan using the standard Gross--Vitells machinery.

Several features of these frameworks are worth highlighting. First, in NPLM each test is individually calibrated via pseudo-experiments before any combination is attempted: each test is converted to a calibrated p-value (or z-score) using its own null distribution estimated from pseudo-experiments, and then the combination is applied at the p-value level. 
This step is necessary since the combined tests follow in general different null distributions, and therefore the raw test statistics cannot be compared directly across configurations. 
Conversely, the MMD approaches calibrate the combined statistic directly as the tests all live in the same domain and are therefore comparable. 
The two routes are complementary: permutation-based calibration (MMDAgg, MMD-FUSE) is computationally cheaper but requires exchangeability of the samples; pseudo-experiment calibration (NPLM) is more expensive but applies to any test statistic, including those with nuisance parameters that break exchangeability. Second, the achieved sensitivity is more uniform across signal types than any single test, approaching that of the best-performing configuration without requiring it to be known a priori --- a property shared by all three methods (MMDAgg, MMD-FUSE, NPLM combination).

\section{Illustrative experiments}\label{sec:experiments}
To illustrate the concepts discussed in the previous sections, we present a set of simple but physically motivated examples that highlight the main strategies for handling the LEE in searches for new physics. 
Their purpose is educational. They are aimed at readers who are not yet familiar with these methods, and are kept deliberately basic, so that each step can be followed and reproduced without specialised software or large computing resources. Despite their simplicity, they retain the essential features of the LEE in a realistic collider search. Readers already familiar with the methods may skip this section, although the examples also bring out practical aspects that are easy to overlook, such as the dependence of the Gross--Vitells estimate on the granularity of the scan (Section~\ref{sec:exp_tf}), and of the trial factor on the width of the signal template (Section~\ref{sec:exp_width}).\footnote{The code used to produce all results in this section is available at \url{https://github.com/serodribe/LEE_review}.} Further details and cross-checks of the examples are collected in Appendix~\ref{app:experiments}.

We divide the experiments into two main categories: single-channel bump-hunt (Section~\ref{subsec:exp1}), which illustrates the continuous-scan corrections of Section~\ref{sec:classical}, and multi-channel searches (Section~\ref{subsec:exp2}), which illustrate the multiple-testing corrections of Section~\ref{sec:multipletesting}. 

\subsection{Single channel bump-hunt}\label{subsec:exp1}
\begin{figure}[h]
  \centering
  \includegraphics[width=0.6\textwidth]{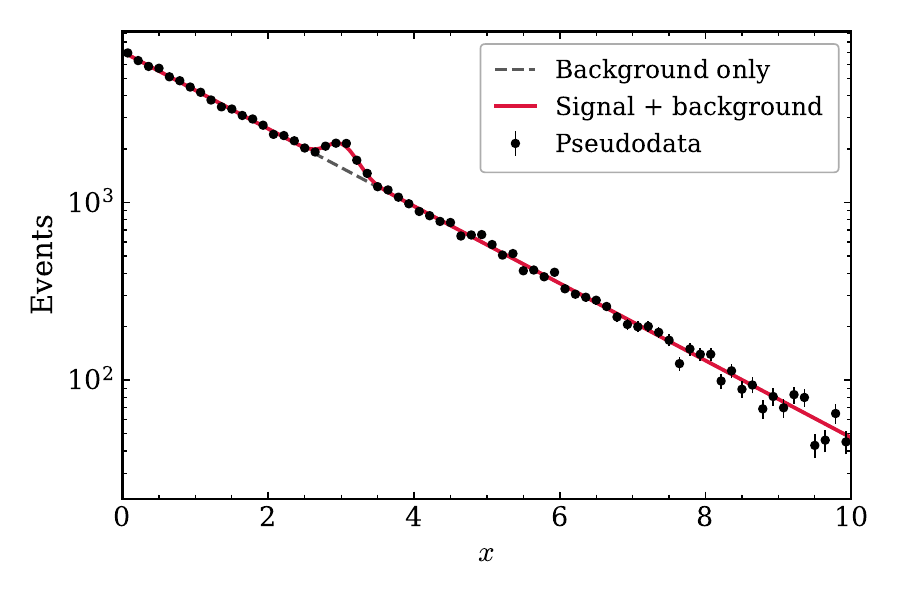}
    \caption{Example pseudo-dataset of $N_\text{ev}=100$k events (with Poisson uncertainties) generated from Eq.~\eqref{eq:toy_model} with a signal injected at $m=3$, $\omega=0.2$ and $\mu=0.02$. The lines show the true background-only (dashed black line) and signal-plus-background (solid red line) densities, scaled to the number of events.}

  \label{fig:single_channel}
\end{figure}

This example illustrates the LEE in a single-channel bump hunt: the search for a narrow Gaussian resonance on top of a smoothly falling background. Events are distributed in an observable $x\in[0,10]$ according to the density
\begin{equation}
  f(x\,|\,\mu,m,\nu;\;\omega)=(1-\mu)\,\frac{\nu\,e^{-\nu x}}{1-e^{-10\nu}}+\mu\,g(x\,|\,m,\omega),
  \label{eq:toy_model}
\end{equation}
where $\mu\geq0$ is the signal fraction, $g(x\,|\,m,\omega)$ is a Gaussian density of mean $m$ and width $\omega$ normalised on $[0,10]$, and $\nu$ is the decay rate of the exponential background. Each pseudo-dataset contains $N_\text{ev}=100$k events, binned in $70$ bins of width $0.14$. The signal width is assumed known and is fixed to $\omega=0.2$ in the fit. The location $m$ is unknown and is scanned over the full range $[0,10]$, and the background decay rate $\nu$ is a free parameter. Figure~\ref{fig:single_channel} shows an example pseudo-dataset generated with $\nu=0.5$ and a signal injected at $m=3$ with $\mu=0.02$. 
All other results in this section are obtained from background-only pseudo-experiments (toys), generated with $\mu=0$.

With this setup, the experiment falls within the framework introduced in Section~\ref{sec:classical}, where $H_0$ is now the exponential background and the continuous parameter $\theta$ that labels the family of alternative hypotheses is the resonance mass $m$.

\subsubsection{Test statistic and $p$-values}
As test statistic we use the profile likelihood ratio for the discovery of a positive signal at location $m$~\cite{Cowan:2010js},
\begin{equation}
    q(m)=
    \begin{cases}
      -2\ln\dfrac{\mathcal{L}(\mu=0,\,\hat{\hat{\nu}})}{\mathcal{L}(\hat{\mu}_m,\,\hat{\nu}_m;\,m)} & \hat{\mu}_m\geq 0,\\[6pt]
      0 & \hat{\mu}_m<0,
    \end{cases}
    \label{eq:qmu}
\end{equation}
where $\hat{\hat{\nu}}$ is the maximum-likelihood estimate of $\nu$ under the background-only hypothesis, and $\hat{\mu}_m$ and $\hat{\nu}_m$ are the joint maximum-likelihood estimates of $\mu$ and $\nu$ at fixed $m$. Setting $q(m)=0$ when $\hat\mu_m<0$, which is equivalent to constraining $\mu\geq0$ in the fit, restricts the test to excesses.

Because the location of the peak is unknown, we effectively perform one hypothesis test for each value of $m$, which is the origin of the LEE. The test statistic of the scan is the maximum
\begin{equation}
  q(\hat m)=\max_{m\in[0,10]} q(m),
\end{equation}
which corresponds to $M$ in Eq.~\eqref{eq:Msup}, with $S\to q$ and $\theta\to m$. At a mass $m_0$ fixed in advance, the null distribution of $q(m_0)$ is asymptotically $\tfrac12\delta(q)+\tfrac12\chi^2_1$~\cite{Cowan:2010js}. We refer to this distribution as $\tfrac12\chi^2_1$ in the following. No such closed form exists for $q(\hat m)$. Its null distribution must either be sampled with pseudo-experiments or approximated with the methods of Section~\ref{sec:classical}, as we do in Section~\ref{sec:exp_tf}.

\begin{figure}[h!]
    \centering
    \includegraphics[width=0.48\textwidth]{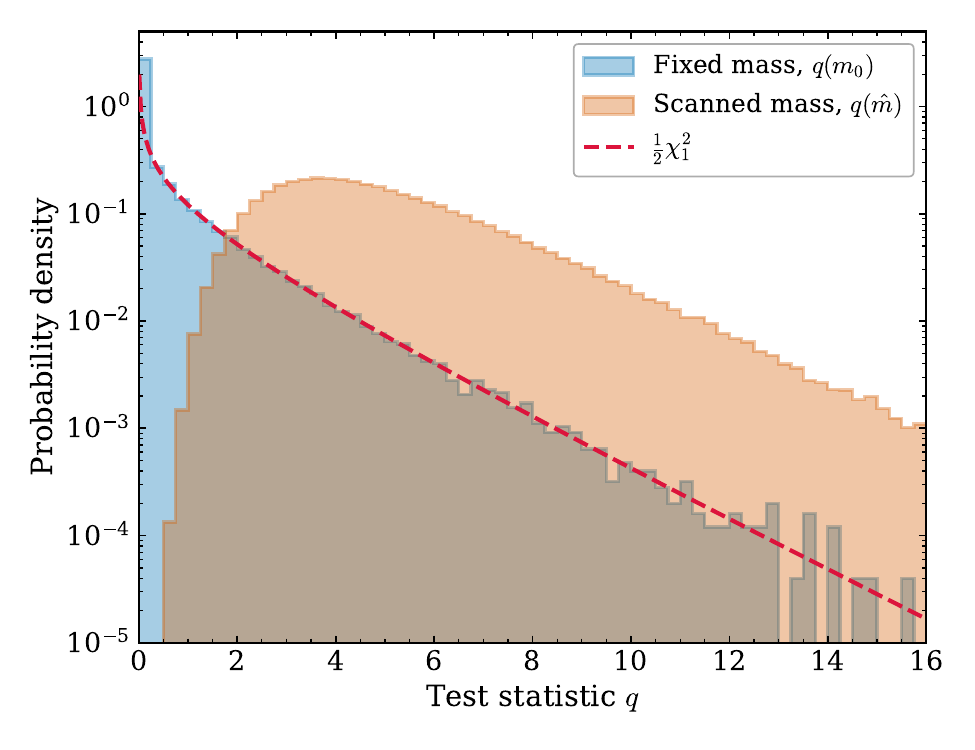}\hfill
    \includegraphics[width=0.48\textwidth]{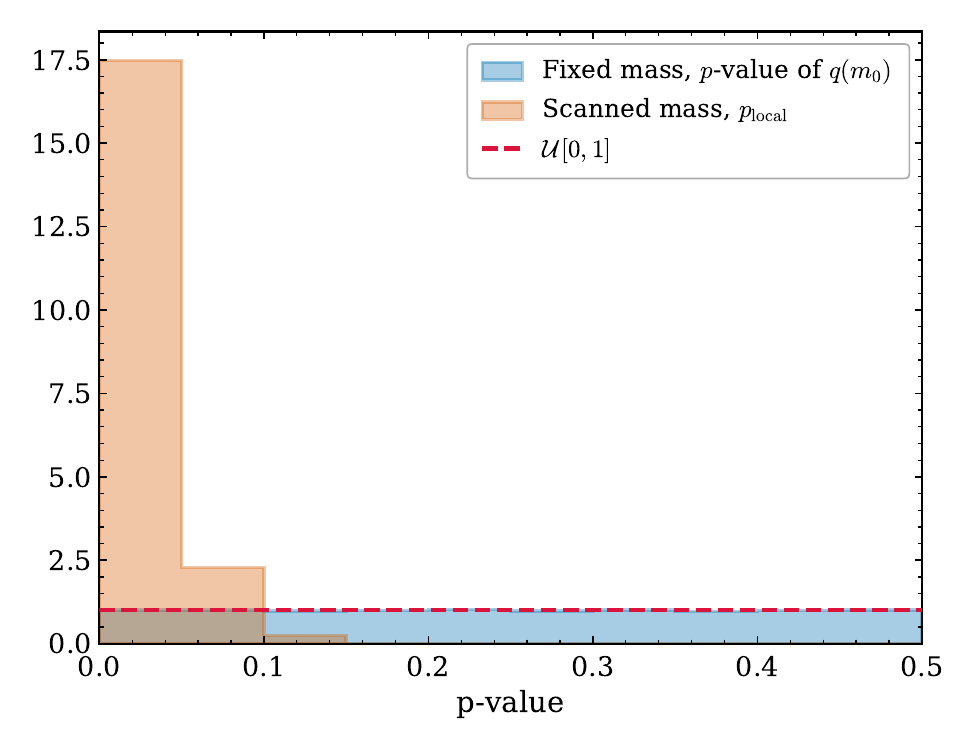}
    \caption{Left: distribution of the test statistic in background-only pseudo-experiments, evaluated at a fixed mass, $q(m_0)$ with $m_0=3$ ($10^5$ pseudo-experiments), and at the best-fit mass, $q(\hat m)=\max_m q(m)$ ($3\times10^5$ pseudo-experiments, scan over $350$ mass points), compared with the asymptotic $\tfrac12\chi^2_1$ density. The first bin includes the entries with $q=0$. Right: $p$-values obtained from the asymptotic $\tfrac12\chi^2_1$ tail probability for the fixed-mass statistic and for the local $p$-value $p_\text{local}$ up to the level of interest $p<0.5$. The dashed line shows the uniform density expected for a calibrated $p$-value.}
    \label{fig:assym_q}
\end{figure}

The left panel of Figure~\ref{fig:assym_q} compares the distributions of $q(m_0)$, with $m_0=3$, and of $q(\hat m)$ in background-only pseudo-experiments with the asymptotic $\tfrac12\chi^2_1$ prediction. The right panel shows the $p$-values obtained by evaluating both statistics with the same asymptotic tail probability. For $q(m_0)$ this gives a valid $p$-value for a test at a pre-specified mass. For $q(\hat m)$ it gives the local $p$-value $p_\text{local}$ of Section~\ref{sec:classical}, which treats the best-fit mass as if it had been fixed in advance. Under the null hypothesis, a calibrated $p$-value satisfies $P(p\leq\alpha)=\alpha$; with the one-sided definition of Eq.~\eqref{eq:qmu}, this holds for all $\alpha<\tfrac12$, which covers every level of interest. The fixed-mass $p$-value satisfies this condition, whereas $p_\text{local}$ is strongly biased towards small values.

\subsubsection{Trial factor estimation}\label{sec:exp_tf}
\begin{figure}[htbp]
    \centering
    \includegraphics[width=0.49\textwidth]{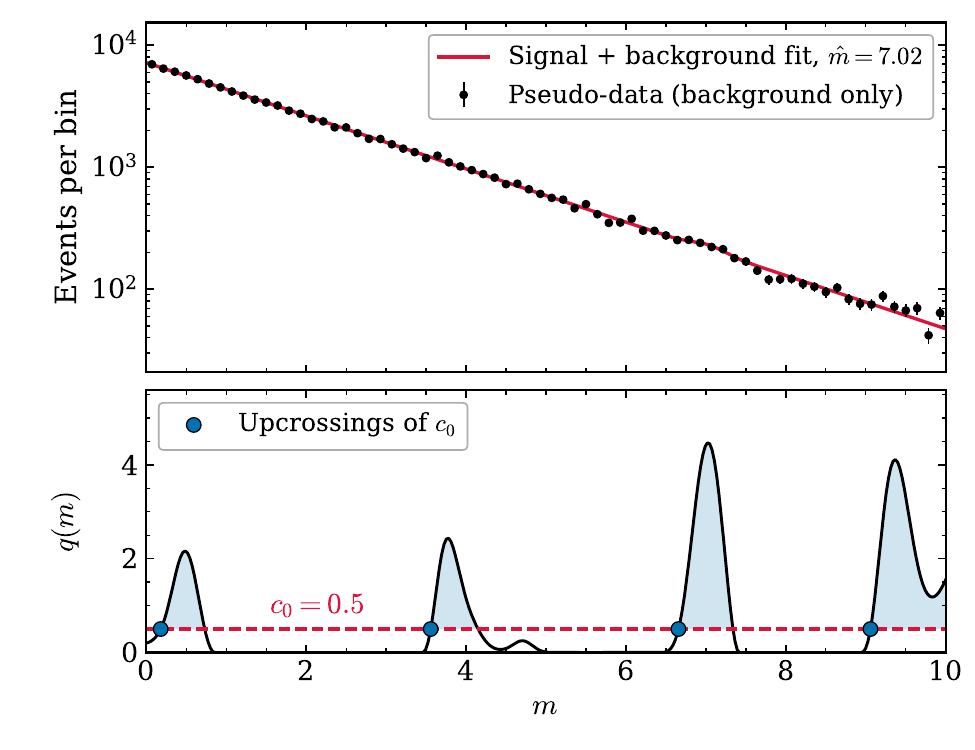}
    \includegraphics[width=0.49\textwidth]{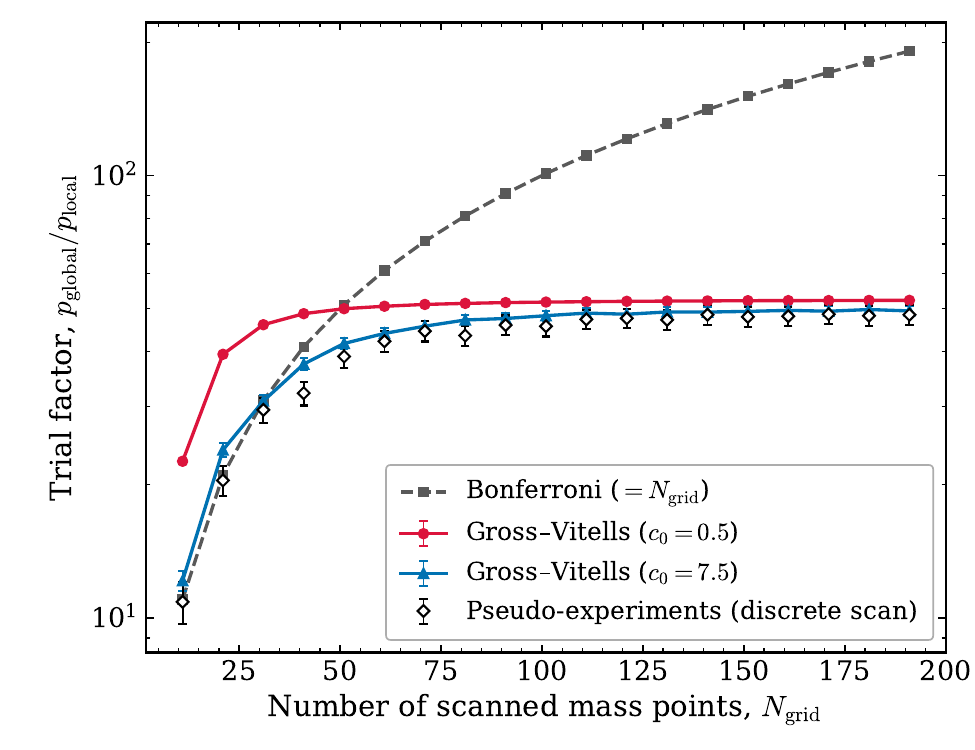}
    \caption{Left: a background-only pseudo-dataset with the fit of $f(x|\mu, m,\nu)$ at the best-fit mass $\hat m=7.02$ (top), and the corresponding test statistic $q(m)$ on the 350-point grid (bottom). The GV procedure is illustrated by showing the upcrossings (circles) of the reference level $c_0=0.5$ (dashed red line): four in this pseudo-dataset, compared with $\langle N(c_0)\rangle\approx4.6$ on average; a similar illustration is given in Ref.~\cite{gross2010}. Right: trial factor at $c=10$ as a function of the number $N_\text{grid}$ of equally spaced mass points in the scan, for the Bonferroni bound ($=N_\text{grid}$), for the GV procedure with $c_0=0.5$ and $c_0=7.5$, and for the discrete scan, estimated directly from the fraction of pseudo-experiments with $\max_i q(m_i)\geq c$ (open diamonds). The signal width $\omega = 0.2$ matches the grid spacing for $N_\text{grid} = 51$ points. All estimates use the same $10^4$ background-only pseudo-experiments.}
    \label{fig:upcrossings-trials_factor}
\end{figure}

The trial factor is the ratio $p_\text{global}/p_\text{local}$, where $p_\text{global}=P(q(\hat m)>c\,|\,H_0)$ and $p_\text{local}=\tfrac12P(\chi^2_1>c)$ for an observed value $c$ of $q(\hat m)$. Estimating $p_\text{global}$ directly from pseudo-experiments is computationally expensive when $c$ is large. A global significance of $5\sigma$ requires $\mathcal{O}(10^7)$ toys. The Gross--Vitells (GV) procedure of Section~\ref{sec:gv} avoids this. It counts the upcrossings of a low reference level $c_0$ by $q(m)$, whose expected number $\langle N(c_0)\rangle$ can be estimated from a small number of toys, and extrapolates to the level of interest with Eq.~\eqref{eq:gv_extrapolation}. Since the test statistic of Eq.~\eqref{eq:qmu} is sensitive only to positive signals, the first term of Eq.~\eqref{eq:gv_asymptotic} must be replaced by the corresponding local p-value defined above. With one fitted signal parameter ($s=1$), this gives
\begin{equation}
  p_\text{global}\approx p_\text{local}+\langle N(c_0)\rangle\,e^{-(c-c_0)/2}.
  \label{eq:gv_onesided}
\end{equation}
The left panel of Figure~\ref{fig:upcrossings-trials_factor} shows the upcrossings of $c_0=0.5$ in a background-only toy.
Although this pseudo-dataset contains no signal, its largest excursion, at $m=7.0$, reaches $q\approx4.5$, a local significance of $2.1\sigma$; once the look-elsewhere effect is accounted for with Eq.~\eqref{eq:gv_onesided}, no significant excess remains (global $p$-value $\approx0.65$).

The right panel of Figure~\ref{fig:upcrossings-trials_factor} shows how the trial factor depends on the number $N_\text{grid}$ of equally spaced mass points in the scan. Two different quantities must be distinguished here. The first is the trial factor of the discrete scan, $P\big(\max_{i=1,\dots,N_{\text{grid}}} q(m_i)>c\,\big|\,H_0\big)/p_\text{local}$, which is the relevant one for an analysis that evaluates $q(m)$ only at the $N_\text{grid}$ grid points. It grows with $N_\text{grid}$ but, since $p_\text{global}\leq N\,p_\text{local}$ for any $N$ tests, it never exceeds the Bonferroni value. The second is the trial factor of the continuous scan, defined as the limit of the first as the grid becomes infinitely dense; this is the quantity the GV procedure is designed to estimate. Pseudo-experiments analysed on the $N_\text{grid}$-point grid estimate the discrete-scan trial factor (open diamonds in the figure). The GV procedure, applied to the same grid, approaches the continuous-scan trial factor instead, and becomes independent of $N_\text{grid}$ once the grid is dense enough to resolve every upcrossing of $c_0$. Depending on the grid spacing, two regimes can therefore be distinguished:

\begin{itemize}
    \item {\bf Coarse grid.} When the grid spacing is larger than the width of the peaks of $q(m)$, which is set by the width $\omega$ of the signal template, the scanned points are effectively independent tests. The discrete-scan trial factor is then close to $N_\text{grid}$, and the Bonferroni value is an accurate estimate of it. The GV estimate, on the other hand, is not: the extrapolation in Eq.~\eqref{eq:gv_onesided} assumes a smooth process, whose tail falls more slowly with $c$ than that of a set of independent points. As a result, the GV estimate overestimates the discrete-scan trial factor, and the overestimate grows with the distance between $c_0$ and $c$. At small $N_\text{grid}$, the $c_0=0.5$ curve lies about a factor of two above the Bonferroni line, whereas for $c_0=7.5$ the extrapolation is short and the GV estimate stays close to the Bonferroni value (Appendix~\ref{app:grid}).

    \item {\bf Dense grid.} When the grid spacing is smaller than both the distance between consecutive upcrossings and the width of the excursions above $c_0$, every upcrossing is resolved and counted. The GV estimate no longer depends on $N_\text{grid}$ and estimates the continuous-scan trial factor, which is much smaller than the Bonferroni value. The value of $N_\text{grid}$ needed to reach this regime depends on $c_0$. The excursions above $c_0$ become narrower as $c_0$ increases (left panel of Figure~\ref{fig:upcrossings-trials_factor}), so a higher reference level requires a denser grid. In this example, the GV estimate is within $5\%$ of its dense-grid value once the grid spacing is below about $1\,\omega$ for $c_0=0.5$, and below about $0.6\,\omega$ for $c_0=7.5$.
   
\end{itemize}

When $c$ is large, the GV procedure has two practical advantages. First, $\langle N(c_0)\rangle$ is estimated at a low reference level, where almost every toy contributes, so far fewer pseudo-experiments are needed than when probing the tail at $c$ directly. Second, the intervals of $m$ in which $q(m)$ exceeds a low level $c_0$ are wider than those in which it exceeds a high level $c$, since only the tips of the tallest peaks rise above $c$. A coarser grid is therefore enough to resolve them, and the GV estimate stabilises at a smaller $N_\text{grid}$ than any method that works directly at $c$.

Table~\ref{tab:trial_factors} quantifies both advantages at $c=10$. It compares three estimates of the trial factor obtained from the same pseudo-experiments: the Davies bound of Eq.~\eqref{eq:davies_bound}, with $\langle N(c)\rangle$ obtained by counting upcrossings directly at $c$; the GV estimate of Eq.~\eqref{eq:gv_onesided} with $c_0=0.5$; and the fully empirical estimate, given by the fraction of pseudo-experiments with $q(\hat m)>c$. The table also reports the Bayer--Seljak estimate, which requires no pseudo-experiments and is discussed at the end of this section.

On the 350-point grid (spacing $\omega/7$), the Davies and empirical estimates agree to within $1.5\%$, which shows that the discrete scan is already close to the continuous limit. The GV estimate is a few per cent higher. This is because, between $c_0$ and $c$, the expected number of upcrossings falls slightly faster than the asymptotic factor $e^{-(c-c_0)/2}$ of Eq.~\eqref{eq:gv_onesided}, so the extrapolation is slightly conservative in this example.

On the 100-point grid (spacing $\omega/2$), the GV estimate changes by only about $1\%$ with respect to the 350-point value, whereas the Davies and empirical estimates both drop by about $5\%$, because the coarser grid does not fully resolve the narrow peaks of $q(m)$ above $c$. Since the empirical estimate is, in effect, the trial factor of the 100-point discrete scan itself, it is the appropriate reference for an analysis performed on that grid; relative to it, the GV estimate is conservative by about $8\%$. The GV estimate is also the most precise: with the same pseudo-experiments, its statistical uncertainty is about twenty times smaller than that of the Davies and empirical estimates. This is mainly because the reference level $c_0$ is crossed more than once in each pseudo-experiment, whereas $c$ is crossed only $ N(c) \approx 0.04$ times on average.

\begin{table}[htbp]
    \centering
    \caption{Trial factor at $c=10$ estimated with four methods on two grids of equally spaced mass points. The first three use the same $3\times10^5$ background-only pseudo-experiments. The Davies and fully empirical estimates refer to the discrete scan on each grid, and the GV estimate to the continuous scan (see text). Their uncertainties are statistical: for the Davies and GV values they are propagated from the uncertainty on the mean number of upcrossings, and for the empirical value they are binomial, from the fraction of pseudo-experiments with $q(\hat{m})\geq c$. 
    The Bayer--Seljak estimate (Eq.~\eqref{eq:bs_exp} and Appendix~\ref{app:bayes}) is computed separately for each dataset, from the height and width of its highest peak. To compare it with the other methods at $c=10$, we use the pseudo-experiments whose highest peak has $8<q(\hat m)<12$, rescale the width of each peak to $q(\hat m)=10$ using $\sigma_m\propto1/\sqrt{q(\hat m)}$, and give the mean trial factor with its statistical uncertainty. The value in brackets is the standard deviation between datasets; it reflects the dependence of the estimate on the width of the observed peak (Appendix~A.2), not a statistical uncertainty of the method.
    }
    \label{tab:trial_factors}
    \begin{tabular}{lccc}
        \toprule
        Method & $N_\text{toys}$ & \multicolumn{2}{c}{$N_\text{grid}$} \\
        \cmidrule(lr){3-4}
         & & 100 ($\omega/2$) & 350 ($\omega/7$) \\
        \midrule
        Davies, Eq.~\eqref{eq:davies_bound} & $3\times10^5$ & $48.36 \pm 0.45$ & $50.81 \pm 0.46$ \\
        GV, Eq.~\eqref{eq:gv_onesided}, $c_0=0.5$ & $3\times10^5$ & $51.56 \pm 0.02$ & $52.08 \pm 0.02$ \\
        Fully empirical & $3\times10^5$ & $47.76 \pm 0.44$ & $50.24 \pm 0.45$ \\
        Bayer--Seljak, Eq.~\eqref{eq:bs_exp} & -- & $51.36 \pm 0.06$ ($\pm10$) & $52.24 \pm 0.06$ ($\pm10$) \\
        \bottomrule
    \end{tabular}

\end{table}

\paragraph{An estimate without pseudo-experiments.}
Section~\ref{sec:bayes} described how Bayer and Seljak~\cite{Bayer:2020pva} use the Bayesian evidence to define the trial factor of a frequentist test statistic, as the ratio of prior to posterior volume of the signal location. In this example the prior range is the scan range, $U-L=10$, and the posterior width $\sigma_m$ is the precision with which the data localise the largest excess, which follows from the curvature of $q(m)$ at its maximum, $-\tfrac12\,q''(\hat m)=1/\sigma_m^2$. The trial factor, the test statistic and the global $p$-value are then
\begin{equation}
\begin{aligned}
  N_B&=\frac{U-L}{\sqrt{2\pi}\,\sigma_m},\\
  q_S&=q(\hat m)-2\ln N_B+\ln\big(2\pi q(\hat m)\big),\\
  p_{\rm global}&\approx 1-\exp\big(-e^{-q_S/2}\big)=1-\exp\!\left(-N_B\,\frac{e^{-q(\hat m)/2}}{\sqrt{2\pi\,q(\hat m)}}\right)
\end{aligned}
\label{eq:bs_exp}
\end{equation}
where, as in Ref.~\cite{Bayer:2020pva}, the local $p$-value is used in its asymptotic form. Both quantities are computed from the observed dataset alone: no pseudo-experiments are needed. Applied to each of the background-only pseudo-experiments of Table~\ref{tab:trial_factors}, the Bayer--Seljak trial factor at $c=10$ agrees on average with the GV estimate to within $1\%$ on both grids. Near a high peak, $N_B$ reproduces the leading term of the Davies bound, and both estimate the same oscillation rate of the process. Since $N_B$ depends on the width of the observed peak, the Bayer–Seljak $p$-value is that of a different test statistic, $q_S$, and its trial factor differs from the GV one by about 20\% from one dataset to the next. This corresponds to a shift of less than $0.1\sigma$ in the global significance at the discovery level, and the resulting global $p$-value is correctly calibrated as a test (Appendix~\ref{app:bayes}). The method relies on the Laplace approximation of Section~\ref{sec:bayes}, i.e.\ on a smooth, approximately Gaussian likelihood peak, and, like the GV procedure, on an explicit parametrisation of the signal, so it does not apply directly to the machine-learning test statistics of Section~\ref{sec:ml}.

\subsubsection{Dependence on the signal-template width}\label{sec:exp_width}
\begin{figure}[h!]
    \centering
    \includegraphics[width=0.6\textwidth]{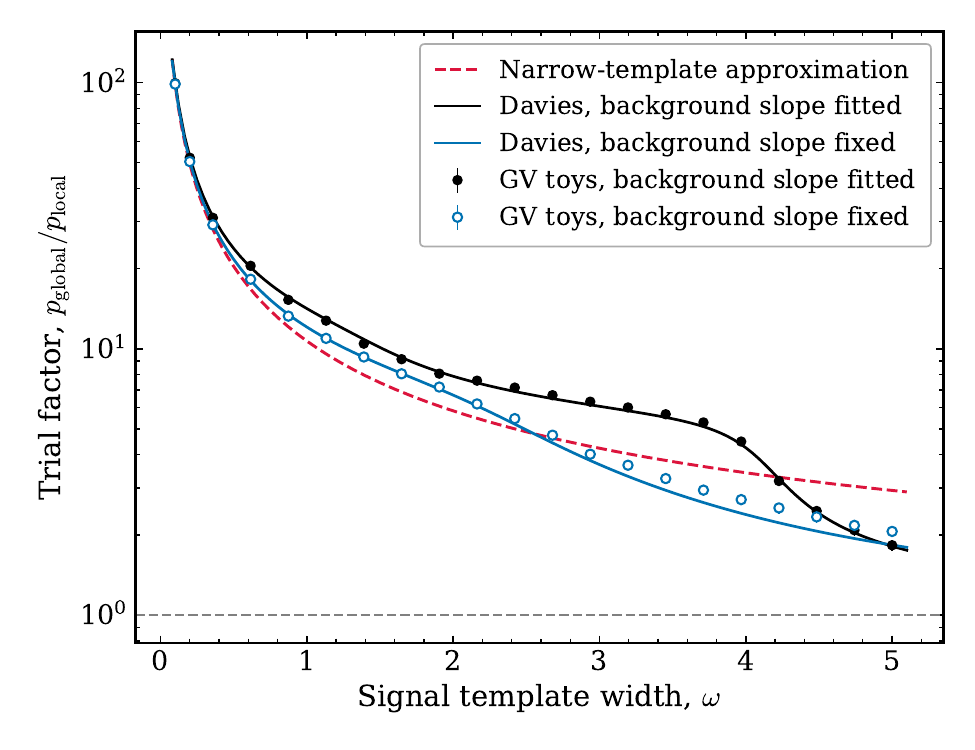}
    \caption{Trial factor at $c=10$ as a function of the width $\omega$ of the signal template used in the fit, from the GV procedure with $c_0=0.5$ applied to $1000$ background-only pseudo-experiments on a grid of $350$ mass points, with the background slope fitted (filled points) or fixed to its true value (open points). The red dashed line shows the narrow-template approximation of Eq.~\eqref{eq:tf_analytic}. The solid lines show the analytical Davies prediction with the correlation of Eq.~\eqref{eq:rho_binned}, with the background slope fitted (black) or fixed (blue). The grey dashed line marks a trial factor of one. The solid lines and the open points are discussed in Appendix~\ref{app:width}.}
    \label{fig:width_dependence}
\end{figure}

The width of the signal template used in the fit sets how rapidly $q(m)$ varies with $m$, and therefore how many upcrossings of a given reference level occur over the scan range. Figure~\ref{fig:width_dependence} shows the GV trial factor at $c=10$ as a function of the template width $\omega$. Since the pseudo-experiments are background-only, their distribution does not depend on the width: only the template used in the fit changes. Broader templates produce $q(m)$ curves that vary more slowly with $m$, with fewer upcrossings and hence a smaller trial factor. This matches the intuition that the effective number of independent tests scales roughly as the ratio of the scan range to the template width. For widths comparable to the scan range, the signal templates at different masses overlap almost entirely. The tests at different masses are then strongly correlated, the scan behaves almost like a single test, and the trial factor approaches one.

We can quantify this intuition with the analytical Davies bound of Section~\ref{subsec:davies}. It applies to the signed square root of the likelihood ratio, $Z(m)$, which is positive for excesses and negative for deficits, satisfies $q(m)=Z(m)^2$ when $Z(m)>0$, and is asymptotically a Gaussian process with unit variance under the null hypothesis (Appendix~\ref{app:width}). Evaluated at the level $\sqrt{c}$ over the scan range $[L,\,U]=[0,10]$, the first term of Eq.~\eqref{eq:davies_normal} is then $p_\text{local}$, and the second the expected number of upcrossings, $\langle N(c)\rangle=\mathcal{N}e^{-c/2}$, with $\mathcal{N}=\frac{1}{2\pi}\int_L^U\sqrt{-\rho_{11}(m)}\,dm$. For a Gaussian template, and neglecting the edges of the range and the fit of the background normalisation and slope, the correlation of the process is $\rho(m',m)=\exp[-(m'-m)^2/4\omega^2]$, independently of the exponential background, so that $\rho_{11}=-1/(2\omega^2)$ and
\begin{equation}
  \mathcal{N}=\frac{U-L}{2\sqrt{2}\,\pi\,\omega}.
  \label{eq:N_analytic}
\end{equation}

This is the functional form of Eq.~\eqref{eq:gv_onesided}, except that $\langle N(c_0)\rangle$ is now computed analytically. For $\omega=0.2$, Eq.~\eqref{eq:N_analytic} gives $\mathcal{N}\approx5.6$ and $\langle N(c_0{=}0.5)\rangle\approx4.4$, compared with 4.6 measured with toys, and a trial factor of about 49 at $c=10$, about $5\%$ below the GV estimate on the dense grid in Table~\ref{tab:trial_factors}. More generally, Eqs.~\eqref{eq:gv_onesided} and~\eqref{eq:N_analytic} give the trial factor at any level $c$,
\begin{equation}
  \frac{p_\text{global}}{p_\text{local}}\approx1+\kappa(c)\,\frac{U-L}{\omega},\qquad
  \kappa(c)=\frac{e^{-c/2}}{2\sqrt{2}\,\pi\,\Phi(-\sqrt{c})}.
  \label{eq:tf_analytic}
\end{equation}
This makes the familiar rule of thumb that the trial factor is the ratio of the scan range to the template width more quantitative. The rule captures the dependence on $\omega$, while the prefactor $\kappa$ grows with the local significance $\sqrt{c}$, as in Eq.~\eqref{eq:trial_factor}. For local significances of $3$, $4$ and $5\sigma$, $\kappa\approx0.93$, $1.2$ and $1.5$, so the rule is numerically accurate only around $3\sigma$ and underestimates the trial factor at the discovery level.

Figure~\ref{fig:width_dependence} compares Eq.~\eqref{eq:tf_analytic} at $c=10$ with the GV estimates. The approximation is accurate for narrow templates, within $1\%$ at $\omega=0.1$ and $5\%$ at $\omega=0.2$. This is the regime of most resonance searches, in which the template is much narrower than both the scan range and the scale over which the background varies. For broader templates, the GV points lie up to $50\%$ above the approximation, and fall below it only for $\omega\gtrsim4.2$, after a sharp drop. The discrepancy does not come from the GV procedure but from the two simplifications behind Eq.~\eqref{eq:N_analytic}: the fit of the background absorbs part of the signal template, and the templates are truncated at the edges of the range. Once both effects are included in the correlation of the process (Appendix~\ref{app:width}), the analytical Davies prediction (solid black line) agrees with the GV estimates to within $4\%$ for all widths, including the sharp drop. The Davies/GV framework therefore describes the trial factor accurately for all widths, once the correlation of the process is computed for the actual fit; Eq.~\eqref{eq:tf_analytic} is a useful approximation for templates much narrower than the scan range and than the scale of the background.

This example shows explicitly that the LEE is not determined by the search range alone, but also by the signal model used in the scan (Section~\ref{sec:model_dependence}). In many high-energy physics searches, the width of the template is set by the detector resolution, so the trial factor is set by the ratio of the scan range to the resolution.
In other cases, such as resonances of unknown intrinsic width, the width becomes a free parameter, and the scan extends over an additional dimension, which further increases the LEE (Section~\ref{sec:multidim}). Since the trial factor grows as $1/\omega$, such a scan is dominated by its narrowest templates, and the lower end of the width range should be motivated physically, for example by the detector resolution, rather than chosen for convenience.

\subsection{Multi-channel search}\label{subsec:exp2}
Many searches in high-energy physics involve several channels, for example different decay modes of the same particle, or event categories defined by detector region or by the number of reconstructed objects. In these cases, the LEE arises not only from the scan over the signal parameters within each channel, but also from the number of channels considered, which is case~(ii) of Section~\ref{sec:case_ii}. In this section, we present two simple examples of multi-channel searches, and illustrate how the multiple-testing corrections of Section~\ref{sec:multipletesting} handle this additional layer.

\subsubsection{Decay channels of a new particle}\label{sec:exp_decay}
We consider a resonance that decays into $N_\text{ch}=6$ channels with unknown branching ratios. Each channel is modelled as in Section~\ref{subsec:exp1}, by Eq.~\eqref{eq:toy_model} with its own background and signal fraction $\mu_i$. The resonance mass is unknown but common to all channels, while the width of the Gaussian signal template differs from channel to channel, reflecting the different detector resolution of each final state. The resonance is narrow, with an intrinsic width $\Gamma=0.02$ much smaller than the channel resolutions $\omega_i\in[0.15,0.40]$, so the observed signal in channel $i$ is a Gaussian of width $\sqrt{\Gamma^2+\omega_i^2}\approx\omega_i$, and each channel is fitted with a template of width $\omega_i$. A signal at $m=3$ is injected in three channels, two with a strong and one with a weak signal, with signal fractions chosen to give an expected local significance of $4.5\sigma$ in channels~1 and~6 and of $3.5\sigma$ in channel~3;\footnote{The expected local significance is $\sqrt{q(m)}$ at the true mass for the Asimov dataset, in which the bin contents are equal to their expected values~\cite{Cowan:2010js}.} the remaining three contain background only. The parameters of each channel, together with the results of the per-channel scans in one pseudo-experiment, are listed in Table~\ref{tab:channels}.

\begin{table}[htbp]
  \centering
  \caption{Setup and results for the multi-channel example. For each channel, the first four columns give the number of events, the background decay rate, the template width (equal to the detector resolution) and the injected signal fraction. The remaining columns show the results for one pseudo-experiment: the best-fit mass, the maximum of the test statistic, and the GV-corrected global $p$-value (clipped at one).}
  \label{tab:channels}
  \begin{tabular}{ccccc|ccc}
    \toprule
    Channel & $N_\text{ev}$ & $\nu$ & $\omega$ & $\mu$ & $\hat m$ & $q(\hat m)$ & $p_\text{GV}$ \\
    \midrule
    1 & $10^5$ & 0.50 & 0.15 & 0.0037 & 3.04 & 26.4 & $1.5\times10^{-5}$ \\
    2 & $10^5$ & 0.30 & 0.25 & 0 & 1.63 & 2.9 & 1.00 \\
    3 & $10^5$ & 0.40 & 0.20 & 0.0035 & 3.18 & 5.5 & 0.39 \\
    4 & $10^5$ & 0.20 & 0.40 & 0 & 0.20 & 2.6 & 0.92 \\
    5 & $10^5$ & 0.60 & 0.30 & 0 & 6.91 & 5.4 & 0.29 \\
    6 & $10^5$ & 0.35 & 0.35 & 0.0062 & 3.04 & 38.1 & $1.9\times10^{-8}$ \\
    \bottomrule
  \end{tabular}
\end{table}

We first treat each channel separately and ask which channels contain a signal. This is the per-channel formulation of Section~\ref{sec:multipletesting}, appropriate when a separate decision is required for each channel. The analysis has two levels, as in the hierarchical searches of Section~\ref{sec:case_ii}. Within each channel, the scan over the resonance mass is corrected with the GV procedure, which gives one global $p$-value per channel. The expected number of upcrossings $\langle N(c_0)\rangle$ is estimated separately for each channel, from $5000$ background-only pseudo-experiments generated with the background of that channel and fitted with its template; it ranges from $6.0$ for the narrowest template ($\omega=0.15$) to $2.5$ for the broadest ($\omega=0.40$), following the dependence on the template width discussed in Section~\ref{sec:exp_width}. Across channels, the LEE from the multi-channel search is accounted for by correcting the threshold applied to these $p$-values, rather than the $p$-values themselves. For this second level, we compare the Bonferroni and \v{S}idák corrections, which control the family-wise error rate (Section~\ref{sec:bonferroni}), with the Benjamini--Hochberg (BH) procedure (Section~\ref{sec:bh}), which controls the false discovery rate. The first two replace the nominal level $\alpha$ by a per-channel threshold, $\alpha/N_\text{ch}$ for Bonferroni and $1-(1-\alpha)^{1/N_\text{ch}}$ for \v{S}idák. BH instead uses rank-dependent thresholds $k\alpha/N_\text{ch}$ for the $k$-th smallest $p$-value. Since the channels are statistically independent, the \v{S}idák threshold is exact here. For illustration, we use a level $\alpha=0.05$ throughout, much looser than a discovery threshold in high energy physics.

\begin{figure}[htbp]
    \centering
    \includegraphics[width=0.85\textwidth]{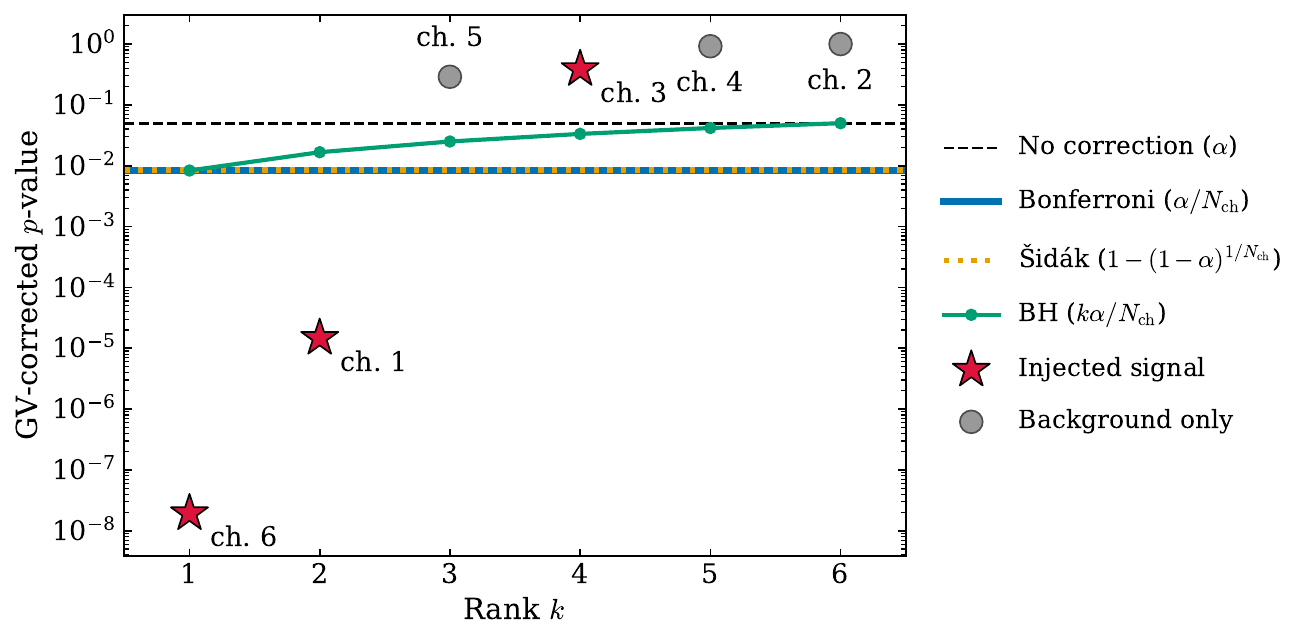}
    \caption{Global $p$-values of the six channels in one pseudo-experiment, corrected with the GV procedure for the scan over the resonance mass, using $5000$ background-only toys per channel, but not for the number of channels. The $p$-values are sorted in increasing order and labelled by channel number. Stars mark channels with an injected signal, and circles background-only channels. The horizontal lines show the per-channel thresholds without correction ($\alpha=0.05$), and with the Bonferroni and \v{S}id\'ak corrections, which are indistinguishable on this scale. The rising line shows the BH threshold $k\alpha/N_\text{ch}$. The error rates of the three procedures over many pseudo-experiments are given in Table~\ref{tab:channel_rates}.}
    \label{fig:decay-pvalues}
\end{figure}

Figure~\ref{fig:decay-pvalues} shows the global $p$-value of each channel after the GV correction for the scan over the resonance mass, sorted in increasing order, together with the uncorrected level $\alpha$ and the per-channel thresholds of the three multi-channel procedures. The background-only channels have large $p$-values, between 0.29 and 1, as expected for background fluctuations once the look-elsewhere effect within each channel is accounted for. The two channels with a strong signal, 1 and 6, lie below every threshold and are identified by all three procedures, and no background-only channel is selected. The weak signal of channel~3 is missed by all of them: in this pseudo-experiment it fluctuates downwards, to $q(\hat m)=5.5$ ($2.35\sigma$ local, against $3.5\sigma$ expected), and its GV-corrected $p$-value, 0.39, is comparable to those of the background-only channels, whereas channel~6 fluctuates upwards ($6.2\sigma$ local, against $4.5\sigma$ expected).

The figure also shows why the three procedures lead to the same decisions here. For six channels, the Bonferroni and \v{S}id\'ak thresholds, $8.33\times10^{-3}$ and $8.51\times10^{-3}$, agree to first order in $\alpha$ and coincide on the scale of the figure. The BH threshold starts at the Bonferroni value and rises linearly with the rank $k$, up to the uncorrected level $\alpha$ at $k=N_\text{ch}$, so that BH identifies more channels than Bonferroni only if, for some $k>1$, the $k$-th smallest $p$-value lies between $\alpha/N_\text{ch}$ and $k\alpha/N_\text{ch}$. Here, channels 1 and 6 lie below $\alpha/N_\text{ch}$ and all the other channels above $\alpha$, so that no $p$-value falls in such an interval. The procedures differ substantially only when many channels do, as in the screening regime of Section~\ref{sec:bh}, which we illustrate in Section~\ref{sec:exp_screening}.

A single pseudo-experiment cannot show the error rates that the procedures control, which are averages over repeated experiments. Table~\ref{tab:channel_rates} reports them for $5000$ pseudo-experiments generated with the same configuration. Without a correction for the number of channels, at least one of the three background-only channels is selected in $15\%$ of the pseudo-experiments, close to $1-(1-\alpha)^3$. The Bonferroni and \v{S}id\'ak corrections reduce this to $2.5\%$ since only $N_0=3$ of the six null hypotheses are true, hence the bound on the FWER is $N_0\alpha/N_\text{ch}=0.025$ rather than $\alpha$. BH controls the FDR at the same level, $N_0\alpha/N_\text{ch}=0.025$, which it reaches here, but lets the FWER rise to $9\%$, above $\alpha$. In exchange, it identifies the weak signal of channel~3 in $65\%$ of the pseudo-experiments, compared with $52\%$ for Bonferroni, and the strong signals in about $90\%$ instead of $85\%$. This is the trade-off discussed in Section~\ref{sec:bh}, FWER control is the appropriate criterion for a discovery claim, and FDR control for a screening stage in which a controlled fraction of false leads is acceptable.
\begin{table}[htbp]
  \centering
  \caption{Error rates and identification probabilities of the multi-channel procedures at $\alpha=0.05$, from $5000$ pseudo-experiments of the six-channel example. The family-wise error rate (FWER) and false discovery rate (FDR) are shown, in addition to the fraction of pseudo-experiments in which each signal channel is identified, and their mean. The statistical uncertainties are below $0.005$ for the FWER, $0.002$ for the FDR and $0.007$ for the identification probabilities. The values controlled by each method are underlined.}
  \label{tab:channel_rates}
  \begin{tabular}{lcccccc}
    \toprule
    Procedure & FWER & FDR & \multicolumn{4}{c}{Identification rate} \\
    \cmidrule(lr){4-7}
     & & & ch.~1 & ch.~3 & ch.~6 & \textbf{Mean} \\
    \midrule
    No correction & 0.146 & 0.043 & 0.92 & 0.74 & 0.96 & \textbf{0.87} \\
    BH & 0.090 & \underline{0.026} & 0.88 & 0.65 & 0.92 & \textbf{0.82} \\
    Bonferroni & \underline{0.025} & 0.008 & 0.82 & 0.52 & 0.88 & \textbf{0.74} \\
    \v{S}id\'ak & \underline{0.026} & 0.009 & 0.82 & 0.53 & 0.88 & \textbf{0.74} \\
    \bottomrule
  \end{tabular}
\end{table}

The example also shows where most of the correction comes from. For channel~1, with $q(\hat m)=26.4$, the trial factor of the mass scan is about 100, much larger than the factor $N_\text{ch}=6$ across channels, so the look-elsewhere effect inside each channel dominates the total correction. The trial factor also differs between channels: for channel~6, whose template is more than twice as broad, it is about 60, even though its $q(\hat m)=38.1$ is larger; at the $q(\hat m)$ of channel~1 it would be about 50.

More fundamentally, these procedures answer the question of \emph{which} channels contain a signal, whereas claiming the discovery of a single resonance requires a single global $p$-value. The most common approach in high-energy physics is a combined likelihood, given by the product of the likelihoods of the individual channels, with a single mass $m$ shared by all channels. Because $m$ is shared, the scan, and hence the look-elsewhere correction, is performed only once, on $\max_m q_\text{comb}(m)$. On the other hand, since the branching ratios are unknown, each channel keeps its own free signal fraction $\mu_i\geq0$. The combination gains sensitivity because channels with weak individual excesses contribute if they occur at the same mass, but it pays for its additional degrees of freedom, as anticipated in Section~\ref{sec:multipletesting}. 

\subsubsection{Screening many event classes}\label{sec:exp_screening}
The previous example is typical of a targeted search: a few channels, one hypothesised particle, and a correction for the number of channels that is small compared with the look-elsewhere effect within each channel. A different regime arises in model-independent general searches, which partition the selected events into many exclusive classes according to their final-state content and scan each class for deviations from the Standard Model prediction (case~(ii) in Section~\ref{sec:case_ii}). Such searches have been performed at HERA~\cite{aaron2009general}, at the Tevatron, with 399 exclusive final states~\cite{CDF:2008voc}, and at the LHC, with several hundred event classes per analysis~\cite{ATLAS:2018zdn,CMS:2020zjg}, and some targeted searches for supersymmetry report their results in more than a hundred exclusive signal regions~\cite{sirunyan2019search}. A general search is primarily a screening tool, whose role is to identify classes that deserve a dedicated follow-up analysis (Section~\ref{sec:summary}). A false lead then costs only the effort of a follow-up study, and the natural criterion is the false discovery rate rather than the family-wise error rate (Section~\ref{sec:bh}). Moreover, a new particle would typically contribute to several final states at once, so several classes may contain a genuine signal. This is the regime in which BH and the family-wise procedures differ most.

We illustrate this regime with a scaled-down general search with $N_\text{ch}=50$ classes, an order of magnitude fewer than in the searches above. Each class is modelled as in Section~\ref{subsec:exp1}, with its own background slope and detector resolution, drawn once from uniform distributions in $[0.2,0.6]$ and $[0.15,0.40]$, respectively, and the mass scan within each class is corrected with the GV procedure as in Section~\ref{sec:exp_decay}. A narrow resonance at $m=3$ is injected in $10$ classes, chosen at random, with signal fractions that give expected local significances evenly spaced between $2.5\sigma$ and $4.5\sigma$, mimicking a new particle that contributes to several final states with different branching ratios. We compare the Bonferroni, \v{S}id\'ak and BH procedures at $\alpha=0.05$ over $3000$ pseudo-experiments, treating the classes as statistically independent; further details of the setup, and a discussion of correlations between classes, are given in Appendix~\ref{app:screening}.

Figure~\ref{fig:screening} shows one pseudo-experiment. Without a correction for the number of classes, 11 classes have a global $p$-value below $\alpha$, 3 of which contain background only. The Bonferroni and \v{S}id\'ak corrections, whose threshold is now $\alpha/N_\text{ch}=10^{-3}$, select 5 classes, all with a signal. BH selects the same 5 and one more signal class, whose $p$-value of $3.5\times10^{-3}$ lies above the Bonferroni threshold but below the BH line at rank 6, $6\alpha/N_\text{ch}=6\times10^{-3}$; the next signal class, at rank 7, misses it narrowly ($7.9\times10^{-3}$ against $7\times10^{-3}$). Here the BH threshold rises enough to matter, because several classes have small $p$-values, as anticipated in Section~\ref{sec:bh}.

The averages over the $3000$ pseudo-experiments are given in Table~\ref{tab:screening_rates} (Appendix~\ref{app:screening}). Without a correction, at least one background-only class is selected in $85\%$ of the pseudo-experiments, and one in five selected classes is a false lead. The Bonferroni and \v{S}id\'ak corrections keep the FWER at $4\%$, consistent with the bound $N_0\alpha/N_\text{ch}=0.04$ for $N_0=40$ background-only classes, but identify on average only $3.4$ of the $10$ signal classes. BH keeps the FDR at $4.1\%$, consistent with the same bound, and identifies on average $4.9$ signal classes, $42\%$ more, at the price of a FWER of $22\%$, well above $\alpha$.

BH is the more useful procedure for a screening search, where a few false leads per hundred selected classes is an acceptable price for following up about 40\% more genuine signals. A discovery claim requires the family-wise criterion, and the classes selected at the screening stage are confirmed with a dedicated analysis (Section~\ref{sec:summary}).

\begin{figure}[htbp]
    \centering
    \includegraphics[width=0.85\textwidth]{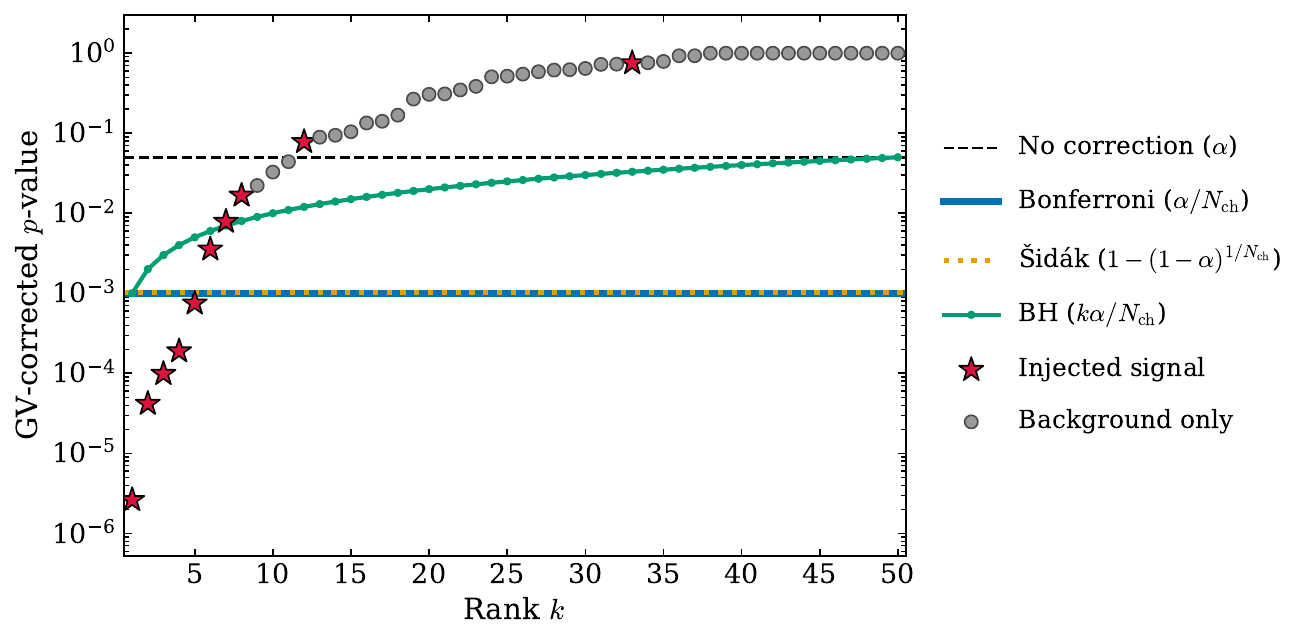}
    \caption{GV-corrected global $p$-values of the $50$ classes in one pseudo-experiment, sorted in increasing order, with the per-class thresholds without correction ($\alpha=0.05$), with the Bonferroni and \v{S}id\'ak corrections, which coincide on this scale, and the BH threshold $k\alpha/N_\text{ch}$. Stars mark classes with an injected signal, and circles background-only classes.}
    \label{fig:screening}
\end{figure}


\section{Summary and recommendations}
\label{sec:summary}
The look-elsewhere effect is a single statistical phenomenon: the inflation of significance when a test statistic is optimised by looking at the data multiple times. However, it manifests differently depending on the structure of the search, the tools used, and the scope of ``elsewhere.'' This article has traced the problem from its classical foundations to its modern incarnations in machine learning, and several overarching themes have emerged.

\paragraph{The LEE depends on both data and model.}
A recurring point throughout this review, already present in Davies' original formulation but often overlooked, is that the trial factor is not a property of the data alone. It depends jointly on the search range (a property of the data and the experimental design), on the resolution at which the search is conducted (a property of the signal template or the ML model), and on the correlation structure of the test statistic across the search space (mainly a property of the experimental resolution when searching for narrow resonances). Two analyses on identical data can have very different trial factors if they use different signal templates or ML models with different effective resolutions. This observation is central to understanding both the classical framework (Section~\ref{sec:model_dependence}) and the ML-specific challenges (Section~\ref{sec:resolution}).

\paragraph{``Elsewhere'' is not uniquely defined.}
The definition of the search space involves scientific judgment, and different choices lead to different trial factors for the same observed excess (Section~\ref{sec:elsewhere}). The three nested cases presented in Section~\ref{sec:elsewhere}--- within a single distribution (case~(i)), across multiple analyses (case~(ii)), and across an entire experimental programme (case~(iii)) --- represent different possible levels of scepticism, from the graduate student's to the laboratory director's. No single definition is universally correct; what matters is that the definition is stated explicitly and that the reader can assess the correction for themselves.

\paragraph{Different problems require different tools.}
The methods reviewed in this paper address distinct statistical problems, and applying the wrong tool leads either to overcorrection (loss of sensitivity) or to conceptual confusion:
\begin{itemize}
\item \textbf{Single search with a scan parameter} (Section~\ref{sec:classical}). One null hypothesis, one test statistic maximised over a nuisance parameter $\theta$. The correction is for the implicit optimisation over $\theta$. The appropriate tool is Davies/GV, which exploits the $\chi^2$ structure and correlation of the process $q(\theta)$. This is the standard setting for resonance searches.

\item \textbf{Multiple independent or quasi-independent searches} (Section~\ref{sec:multipletesting}). Many null hypotheses, each tested separately. The correction is for the multiplicity of tests. Bonferroni controls the FWER under arbitrary dependence, \v{S}id\'{a}k under independence; BH controls the FDR and is appropriate for screening. This is the natural setting for case~(ii): combining results across multiple final states or analyses.

\item \textbf{Sequential monitoring} (Section~\ref{sec:evalues}). Data arrive over time and the analyst wishes to assess significance repeatedly without penalty. E-values and e-processes provide anytime-valid inference without a pre-specified stopping rule or significance level. Their natural home is in online monitoring and evidence accumulation across successive data-taking periods, rather than in standard offline analyses where properly calibrated p-values remain more powerful.

\item \textbf{ML-based searches} (Section~\ref{sec:ml}). These can fall into any of the above categories depending on the setup. A single ML model scanning the full phase space is closest to the first setting (one test, implicit scan), but with the complication that the $\chi^2$ assumptions of Davies/GV typically fail and the effective resolution is emergent rather than specified. Multiple ML models with different hyperparameters on the same data are closer to the second setting (multiple correlated tests). The implicit LEE of an ML model can be addressed by reducing it (regularisation), avoiding its contamination of the test (data splitting), or measuring and correcting for it (pseudo-experiment calibration). These strategies are complementary, and their relative merits depend on the computational budget and the model complexity.
\end{itemize}

\paragraph{Calibration is the central challenge for ML.}
The conceptual problem posed by ML-based searches is not new --- it is the same local-versus-global p-value confusion that Davies identified in 1977, now occurring inside the model's optimisation rather than in an explicit parameter scan. What is new is the \emph{computational} cost of the standard remedy. Pseudo-experiment calibration, which absorbs the full LEE by construction, requires retraining the ML model for each toy --- an expense that scales with model complexity and can be prohibitive for large networks. The Gross-Vitells extrapolation trick, which makes classical calibration cheap, does not apply because the null distribution of the ML test statistic is generally not $\chi^2$. Developing analogous shortcuts for ML test statistics --- or designing ML models whose calibration is cheap by construction (e.g., through aggressive regularisation or sparsity constraints) --- is one of the most important open problems in this area.

\paragraph{Screening and discovery should be separated.}
Signal-agnostic ML searches serve a fundamentally different scientific function from targeted analyses: they identify anomalies for follow-up, rather than standalone physics discoveries. 
A two-tier approach --- screening at a lower threshold with a properly calibrated global p-value, followed by targeted confirmation on fresh data at the standard discovery threshold --- separates the two scientific functions and applies the appropriate statistical standard to each. 
We note that, for the screening tier to be useful in practice, the screening method must do more than return a global significance: it must offer some handle on \emph{why} a region was flagged. Concretely, the ability to localise where in feature space an anomaly manifests, and to quantify how much of it can be absorbed by systematic-uncertainty, is what allows uninteresting flags --- those attributable to background mismodelling rather than new physics --- to be triaged and discarded without committing to a full targeted follow-up. Methods that incorporate systematic uncertainties~\cite{dAgnolo:2021aun} and help characterising the detected anomaly~\cite{Chakravarti:2021svb,DAgnolo:2018cun,Cappelli:2025myc,grosso2025sparse} provide exactly this kind of information. Screening is therefore most effective when paired with interpretable test statistics: interpretability is not a peripheral feature but a prerequisite for the two-tier strategy, since it determines whether the follow-up stage is applied only to genuinely promising anomalies or squandered on modelling artefacts.

\paragraph{Scope.} We wish to bring to the reader's attention that, while extensive, this article does not cover all the literature that connects with the LEE problem. We do not include any discussion about whether extra penalties should be imposed on analyses that are dominated by systematics and/or are not blind analyses. This is a topic worthy of further consideration. We also touch the look ``elsewhen'' effect only briefly, and we  leave an in-depth exposition of anytime valid statistics to future work.

\vspace{6pt}

Based on these considerations, we offer the following recommendations for community discussion. They are intended as starting points, not settled standards, and may need to be adapted to the circumstances of a particular search.

\begin{tcolorbox}[breakable, colback=figblue!5, colframe=figblue!60, title=Recommendations for the High Energy Physics community]
\paragraph{1. Define and report \textit{elsewhere}.}
The trial factor can be quantified at the level of a single analysis, and every analysis should report a global p-value at that level alongside the local p-value for discovery claims, making the cost of the search explicit and comparable across analyses. The global p-value should be accompanied by the definition of ``elsewhere'' used to compute it: what was scanned (parameters, ranges, channels or ML configurations) and why.

\paragraph{2. Adapted discovery threshold convention.}
We recommend changing the convention on the significance threshold to reflect the difference between two main classes of statistical analyses:
\begin{itemize}
    \item \textbf{BSM Discovery.} The convention of reporting $5\sigma$ on the local p-value for discovery claims ignores the analysis-level LEE correction. \textbf{\textit{We propose a threshold of $5\sigma$ on the global p-value reported by an analysis.}}
    \footnote{For example, with the trial factor of $\approx 12$ of the CMS Higgs discovery over the mass range 110--145~GeV~\cite{CMS:2012qbp}, a $5\sigma$ global threshold requires a local significance of about $5.5\sigma$.}
    \item \textbf{SM confirmation.} Analyses which merely confirm SM predictions (e.g. HH production or rare H decay modes) might be technically very impressive, but are not ``discoveries''. For most of these cases, global and local p-value are usually the same. As such, \textbf{\textit{we propose a $3.5\sigma$ global threshold}}. 
    
    \item \textbf{Screening.} Signal-agnostic searches that operate like Goodness-of-Fit tests of the SM identify statistical anomalies for follow-up, not stand-alone physics discoveries, and should be treated as such. We recommend a two-step approach:\\ 
    \emph{(a)~screening}, where the search reports the analysis-level global p-value, and \textit{\textbf{anomalies above a threshold of 3$\sigma$ global significance are flagged for targeted follow-up}}\footnote{Since a flagged anomaly is not a claim, the threshold carries little risk: it merely sets the workload of follow-up studies, trading the number of false leads against the chance of missing a real signal}; \\
    \emph{(b)~confirmation}, where a dedicated analysis on a fresh set of data using a specific signal hypothesis motivated by the anomaly is held to the threshold of a ``BSM discovery''. Screening-stage detections should be publicly reported to the community to enable follow-up, but no stand-alone discovery claim should be based solely on a screening search without targeted confirmation.
\end{itemize}

\paragraph{3. Calibrate and interpret ML-based searches appropriately.}
ML-based strategies whose model is trained on, or otherwise adapted to, the observed data introduce an implicit LEE and require calibration of the null distribution. This ranges from signal-dependent methods, such as weakly supervised resonant anomaly detection, to signal-agnostic methods, such as ML-based goodness-of-fit tests.
\begin{itemize}
  \item \textbf{GOF tests return a global p-value.} Model-agnostic searches that operate like Goodness-of-Fit tests of the SM return a global p-value that, after proper calibration, is already corrected for the look-elsewhere effect within the space of input features analysed.
  \item \textbf{Multiple tests.} Running several ML strategies on the same data --- differing in algorithm, architecture, or hyperparameter choices --- amounts to performing several distinct searches, and the significance of the most anomalous result must be corrected for this multiplicity. When feasible, we recommend that analyses combining multiple ML strategies report a single global p-value calibrated by running all strategies on the same background-only pseudo-experiments, which accounts for the correlation automatically. When the strategies are themselves scanned over a physics parameter (e.g.\ the resonance mass in ML-enhanced bump hunting), the discrete multiplicity and the continuous scan LEE compound and must be corrected jointly.
  \item \textbf{Train/test splitting controls contamination.} Where retraining on many toys is prohibitive, train/test splitting ensures the score function is independent of the search data, removing the post-selection bias at the cost of scoring on just a fraction of the data. For statistics with an analytic null on the held-out fold this yields a toy-free route; for intractable-null statistics it reduces but does not eliminate the calibration burden ($k$-folding leaves a residual bias whose tail reliability is unestablished).
\end{itemize}

\paragraph{4. Additional parameters of interest.}
A signal may be characterised by parameters beyond its mass, notably its width or lifetime. How these enter the LEE depends on a prior physics decision: whether the additional parameter is itself a parameter of interest or a nuisance parameter. Both mass and any continuous companion parameter should be sampled on a sufficiently fine scale.
\begin{itemize}
  \item \textbf{Continuous parameter, profiled (default).} When the extra parameter is continuous, present only under the signal hypothesis (e.g.\ width or lifetime), and not itself of interest, we recommend treating it as a nuisance parameter profiled at each mass point.
  \item \textbf{Joint parameters of interest.} When both mass and the additional parameter are of interest, the LEE is defined over a 2-D region of parameter space and should be evaluated by toys or by the multi-dimensional Vitells--Gross method.
  \item \textbf{Widely-separated discrete values.} When the parameter takes a small number of physically distinct, well-separated values analysed independently (e.g.\ lifetime regimes probed by different detector subsystems in LLP searches), these are better treated as separate searches, with the correction following the multiple-testing prescriptions of Section~\ref{sec:multipletesting}.
\end{itemize}
\end{tcolorbox}
\section*{Acknowledgments} 
The authors are grateful to the members of the CMS Statistics Committee---in particular Bob Cousins, Francisco Matorras, Louis Moureaux, and Nicholas Wardle---for their valuable input and insightful discussions. We also thank Richard Lockhart for helpful conversations on discovery thresholds, Tobias Golling for his insights on online anomaly detection, and Aaditya Ramdas for his extremely helpful feedback on e-values. We are further grateful to David Bailey, Wolfgang Rolke, Phil Harris, Adrian Bayer and Roberto Trotta for valuable discussions, and to the CMS Physics Coordinators and the conveners of the Higgs, Exotica, and SUSY physics analysis groups for their support and feedback.
G.G. is supported by the National Science Foundation under Cooperative Agreement PHY-2019786 (The NSF AI Institute for Artificial Intelligence and Fundamental Interactions, http://iaifi.org/). S.R.B. is supported by the Spanish Research Agency (Agencia Estatal de Investigación) through projects PID2022-142545NB-C21, PID2025-168810NB-I00, CEX2020-001007-S, CEX2025-001574-S (research lines BSM, Quantum) and grant PRE2022-101860 funded by MCIN/AEI/10.13039/501100011033.

\appendix
\section{Additional details on the illustrative experiments}\label{app:experiments}
This appendix collects details and cross-checks of the examples of Section~\ref{sec:experiments} that complement the discussion in the main text but are not needed to follow it.

\subsection{Grid granularity and the Gross--Vitells estimate}\label{app:grid}
This section complements the discussion of the right panel of Figure~\ref{fig:upcrossings-trials_factor} and of Table~\ref{tab:trial_factors} in Section~\ref{sec:exp_tf}.
On a coarse grid, where the scanned points are effectively independent tests, the trial factor of the discrete scan is close to $N_\text{grid}$. The GV estimate is not, because the extrapolation in Eq.~\eqref{eq:gv_onesided} assumes a smooth process: near a peak of a smooth $q(m)$, maximising over $m$ adds one degree of freedom, which is the origin of the $\chi^2_{s+1}$ term in Eq.~\eqref{eq:gv_asymptotic}. A set of independent points lacks this extra degree of freedom, so its tail falls faster with $c$ than the extrapolation assumes. Because the GV procedure measures the upcrossing rate at $c_0$ and rescales it to $c$, this mismatch grows with the gap between the two levels: the further $c$ lies above $c_0$, the more the GV estimate overshoots the trial factor of the discrete scan. For independent points, about $18\%$ of the steps of the scan upcross $c_0=0.5$, and extrapolating this rate to $c=10$ gives an estimate of about $1+2N_\text{grid}$, roughly twice the trial factor of the discrete scan. This is the behaviour of the $c_0=0.5$ curve at small $N_\text{grid}$, which lies above the Bonferroni line. For $c_0=7.5$ the extrapolation is short, and the GV estimate follows the Bonferroni value, but it requires a denser scan in order to correctly approximate the trial factor of the continuous scan.

\subsection{The Bayer--Seljak estimate}\label{app:bayes}
This section gives the details of the Bayer--Seljak estimate of Eq.~\eqref{eq:bs_exp}, summarised at the end of Section~\ref{sec:exp_tf}. As in Ref.~\cite{Bayer:2020pva}, the local $p$-value in Eq.~\eqref{eq:bs_exp} is used in its asymptotic form. With this form, $N_B$ times the local $p$-value equals the second term of the Davies bound, Eq.~\eqref{eq:davies_normal}, in which the oscillation rate of the process is estimated from the curvature of the observed peak. With the exact local $p$-value, $\tfrac12P(\chi^2_1>c)$, the estimate at $c=10$ would be about $8\%$ lower. We compute $q''(\hat m)$ by repeating the fit at $\hat m\pm0.02$, so that the estimate does not depend on the spacing of the scan.

To test it, we apply Eq.~\eqref{eq:bs_exp} to each of the background-only pseudo-experiments of Table~\ref{tab:trial_factors}. On average, the Bayer--Seljak trial factor at $c=10$ agrees with the GV estimate to within $1\%$ on both grids, and, like it, lies a few per cent above the Davies and empirical estimates on the 350-point grid. The agreement with GV is expected. Near a high peak of local significance $Z$, the posterior width is $\sigma_m=1/(Z\sqrt{-\rho_{11}})$, so that $N_B$ reproduces the leading term of the Davies bound, and both estimate the same oscillation rate of the process: the value of $\sqrt{-\rho_{11}}$ inferred from the curvature of the peaks, $3.67$, agrees to within 2\% with that inferred from the upcrossing rate measured in the GV procedure, $3.73$.

\begin{figure}[htbp]
    \centering
    \includegraphics[width=\textwidth]{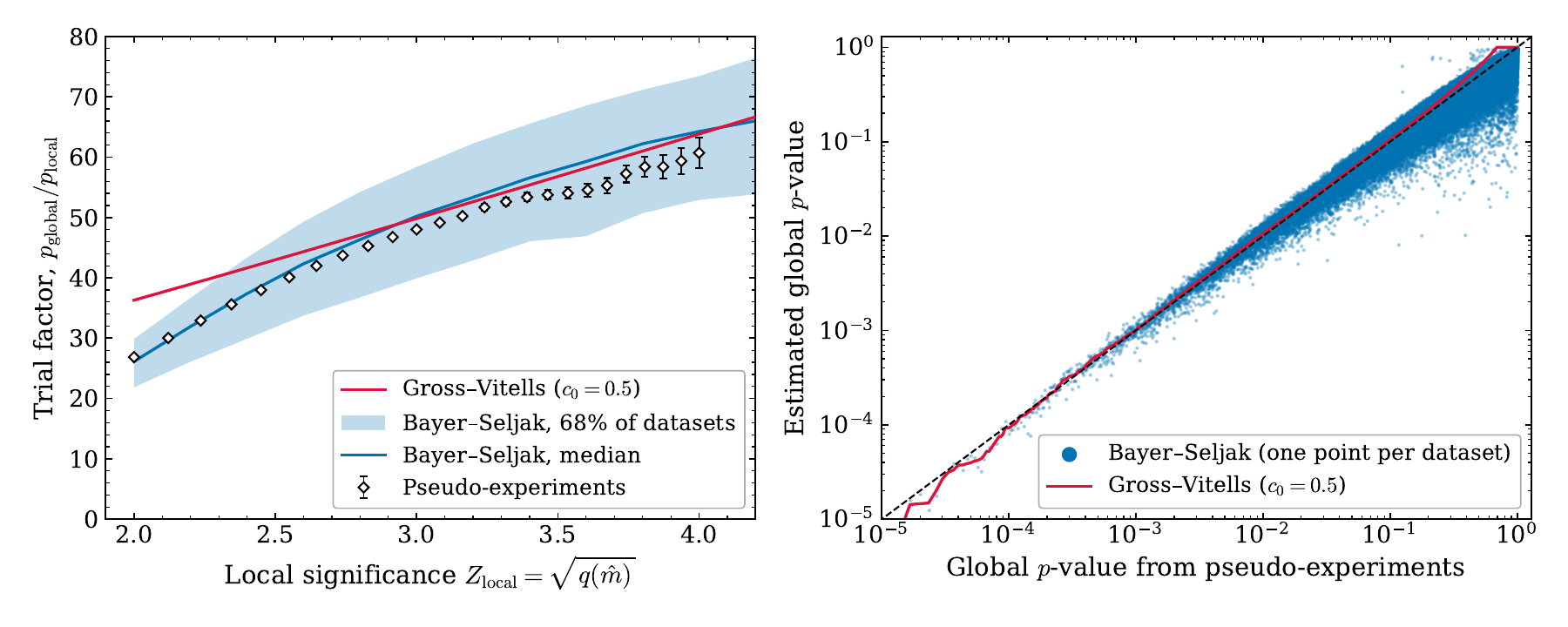}
    \caption{Comparison of the Bayer--Seljak estimate, Eq.~\eqref{eq:bs_exp}, computed from each dataset without pseudo-experiments, with the GV procedure and with the pseudo-experiments, for the 350-point scan. Left: trial factor as a function of the local significance of the highest peak, $Z_\text{local}=\sqrt{q(\hat m)}$, from the fraction of pseudo-experiments above each level (open diamonds), from the GV extrapolation with $c_0=0.5$ (red line), and from the Bayer--Seljak estimate of each dataset (median and central $68\%$ of the datasets in bins of $Z_\text{local}$, blue). Right: global $p$-value of each dataset estimated with the Bayer--Seljak method (one point per dataset) and with the GV procedure (red line), as a function of the global $p$-value of $q(\hat m)$ obtained from the full set of pseudo-experiments; the dashed line marks equality. Since the Bayer--Seljak $p$-value belongs to a different statistic, $q_S$, individual points are not expected to lie on the diagonal. Both panels use $3\times10^5$ background-only pseudo-experiments.}
    \label{fig:bs}
\end{figure}

The two approaches differ in the test statistic. The GV and Davies estimates give the $p$-value of $q(\hat m)$, whereas the Bayer--Seljak $p$-value is a function of $q_S$, in which $N_B$ is estimated from the single observed peak, whose curvature fluctuates from one dataset to the next. The left panel of Figure~\ref{fig:bs} shows the trial factor as a function of the local significance. The Bayer--Seljak estimate follows the growth with $Z_\text{local}$ of the GV extrapolation and of the pseudo-experiments for local significances between $2.5$ and $4\sigma$, and at fixed $Z_\text{local}$ it differs from the GV value by about 20\% from one dataset to the next. Below about $2.5\sigma$ it also follows the pseudo-experiments, where the GV formula, an approximation for the tail, overestimates the trial factor. The reason is that Eq.~\eqref{eq:bs_exp} has the form of a \v{S}id\'ak correction, so the global $p$-value it gives saturates at one, while the GV estimate, based on the expected number of upcrossings, can exceed one. The right panel compares, dataset by dataset, the estimated global $p$-values with those of $q(\hat m)$ obtained from the full set of pseudo-experiments. The GV $p$-value is a fixed function of $q(\hat m)$ and lies on a single curve, whereas the Bayer--Seljak $p$-values scatter around the diagonal. This scatter is not an error of the estimate: $q_S$ and $q(\hat m)$ are different test statistics, whose $p$-values need not agree dataset by dataset even when both are exact. The relevant check is the calibration: in background-only pseudo-experiments, $p<\alpha$ occurs in a fraction $0.051$ of cases for $\alpha=0.05$ and $0.010$ for $\alpha=0.01$, so that it is correctly calibrated as a test.\footnote{For the right panel we follow the prescription of Ref.~\cite{Bayer:2020pva}, which selects, among all local maxima of $q(m)$, the one with the largest $q_S$, i.e. the smallest global $p$-value} in most datasets this is the highest peak.

The Bayer--Seljak estimate thus provides, from a single fit around the observed excess, a correctly calibrated global $p$-value; its trial factor differs from the GV one by about 20\% dataset by dataset, which corresponds to a shift of less than $0.1\sigma$ in the global significance at the discovery level.\footnote{At $Z_\text{local}=5$, $p_\text{local}=2.9\times10^{-7}$; the GV trial factor of this example, $78$, gives $Z_\text{global}=4.08$, and a trial factor $20\%$ larger or smaller gives $4.04$ or $4.13$.} In addition to the requirements listed in Section~\ref{sec:exp_tf}, the method uses the width of the observed peak to represent the whole scan range, which becomes inaccurate if the resolution varied strongly across the range.

\subsection{Trial factor for broad signal templates}\label{app:width}
This section gives the derivation of the analytical Davies prediction of Section~\ref{sec:exp_width} and explains the behaviour of the trial factor for broad templates in Figure~\ref{fig:width_dependence}.

To apply the analytical Davies bound of Section~\ref{subsec:davies}, we write the test statistic in terms of the signed square root of the likelihood ratio,
\begin{equation}
  Z(m)=\operatorname{sign}(\hat\mu_m)\,\sqrt{-2\ln\frac{\mathcal{L}(\mu=0,\,\hat{\hat{\nu}})}{\mathcal{L}(\hat\mu_m,\,\hat\nu_m;\,m)}}\,,
  \label{eq:Zm}
\end{equation}
in which the likelihood ratio of Eq.~\eqref{eq:qmu} is evaluated for either sign of $\hat\mu_m$. Excesses give $Z(m)>0$ and deficits $Z(m)<0$, and the two statistics are related by $q(m)=Z(m)^2$ when $Z(m)>0$ and $q(m)=0$ otherwise. Under the null hypothesis, $Z(m)$ follows a standard normal distribution at each mass and, as a function of $m$, is asymptotically a Gaussian process with unit variance. It therefore plays the role of $S(\theta)$ in Eq.~\eqref{eq:davies_normal}, with $\theta\to m$ and $[L,\,U]=[0,10]$. Since $q(\hat m)>c$ exactly when $\max_m Z(m)>\sqrt{c}$, Eq.~\eqref{eq:davies_normal} is evaluated at the level $\sqrt{c}$. Its first term then becomes $\Phi(-\sqrt{c})=\tfrac12P(\chi^2_1>c)=p_\text{local}$, and its second term the expected number of upcrossings, $\langle N(c)\rangle=\mathcal{N}e^{-c/2}$, with $\mathcal{N}=\frac{1}{2\pi}\int_{L}^{U}\sqrt{-\rho_{11}(m)}\,dm$.

The only model-dependent ingredient of $\mathcal{N}$ is $\rho_{11}(m)$. It is defined from the correlation between the values of $Z$ at two different masses, $\rho(m',m)=\operatorname{corr}[Z(m'),Z(m)]$, as
\[
  \rho_{11}(m)=\frac{\partial^2\rho(m',m)}{\partial m'^2}\bigg|_{m'=m}\,,
\]
and it measures how fast the correlation drops when $m'$ moves away from $m$: the faster it drops, the more upcrossings. In general $\rho_{11}$ depends on $m$, which is why it is integrated in $\mathcal{N}$. We derive $\rho$ for a Gaussian template $g(x\,|\,m,\omega)$, first neglecting the edges of the range and the fit of the background normalisation and slope; in that case $\rho_{11}$ turns out to be independent of $m$.

\emph{Step 1: $Z(m)$ as a score.} Divide the $x$ range into narrow bins $i$, and let $n_i$ be the number of observed events in each. Let $b_i$ be the expected number of background events in bin $i$, and $g_i=g_i(m)$ the expected number of signal events for $\mu=1$ with the template of mass $m$ and width $\omega$. The log-likelihood of the signal strength is then Poissonian,
\begin{equation}
  \ln\mathcal{L}(\mu)=\sum_i\big[n_i\ln(b_i+\mu g_i)-(b_i+\mu g_i)\big]+\text{const.}
\end{equation}
Under the null hypothesis $\mu=0$ the estimate $\hat\mu_m$ is small, so we expand $\ln\mathcal{L}$ to second order around $\mu=0$:
\begin{equation}
  \ln\mathcal{L}(\mu)\simeq\ln\mathcal{L}(0)+\mu\,u(m)-\tfrac12\mu^2 I(m),
\end{equation}
with
\begin{equation}
  u(m)=\frac{\partial\ln\mathcal{L}}{\partial\mu}\bigg|_{\mu=0}=\sum_i g_i\,\frac{n_i-b_i}{b_i},\qquad
  I(m)=-\frac{\partial^2\ln\mathcal{L}}{\partial\mu^2}\bigg|_{\mu=0}\simeq\sum_i\frac{g_i^2}{b_i}.
\end{equation}
Here $u$ is the score, the standard name for the derivative of the log-likelihood with respect to a parameter, evaluated at its null value: it is the slope of $\ln\mathcal{L}$ at $\mu=0$, and measures how strongly the data pull $\mu$ away from zero. For $I$, the Fisher information, the observed value $\sum_i g_i^2\,n_i/b_i^2$ has been replaced by its expectation under the null hypothesis, where $\langle n_i\rangle=b_i$. The maximum of the parabola is at $\hat\mu_m=u/I$, where $\ln\mathcal{L}(\hat\mu_m)-\ln\mathcal{L}(0)=u^2/2I$. The likelihood ratio in Eq.~\eqref{eq:Zm} is therefore $-2\ln[\mathcal{L}(0)/\mathcal{L}(\hat\mu_m)]=u^2/I$, and
\begin{equation}
  Z(m)=\frac{u(m)}{\sqrt{I(m)}},
\end{equation}
since $\hat\mu_m$ and $u$ have the same sign.

To find the distribution of $Z(m)$ under the null hypothesis, note that the events follow a Poisson process, so the $n_i$ are independent Poisson variables with mean and variance
\begin{equation}
  \langle n_i\rangle=\operatorname{Var}[n_i]=b_i .
\end{equation}
The score is therefore a linear combination of independent variables with zero mean, $u=\sum_i g_i\,(n_i-b_i)/b_i$. Then $\langle u(m)\rangle=0$ and, since the cross terms vanish by independence,
\begin{equation}
  \operatorname{Var}[u(m)]=\sum_i\frac{g_i^2}{b_i^2}\,b_i=\sum_i\frac{g_i^2}{b_i}=I(m).
\end{equation}
The variance of the score is thus the Fisher information, and $Z(m)=u(m)/\sqrt{I(m)}$ has zero mean and unit variance. In addition, when the expected number of events is large, $u(m)$ is a sum of many independent contributions, so by the central limit theorem it is asymptotically Gaussian, as stated above. The same holds jointly for any set of masses, since all the $u(m)$ are linear in the same $n_i$, which makes $Z(m)$ a Gaussian process.

\emph{Step 2: correlation.} The same argument gives the covariance of the scores at two masses. With $g'_i=g_i(m')$, only the terms with the same bin survive,
\begin{equation}
  \operatorname{cov}[u(m'),u(m)]=\sum_i\frac{g'_i\,g_i}{b_i^2}\,b_i=\sum_i\frac{g'_i\,g_i}{b_i}.
\end{equation}
Dividing by the normalisations $\sqrt{I(m')\,I(m)}$ gives
\begin{equation}
  \rho(m',m)=\frac{\langle g(m'),g(m)\rangle}{\lVert g(m')\rVert\,\lVert g(m)\rVert},\qquad
  \langle v,v'\rangle=\sum_i\frac{v_i\,v'_i}{b_i},
\end{equation}
where $g(m)$ denotes the vector of the $g_i(m)$, and $\lVert v\rVert=\langle v,v\rangle^{1/2}$.
This is Eq.~\eqref{eq:rho_binned} below, without the projections that account for the background fit. For narrow bins, $b_i=b(x_i)\,\Delta x$ and $g_i=g(x_i\,|\,m,\omega)\,\Delta x$, the common factor $\Delta x$ cancels in $\rho$ and the sums become integrals,
\begin{equation}
  \rho(m',m)=\frac{\int g(x\,|\,m',\omega)\,g(x\,|\,m,\omega)\,b(x)^{-1}\,dx}{\sqrt{I(m')\,I(m)}},\qquad I(m)=\int\frac{g(x\,|\,m,\omega)^2}{b(x)}\,dx ,
\end{equation}
where now $b(x)$ is the background density.

\emph{Step 3: Gaussian template and exponential background.} For $b(x)\propto e^{-\nu x}$, the product of the two templates is
$g(x\,|\,m',\omega)\,g(x\,|\,m,\omega)\propto\exp[-(x-\bar m)^2/\omega^2]\,\exp[-(m'-m)^2/4\omega^2]$, with $\bar m=(m+m')/2$. Multiplying by $1/b\propto e^{\nu x}$ and integrating over $x$ gives, in the numerator, $\exp[-(m'-m)^2/4\omega^2]$ times a factor $e^{\nu\bar m+\nu^2\omega^2/4}$. Setting $m'=m$ shows that $\sqrt{I(m')\,I(m)}$ contains the same factor, with the same constants. The dependence on the background therefore cancels, leaving
\begin{equation}
  \rho(m',m)=\exp\!\left[-\frac{(m'-m)^2}{4\omega^2}\right],\qquad \rho_{11}=-\frac{1}{2\omega^2},
\end{equation}
independently of the slope $\nu$. Inserting $\rho_{11}$ in $\mathcal{N}$ gives Eq.~\eqref{eq:N_analytic}.

As noted in Section~\ref{sec:exp_width}, for broad templates the GV points in Figure~\ref{fig:width_dependence} deviate from the narrow-template prediction of Eq.~\eqref{eq:tf_analytic}, because Eq.~\eqref{eq:N_analytic} neglects two effects. First, the background fit absorbs the part of the signal template that mimics a change in the background normalisation or slope. For broad templates, the part that remains varies more rapidly with $m$ than the full template, which shortens the correlation length and increases the number of upcrossings relative to Eq.~\eqref{eq:tf_analytic}. Second, templates near the edges of the range are truncated, which reduces the number of upcrossings. The two effects act in opposite directions, and both are accounted for when the correlation function of $Z(m)$ is computed directly from the binned templates, with the background fit included,

\begin{equation}
  \rho(m',m)=\frac{\langle \tilde g(m'),\tilde g(m)\rangle}{\lVert \tilde g(m')\rVert\,\lVert \tilde g(m)\rVert},\qquad
  \langle v,v'\rangle=\sum_i\frac{v_i\,v'_i}{b_i},
  \label{eq:rho_binned}
\end{equation}
where $b_i$ is the background in bin $i$, and the template $g(m)$ of Step 2 is replaced by $\tilde g(m)$, the part of it that the background fit cannot mimic. To see which part this is, regard the expected counts in all bins as a vector, $(b_1,b_2,\dots)$. To first order, a small change $\delta A$ of the background normalisation and $\delta\nu$ of the slope change the expected counts by
\begin{equation}
  \delta b_i\simeq\frac{\delta A}{A}\,b_i+\delta\nu\,\frac{\partial b_i}{\partial\nu}.
\end{equation}
A change of normalisation thus rescales all bins in proportion to $b_i$ itself, and a change of slope tilts the spectrum, moving events between low and high $x$ in proportion to $\partial b_i/\partial\nu$. The background fit can therefore reproduce any small excess or deficit with the shape of a combination of these two vectors, and any such component of the signal template is absorbed by the fit and carries no evidence for a signal. Accordingly, $\tilde g(m)$ is the template minus its projection onto the two vectors $b_i$ and $\partial b_i/\partial\nu$, with the same inner product; equivalently, it is the residual of a weighted least-squares fit of $g_i(m)$ by $b_i$ and $\partial b_i/\partial\nu$, with weights $1/b_i$. For a narrow template the projection is negligible, $\tilde g(m)\simeq g(m)$, and Step 2 is recovered.

Evaluating $\mathcal{N}$ numerically with this correlation function gives the solid black line in Figure 4, which agrees with the GV estimates to within $4\%$ for all widths, including the sharp drop near $\omega\approx4$. This drop arises because, for $2\lesssim\omega\lesssim4$, there is a mass inside the scan range at which the template is almost entirely absorbed by the background fit. Around this mass the residual shape changes rapidly with $m$, which adds upcrossings. As $\omega$ increases, this mass moves towards the lower edge of the range and leaves it just above $\omega=4$, so its contribution to $\mathcal{N}$ disappears. Repeating the GV procedure with the background slope fixed to its true value (open points) removes the drop; the points agree with the corresponding Davies prediction, in which only the normalisation is projected out (solid blue line), to within 5\% for $\omega\lesssim2.5$, and lie up to 14\% above it for broader templates.

\subsection{Screening search: setup details and identification rates}\label{app:screening}
This section complements the screening example of Section~\ref{sec:exp_screening}. The mass scan within each class is corrected with the GV procedure, with $\langle N(c_0)\rangle$ estimated for each class from $2000$ background-only pseudo-experiments; it ranges from $2.5$ to $5.7$, depending on the template width. For each procedure we record the family-wise error rate, the false discovery rate, and the fraction of pseudo-experiments in which each signal class is identified, which are given in Table~\ref{tab:screening_rates}.

The classes are treated as statistically independent. In a real general search they are exclusive, but their predictions share systematic uncertainties, and migrations between neighbouring classes, for example in jet multiplicity, can anti-correlate them. BH is guaranteed to control the false discovery rate under positive dependence, while the Benjamini--Yekutieli variant does so under arbitrary dependence, at a cost in power (Section~\ref{sec:bh}).

On average, BH selects $0.26$ background-only classes per pseudo-experiment. In return, it identifies more of the genuine signals, and the relative gain over Bonferroni is largest for the weakest ones: a class with an expected local significance of $2.5\sigma$ is identified in $15\%$ of the pseudo-experiments by BH and in $7\%$ by Bonferroni. Even the strongest signals, at $4.5\sigma$, are identified in only $80\%$ (BH) and $66\%$ (Bonferroni) of the pseudo-experiments. This is because, once the look-elsewhere effect within each class and across the $50$ classes is included, a Bonferroni-corrected discovery at $\alpha=0.05$ requires a local significance of about $4.1\sigma$.

\begin{table}[htbp]
  \centering
  \caption{Error rates and identification probabilities of the multi-channel procedures at $\alpha=0.05$, from $3000$ pseudo-experiments of the 50-class general search. The family-wise error rate (FWER) and false discovery rate (FDR) are shown, in addition to the fraction of pseudo-experiments in which each of the 10 signal classes is identified, and its average over the signal classes. The statistical uncertainties are below $0.008$ for the FWER, $0.002$ for the FDR and $0.01$ for the identification probabilities. The values controlled by each method are underlined.}
  \label{tab:screening_rates}
  \setlength{\tabcolsep}{3.4pt}
  \begin{tabular}{lccccccccccccc}
    \toprule
    Procedure & FWER & FDR & \multicolumn{11}{c}{Identification rate} \\
    \cmidrule(lr){4-14}
     & & & cl.~3 & cl.~10 & cl.~16 & cl.~19 & cl.~22 & cl.~24 & cl.~29 & cl.~32 & cl.~46 & cl.~47 & \textbf{Mean} \\
    \midrule
    No correction & 0.854 & 0.199 & 0.94 & 0.91 & 0.36 & 0.59 & 0.43 & 0.79 & 0.66 & 0.90 & 0.59 & 0.87 & \textbf{0.70} \\
    BH & 0.220 & \underline{0.041} & 0.80 & 0.75 & 0.15 & 0.33 & 0.20 & 0.56 & 0.40 & 0.72 & 0.32 & 0.64 & \textbf{0.49} \\
    Bonferroni & \underline{0.043} & 0.011 & 0.66 & 0.60 & 0.07 & 0.20 & 0.09 & 0.37 & 0.25 & 0.54 & 0.18 & 0.46 & \textbf{0.34} \\
    \v{S}id\'ak & \underline{0.043} & 0.011 & 0.66 & 0.60 & 0.07 & 0.20 & 0.09 & 0.38 & 0.25 & 0.55 & 0.19 & 0.47 & \textbf{0.34} \\
    \bottomrule
  \end{tabular}
\end{table}
\clearpage 

\section{Notation and abbreviations}\label{app:symbols}
This appendix collects the mathematical symbols (Table~\ref{tab:symbols}) and the abbreviations (Table~\ref{tab:abbreviations}) used in the paper. Symbols are grouped by the part of the paper in which they are introduced, and the last column gives the section or equation where each is defined. The general discussion is written in terms of a generic scan parameter $\theta$. In the illustrative experiments the scan parameter is the mass $m$, and the quantities defined for $\theta$ are used with $\theta\to m$, e.g.\ $q(m)$ and $\hat m$. A few letters, notably $\sigma$, $c$, $k$, $M$, $N$, $V$, $\lambda$, $s$ and $t$, carry different meanings in different parts of the paper; each meaning is listed in the group where it is used, and the intended one is clear from the context.

\newcommand{\symgroup}[1]{\addlinespace[6pt]\multicolumn{3}{@{}l}{\textbf{#1}}\\\addlinespace[2pt]}
\newcommand{\symref}[1]{\footnotesize #1}

{\small
\renewcommand{\arraystretch}{1.2}
\begin{longtable}{@{}>{\raggedright\arraybackslash}p{0.20\textwidth}>{\raggedright\arraybackslash}p{0.57\textwidth}>{\raggedright\arraybackslash}p{0.17\textwidth}@{}}
\caption{Mathematical symbols used in the paper.}\label{tab:symbols}\\
\toprule
Symbol & Meaning & Defined in \\
\midrule
\endfirsthead
\multicolumn{3}{@{}l}{}\\
\toprule
Symbol & Meaning & Defined in \\
\midrule
\endhead
\midrule
\multicolumn{3}{r@{}}{}\\
\endfoot
\bottomrule
\endlastfoot
\symgroup{General statistical notation}
$H_0$, $H_1$ & Null hypothesis (typically background only) and alternative hypothesis (signal present). & \symref{Sec.~\ref{sec:classical}} \\
$x$; $p(x\,|\,H)$ & Observed data; probability density of the data under the hypothesis $H$, which is a marginal likelihood when the parameters of $H$ are averaged over their prior. & \symref{Secs.~\ref{sec:bayes}, \ref{sec:evalues}} \\
$p$ & p-value: probability under $H_0$ of a result at least as extreme as the one observed. & \symref{Sec.~\ref{sec:thresholds}} \\
$\plocal$ & Local p-value: p-value of the most significant point of a scan, computed as if that point had been fixed in advance. & \symref{Sec.~\ref{sec:classical}} \\
$\pglobal$ & Global p-value: probability under $H_0$ of an excess at least as significant anywhere in the search range. & \symref{Eq.~\eqref{eq:pglobal}} \\
$\pglobal/\plocal$ & Trial factor. & \symref{Sec.~\ref{sec:bayes}} \\
$Z$; $Z_\mathrm{local}$, $Z_\mathrm{global}$ & Gaussian significance, $Z=\Phi^{-1}(1-p)$, one-sided unless stated otherwise and quoted in units of $\sigma$; local and global significance. & \symref{Sec.~\ref{sec:thresholds}} \\
$\Phi$ & Cumulative distribution function of the standard normal distribution. & \symref{Sec.~\ref{sec:thresholds}} \\
$\alpha$ & Significance level (type~I error rate); in multiple testing, the target family-wise error rate or false discovery rate. & \symref{Secs.~\ref{sec:exclusion}, \ref{sec:bh}} \\
$\chi^2_s$ & Chi-squared distribution with $s$ degrees of freedom. & \symref{Eq.~\eqref{eq:gv_asymptotic}} \\
$\tfrac12\chi^2_1$ & Shorthand for the mixture $\tfrac12\delta(q)+\tfrac12\chi^2_1$, with $\delta$ the Dirac delta: the asymptotic null distribution of the one-sided test statistic at a fixed mass. & \symref{Sec.~\ref{subsec:exp1}} \\
$P(\cdot)$; $\mathbb{E}[\cdot]$ & Probability; expectation value. & -- \\
$\operatorname{Var}$, $\operatorname{cov}$, $\operatorname{corr}$ & Variance, covariance and correlation. & -- \\
$\mathbf{1}[\cdot]$ & Indicator function: $1$ if the condition holds and $0$ otherwise. & \symref{Eq.~\eqref{eq:nplm_pvalue}} \\
$\mu$; $\hat\mu$ & Signal strength and its maximum-likelihood estimate & \symref{Sec.~\ref{sec:bayes}, Eq.~\eqref{eq:toy_model}, Sec.~\ref{sec:exp_decay}} \\
\symgroup{Scope of the search and discovery thresholds}
$m_H$ & Mass of the Higgs boson. & \symref{Sec.~\ref{sec:case_i}} \\
$p_T$; $\sum p_T$ & Transverse momentum; scalar sum of the transverse momenta of the objects in an event. & \symref{Sec.~\ref{sec:case_ii}} \\
$\tilde p$ & Local p-value of one distribution in a general search, already corrected for the scan within that distribution. & \symref{Sec.~\ref{sec:case_ii}} \\
$k$ & Number of null hypotheses that can be simultaneously true; it bounds the trial factor relevant for exclusion limits. & \symref{Sec.~\ref{sec:exclusion}} \\
$m_0$ & Mass at which an exclusion limit is set. & \symref{Sec.~\ref{sec:exclusion}} \\
$\sigma$ & Signal cross-section, in the context of exclusion limits. & \symref{Sec.~\ref{sec:exclusion}} \\
\symgroup{Single search with a scan parameter}
$\theta$ & Scan parameter defined only under the alternative hypothesis, such as the mass of a resonance. & \symref{Sec.~\ref{sec:classical}} \\
$\theta_0$ & Value of the scan parameter fixed in advance, used as the reference point of the local p-value. & \symref{Sec.~\ref{sec:classical}} \\
$\hat\theta$ & Value of the scan parameter at which the test statistic is maximal. & \symref{Eq.~\eqref{eq:gv_asymptotic}} \\
$[L,\,U]$ & Range over which the scan parameter is scanned. & \symref{Eq.~\eqref{eq:Msup}} \\
$S(\theta)$ & Test statistic at fixed $\theta$, as a function of $\theta$; under $H_0$, a Gaussian or chi-squared process. & \symref{Sec.~\ref{sec:classical}} \\
$q(\theta)$ & Profile likelihood ratio test statistic, as a function of the scan parameter. & \symref{Sec.~\ref{sec:gv}} \\
$M$; $q(\hat\theta)$ & Supremum of $S(\theta)$ over the scan range, which is the test statistic of the scan; for the profile likelihood ratio, $q(\hat\theta)=\max_\theta q(\theta)$. & \symref{Eqs.~\eqref{eq:Msup}, \eqref{eq:gv_asymptotic}} \\
$c$ & Level of the test statistic at which upcrossings are counted: a level of the Gaussian process $S(\theta)$ in Eq.~\eqref{eq:davies_normal}, and of the chi-squared statistic $q$ from Eq.~\eqref{eq:gv_extrapolation} on, so that the corresponding Gaussian level is $\sqrt{c}$ for $s=1$. When a result is assessed, the observed value of $q(\hat\theta)$. &\symref{Sec.~\ref{subsec:davies}, Eq.~\eqref{eq:gv_extrapolation}} \\
$c_0$ & Low reference level of the Gross--Vitells procedure, typically $c_0=0.5$. & \symref{Sec.~\ref{sec:gv}} \\
$\Nc$ & Expected number of upcrossings of the level $c$ by $S(\theta)$ over the scan range. & \symref{Eq.~\eqref{eq:davies_bound}} \\
$\Ncz$ & Expected number of upcrossings of the reference level $c_0$, estimated from pseudo-experiments. & \symref{Eq.~\eqref{eq:gv_extrapolation}} \\
$\rho(\theta',\theta)$ & Correlation of the process at two values of the scan parameter, $\operatorname{corr}[S(\theta'),S(\theta)]$. & \symref{Eq.~\eqref{eq:davies_normal}} \\
$\rho_{11}(\theta)$ & Second derivative of $\rho(\theta',\theta)$ with respect to $\theta'$ at $\theta'=\theta$; $-\rho_{11}$ sets how fast the process decorrelates. & \symref{Eq.~\eqref{eq:davies_normal}} \\
$s$ & Number of degrees of freedom of a chi-squared process, equal to the number of fitted signal parameters. & \symref{Sec.~\ref{subsec:davies}} \\
$Z_1,\dots,Z_s$ & Gaussian components of a chi-squared process, $S=Z_1^2+\dots+Z_s^2$. & \symref{Sec.~\ref{subsec:davies}} \\
$\lambda_1(\theta),\dots,\lambda_s(\theta)$ & Eigenvalues that generalise $-\rho_{11}$ to a chi-squared process. & \symref{Sec.~\ref{subsec:davies}} \\
$V$ & Total variation of the observed curve $S(\theta)$. & \symref{Eq.~\eqref{eq:total_variation}} \\
$\theta_1,\dots,\theta_n$ & Successive turning points of $S(\theta)$. & \symref{Eq.~\eqref{eq:total_variation}} \\
$\mathcal{N}$ & Effective number of independent search regions in the scan range: coefficient of the asymptotic Gross--Vitells formula; for $s=1$, $\Nc=\mathcal{N}e^{-c/2}$. & \symref{Eqs.~\eqref{eq:gv_asymptotic}, \eqref{eq:N_analytic}} \\
$Z(\theta)$ & Signed square root of the likelihood ratio: under $H_0$, a Gaussian process of unit variance, with $q=Z^2$ for excesses ($s=1$). & \symref{Sec.~\ref{sec:gv}} \\
$Z'$, $Z''$; $\lambda$; $\delta$ & First and second derivatives of $Z(\theta)$; $\lambda=\operatorname{Var}[Z']$; shift of the scan parameter around the maximum. & \symref{Sec.~\ref{sec:gv}, footnote} \\
$A_c$ & Excursion set $\{\theta: S(\theta)>c\}$ of a multi-dimensional search. & \symref{Sec.~\ref{sec:multidim}} \\
$\phi(A_c)$ & Euler characteristic of the excursion set $A_c$. & \symref{Eq.~\eqref{eq:euler_2d}} \\
$N_1$, $N_2$ & Coefficients of the expected Euler characteristic, set by the geometry of the search region and the correlation of the field. & \symref{Eq.~\eqref{eq:euler_2d}} \\
$m_X$, $m_Y$ & Masses of the particles $X$ and $Y$ in a two-dimensional search for $X\to YH$. & \symref{Sec.~\ref{sec:multidim}} \\
$c\tau$ & Proper decay length of a long-lived particle. & \symref{Sec.~\ref{sec:multidim}} \\
\symgroup{Bayesian view of the trial factor}
$B_{10}$ & Bayes factor, $p(x\,|\,H_1)/p(x\,|\,H_0)$: ratio of the marginal likelihoods of the two hypotheses. & \symref{Sec.~\ref{sec:bayes}} \\
$\pi(\theta)$ & Prior density of the scan parameter, flat ($1/\Delta$) in Section~\ref{sec:bayes}; in the e-value framework, the weighting density over the scan range, fixed in advance, that defines a mixture e-value. & \symref{Sec.~\ref{sec:bayes}, Eq.~\eqref{eq:mixture_evalue}} \\
$\Delta$ & Prior range of a parameter: the scan range $U-L$ for $\theta$, or the prior range of $\mu$. & \symref{Sec.~\ref{sec:bayes}} \\
$\sigma_\theta$ & Width of the likelihood peak in the scan parameter, i.e.\ the precision with which the data localise the excess (written $\sigma$ in Eq.~\eqref{eq:trotta}). & \symref{Eq.~\eqref{eq:trotta}, Sec.~\ref{sec:bayes}} \\
$\sigma_\mu$ & Measurement uncertainty of the signal strength, with $\hat\mu\approx\sqrt{q}\,\sigma_\mu$; Bayer and Seljak set the width of the prior on $\mu$ from it. & \symref{Sec.~\ref{sec:bayes}} \\
$N_B$ & Bayer--Seljak trial factor, $(U-L)/(\sqrt{2\pi}\,\sigma_\theta)$: the number of peak widths in the scan range. & \symref{Sec.~\ref{sec:bayes}, Eq.~\eqref{eq:bs_exp}} \\
$q_S$ & Test statistic whose global p-value is given by the Bayer--Seljak formula, $q_S=Z^2-2\ln N_B+\ln(2\pi Z^2)$. & \symref{Sec.~\ref{sec:bayes}, Eq.~\eqref{eq:bs_exp}} \\
\symgroup{Multiple searches}
$N$ & Number of tests (hypotheses, channels or grid points). & \symref{Eq.~\eqref{eq:bonferroni}} \\
$p_{\min}$ & Smallest of the $N$ observed p-values. & \symref{Eq.~\eqref{eq:bonferroni}} \\
$p_{(1)}\le\dots\le p_{(N)}$ & Observed p-values sorted in increasing order. & \symref{Eq.~\eqref{eq:bh}} \\
$k$ & Rank of a p-value in the Benjamini--Hochberg procedure; index of a channel. & \symref{Sec.~\ref{sec:multipletesting}, Eq.~\eqref{eq:bh}} \\
$R$ & Number of rejected null hypotheses (discoveries). & \symref{Sec.~\ref{sec:bh}} \\
$V$ & Number of false rejections, i.e.\ of true null hypotheses that were rejected. & \symref{Sec.~\ref{sec:bh}} \\
$c(N)$ & Harmonic factor $\sum_{i=1}^N 1/i\approx\ln N$ of the Benjamini--Yekutieli procedure. & \symref{Sec.~\ref{sec:bh}} \\
$\epsilon_s$, $\epsilon_b$ & Signal efficiency, $P(\text{tag}\,|\,\text{signal})$, and background mis-identification rate, $P(\text{tag}\,|\,\text{background})$, of a tagger. & \symref{Sec.~\ref{sec:multipletesting}} \\
$N_s/N_b$ & Ratio of signal to background objects in a sample. & \symref{Sec.~\ref{sec:multipletesting}} \\
\symgroup{E-values and anytime-valid inference}
$E$ & E-value: non-negative random variable with $\mathbb{E}[E\,|\,H_0]\le1$. & \symref{Eq.~\eqref{eq:evalue_def}} \\
$E_i$; $w_i$ & E-values and the non-negative weights, with $\sum_i w_i=1$, of their weighted average. & \symref{Sec.~\ref{sec:evalues}} \\
$(M_t)_{t\ge1}$ & E-process: sequence of e-values updated as data arrive; the capital of a gambler betting against $H_0$. & \symref{Eq.~\eqref{eq:ville}} \\
$t$; $\tau$ & Time (batch) index; stopping time. & \symref{Eq.~\eqref{eq:ville}} \\
$p_t$ & Anytime-valid p-value, $\min\big(1,1/\max_{s\le t}M_s\big)$. & \symref{Sec.~\ref{sec:evalues}} \\
$E_\theta$; $E_\pi$ & E-value for the background-only hypothesis against a signal at $\theta$; mixture e-value $\int E_\theta\,\pi(\theta)\,d\theta$. & \symref{Eq.~\eqref{eq:mixture_evalue}} \\
$\mathbb{E}[\log E\,|\,H_1]$ & Expected growth rate of an e-value under the alternative hypothesis. & \symref{Sec.~\ref{sec:evalues}} \\
\symgroup{Machine learning}
$t$; $t_\mathrm{obs}$ & Value of the test statistic in a pseudo-experiment; observed value. & \symref{Eq.~\eqref{eq:nplm_pvalue}} \\
$\mathcal{T}_R$ & Set of test-statistic values from pseudo-experiments generated under the reference hypothesis. & \symref{Eq.~\eqref{eq:nplm_pvalue}} \\
$\sigma$ & Kernel width (bandwidth) of a kernel-based test, an explicit resolution parameter. & \symref{Secs.~\ref{sec:resolution}, \ref{sec:multiple_ml}} \\
$k$ & Number of folds in $k$-folding. & \symref{Sec.~\ref{subsec:splitting}} \\
\symgroup{Illustrative experiments}
$x$ & Observable, $x\in[0,10]$. & \symref{Eq.~\eqref{eq:toy_model}} \\
$m$ & Mass of the resonance, i.e.\ the location of the Gaussian signal: the scan parameter $\theta$ of the experiments. Quantities defined for $\theta$ are used with $\theta\to m$, e.g.\ $\hat m$, $q(m)$, $Z(m)$ and $\sigma_m$. & \symref{Eq.~\eqref{eq:toy_model}, Sec.~\ref{subsec:exp1}} \\
$\mu$; $\hat\mu$ & Signal fraction and its maximum-likelihood estimate & \symref{Sec.~\ref{sec:bayes}, Eq.~\eqref{eq:toy_model}, Sec.~\ref{sec:exp_decay}} \\
$f(x\,|\,\mu,m,\nu;\;\omega)$ & Probability density of the toy model: exponential background plus Gaussian signal. & \symref{Eq.~\eqref{eq:toy_model}} \\
$g(x\,|\,m,\omega)$ & Gaussian signal template of mean $m$ and width $\omega$, normalised on $[0,10]$. & \symref{Eq.~\eqref{eq:toy_model}} \\
$\nu$ & Decay rate (slope) of the exponential background. & \symref{Eq.~\eqref{eq:toy_model}} \\
$\omega$; $\omega_i$ & Width of the signal template; template width (detector resolution) of channel $i$. & \symref{Secs.~\ref{subsec:exp1}, \ref{sec:exp_decay}} \\
$\Gamma$ & Intrinsic width of the resonance. & \symref{Sec.~\ref{sec:exp_decay}} \\
$\mathcal{L}(\mu,\nu;m)$ & Likelihood of the toy model. & \symref{Eq.~\eqref{eq:qmu}} \\
$\hat{\hat\nu}$ & Maximum-likelihood estimate of $\nu$ under the background-only hypothesis. & \symref{Eq.~\eqref{eq:qmu}} \\
$\hat\mu_m$, $\hat\nu_m$ & Joint maximum-likelihood estimates of $\mu$ and $\nu$ at fixed $m$. & \symref{Eq.~\eqref{eq:qmu}} \\
$N_\mathrm{ev}$ & Number of events per pseudo-dataset. & \symref{Sec.~\ref{subsec:exp1}} \\
$N_\mathrm{grid}$ & Number of equally spaced mass points in the scan. & \symref{Sec.~\ref{sec:exp_tf}} \\
$N_\mathrm{toys}$ & Number of pseudo-experiments. & \symref{Table~\ref{tab:trial_factors}} \\
$N_\mathrm{ch}$ & Number of channels or event classes. & \symref{Sec.~\ref{sec:exp_decay}} \\
$N_0$ & Number of true null hypotheses (background-only channels). & \symref{Sec.~\ref{sec:exp_decay}} \\
$p_\mathrm{GV}$ & Global p-value of a channel, corrected with the Gross--Vitells procedure for the scan over the mass. & \symref{Table~\ref{tab:channels}} \\
$q_\mathrm{comb}(m)$ & Test statistic of the combined likelihood of all channels, with a common mass. & \symref{Sec.~\ref{sec:exp_decay}} \\
$\kappa(c)$ & Prefactor of the narrow-template approximation of the trial factor. & \symref{Eq.~\eqref{eq:tf_analytic}} \\
$n_i$; $b_i$ & Observed number of events and expected number of background events in bin $i$. & \symref{App.~\ref{app:width}} \\
$g_i(m)$; $g(m)$ & Expected number of signal events in bin $i$ for $\mu=1$ and a template at mass $m$; vector of these numbers. & \symref{App.~\ref{app:width}} \\
$b(x)$ & Background density. & \symref{App.~\ref{app:width}} \\
$u(m)$ & Score: derivative of $\ln\mathcal{L}$ with respect to $\mu$ at $\mu=0$. & \symref{App.~\ref{app:width}} \\
$I(m)$ & Fisher information for $\mu$ at $\mu=0$, equal to the variance of the score. & \symref{App.~\ref{app:width}} \\
$\langle v,v'\rangle$; $\lVert v\rVert$ & Inner product $\sum_i v_i v'_i/b_i$ of two vectors of bin contents; associated norm. & \symref{App.~\ref{app:width}} \\
$\tilde g(m)$ & Part of the template $g(m)$ that the fit of the background cannot mimic. & \symref{Eq.~\eqref{eq:rho_binned}} \\
$A$; $\delta A$, $\delta\nu$ & Background normalisation; small variations of the normalisation and of the slope. & \symref{App.~\ref{app:width}} \\
$\bar m$; $\Delta x$ & Mean mass $(m+m')/2$; bin width. & \symref{App.~\ref{app:width}} \\
\end{longtable}
}

{\small
\renewcommand{\arraystretch}{1.15}
\begin{longtable}{@{}>{\raggedright\arraybackslash}p{0.18\textwidth}>{\raggedright\arraybackslash}p{0.78\textwidth}@{}}
\caption{Abbreviations used in the paper.}\label{tab:abbreviations}\\
\toprule
Abbreviation & Meaning \\
\midrule
\endfirsthead
\multicolumn{2}{@{}l}{}\\
\toprule
Abbreviation & Meaning \\
\midrule
\endhead
\midrule
\multicolumn{2}{r@{}}{}\\
\endfoot
\bottomrule
\endlastfoot
BDT & Boosted decision tree \\
BH & Benjamini--Hochberg (procedure) \\
BSM & Beyond the Standard Model \\
CLs & Modified frequentist method for exclusion limits, based on the ratio of the confidence levels of the signal-plus-background and background-only hypotheses \\
CWoLa & Classification without labels \\
e-BH & E-value analogue of the Benjamini--Hochberg procedure \\
FDR & False discovery rate \\
FWER & Family-wise error rate \\
GOF & Goodness of fit \\
GV & Gross--Vitells (procedure) \\
GWAS & Genome-wide association study \\
HEP & High-energy physics \\
HERA & Hadron--Elektron--Ringanlage (electron--proton collider at DESY) \\
HH & Higgs boson pair production \\
LEE & Look-elsewhere effect \\
LEP & Large Electron--Positron collider \\
LHC & Large Hadron Collider \\
LLP & Long-lived particle \\
ML & Machine learning \\
MLP & Multilayer perceptron \\
MMD & Maximum mean discrepancy \\
MUSiC & Model Unspecific Search in CMS \\
NPLM & New Physics Learning Machine \\
PoSI & Post-selection inference (in the sense of Berk et al.) \\
SM & Standard Model \\
SNP & Single-nucleotide polymorphism \\
SUSY & Supersymmetry \\
\end{longtable}
}

\bibliographystyle{ieeetr}
\bibliography{bibliography}


\end{document}